\documentclass[twocolumn]{aastex631}

\definecolor{red}{RGB}{190,0,0}

\DeclareRobustCommand{\ion}[2]{%
\relax\ifmmode
\ifx\testbx\f@series
{\mathbf{#1\,\mathsc{#2}}}\else
{\mathrm{#1\,\mathsc{#2}}}\fi
\else\textup{#1\,{\mdseries\textsc{#2}}}%
\fi}

\usepackage[flushleft]{threeparttable}

\begin{document}

\title{ODIN: Searching for LyC emission from Lyman-$\alpha$ emitters at $z=4.5$ in the E-COSMOS and XMM-LSS fields}

\author[0009-0007-1810-5117]{Eunsuk Seo}
\affiliation{Department of Astronomy and Space Science, Chungnam National University, Daejeon 34134, Republic of Korea}

\author[0000-0002-4362-4070]{Hyunmi Song}
\affiliation{Department of Astronomy and Space Science, Chungnam National University, Daejeon 34134, Republic of Korea}

\author[0000-0002-4902-0075]{Lucia Guaita}
\affiliation{Universidad Andres Bello, Facultad de Ciencias Exactas, Departamento de Fisica y Astronomia, Instituto de Astrofisica, Fernandez Concha 700, Las Condes, Santiago RM, Chile}
\affiliation{Millennium Nucleus for Galaxies (MINGAL)}

\author[0000-0003-3004-9596]{Kyoung-Soo Lee}
\affiliation{Department of Physics and Astronomy, Purdue University, 525 Northwestern Avenue, West Lafayette, IN 47907, USA}

\author[0000-0003-1530-8713]{Eric Gawiser}
\affiliation{Department of Physics and Astronomy, Rutgers, the State University of New Jersey, Piscataway, NJ 08854, USA}

\author[0000-0002-1328-0211]{Robin Ciardullo}
\affiliation{Department of Astronomy \& Astrophysics, The Pennsylvania
State University, University Park, PA 16802, USA}
\affiliation{Institute for Gravitation and the Cosmos, The Pennsylvania State University, University Park, PA 16802, USA}

\author[0000-0002-4928-4003]{Arjun Dey}
\affiliation{NSF's National Optical-Infrared Astronomy Research Laboratory, 950 N. Cherry Avenue, Tucson, AZ 85719, USA}

\author[0000-0002-0112-5900]{Seok-Jun Chang}
\affiliation{Max-Planck-Institut f\"{u}r Astrophysik, Karl-Schwarzschild-Stra$\beta$e 1, 85748 Garching b. M\"{u}nchen, Germany}

\author[0000-0002-9811-2443]{Nicole Firestone}
\affiliation{Department of Physics and Astronomy, Rutgers, the State University of New Jersey, Piscataway, NJ 08854, USA}

\author[0000-0001-8221-8406]{Stephen Gwyn}
\affiliation{Herzberg Astronomy and Astrophysics Research Centre, National Research Council of Canada, Victoria, British Columbia, Canada}

\author[0000-0003-3428-7612]{Ho Seong Hwang}
\affiliation{Astronomy Program, Department of Physics and Astronomy, Seoul National University, 1 Gwanak-ro, Gwanak-gu, Seoul 08826, Republic of Korea}
\affiliation{SNU Astronomy Research Center, Seoul National University, 1 Gwanak-ro, Gwanak-gu, Seoul 08826, Republic of Korea}

\author[0000-0001-9991-8222]{Sungryong Hong}
\affiliation{Korea Astronomy and Space Science Institute, 776 Daedeokdae-ro, Yuseong-gu, Daejeon 34055, Republic of Korea}

\author[0009-0003-9748-4194]{Sang Hyeok Im}
\affiliation{Astronomy Program, Department of Physics and Astronomy, Seoul National University, 1 Gwanak-ro, Gwanak-gu, Seoul 08826, Republic of Korea}
\affiliation{Korea Institute for Advanced Study,
85 Hoegi-ro, Dongdaemun-gu, Seoul 02455, Republic of Korea}

\author[0000-0002-2770-808X]{Woong-Seob Jeong}
\affiliation{Korea Astronomy and Space Science Institute, 776 Daedeokdae-ro, Yuseong-gu, Daejeon 34055, Republic of Korea}

\author[0000-0002-6810-1778]{Jaehyun Lee}
\affiliation{Korea Astronomy and Space Science Institute, 776 Daedeokdae-ro, Yuseong-gu, Daejeon 34055, Republic of Korea}

\author[0000-0001-5342-8906]{Seong-Kook Lee}
\affiliation{Department of Physics and Astronomy, Seoul National University, 1 Gwanak-ro, Gwanak-gu, Seoul 08826, Republic of Korea}
\affiliation{SNU Astronomy Research Center, Seoul National University, 1 Gwanak-ro, Gwanak-gu, Seoul 08826, Republic of Korea}

\author[0000-0001-9521-6397]{Chanbom Park}
\affiliation{Korea Institute for Advanced Study, 85 Hoegi-ro, Dongdaemun-gu, Seoul 02455, Republic of Korea}

\author[0000-0002-9176-7252]{Vandana Ramakrishnan}
\affiliation{Department of Physics and Astronomy, Purdue University, 525 Northwestern Avenue, West Lafayette, IN 47907, USA}

\author[0000-0002-7712-7857]{Marcin Sawicki}
\affiliation{Institute for Computational Astrophysics and Department of Astronomy \& Physics, Saint Mary's University, 923 Robie Street, Halifax, Nova Scotia, B3H 3C3, Canada}

\author[0000-0003-3078-2763]{Yujin Yang}
\affiliation{Korea Astronomy and Space Science Institute, 776 Daedeokdae-ro, Yuseong-gu, Daejeon 34055, Republic of Korea}

\author[0000-0001-6047-8469]{Ann Zabludoff}
\affiliation{Steward Observatory, University of Arizona, 933 North Cherry Avenue, Tucson, AZ 85721, USA}

\correspondingauthor{Hyunmi Song}
\email{hmsong@cnu.ac.kr}

\begin{abstract}
We investigated Lyman-continuum (LyC) emission from Lyman-$\alpha$ emitters (LAEs) at $z=4.5$, identified in the One-hundred-deg$^2$ DECam Imaging in Narrowbands (ODIN) survey.
Of the 7,498 LAEs (4,101 in COSMOS and 3,397 in XMM-LSS), we excluded LAEs that are either likely low-z objects or contaminated by neighboring sources.
Additional background modeling process with thorough quality assessments leaves a final sample of 851 galaxies.
We then performed forced photometry on $u/u^*$-band images from the CFHT large area $u$-band deep survey (CLAUDS) to measure their LyC fluxes.
This represents the largest sample of $z=4.5$ LAEs searched for such a purpose.
Within this sample, we identified 12 `gold' and 39 `silver' LyC-emitting candidates, with LyC fluxes detected of $>3\sigma$ and between $2\sigma$ and $3\sigma$, respectively, in the range of 5.16--55.29 nJy.
No LyC signal is detected in the weighted mean stack of the final sample ($0.20 \pm 0.37$ nJy).
Given the UVC magnitudes of LAEs in our sample, the expected LyC emission is likely below the detection limit even when stacking the full sample of ODIN LAEs.
Nevertheless, having a large sample of LAEs remains valuable for identifying individual LyC leaker candidates.
Among the gold and silver candidates, the LyC flux appears to correlate positively with UVC flux and negatively with Ly$\alpha$ equivalent width, although the correlations are weak.
A larger sample of LyC leakers will allow a more robust confirmation of these trends and provide better insights into their physical origins.
\end{abstract}

\section{Introduction}\label{sec:intro}

The intergalactic medium (IGM) of the present universe remains ionized.
This state has persisted since the Epoch of Reionization (EoR) 
\citep[$6 \lesssim z \lesssim 14$,][]{Fan2006}, when radiation shortward of 912~\AA, otherwise known as the Lyman continuum (LyC), ionized most of the universe's neutral hydrogen.
This radiation is believed to be produced by star-forming galaxies (SFGs) and active galactic nuclei (AGN) \citep{Fan2006,Bouwens2007,Ouchi2009a,Ouchi2009b,Jia2011,Georgakakis2015,Giallongo2015,Wise2019}.
While the dominant high-redshift source remains uncertain, SFGs, which are more numerous than AGN, 
are likely to play a crucial role in cosmic reionization \citep{Pawlik2009,Vanzella2012,Mitra2015,Giallongo2015,Robertson2015,Madau2015,Feng2016}.
Thus, it is essential to evaluate the ionizing photon budget emitted by individual SFGs and their escape fraction.

Owing to the severe IGM attenuation and contamination from foreground sources for high-$z$ LyC emission, many studies have focused on low-redshift analogs rather than on LyC leakers from the epoch of reionization.
Green pea galaxies, which are identified as low-$z$ analogues of high-$z$ SFGs \citep{Cardamone2009}, exhibit a high chance of leaking LyC \citep[e.g., four out of nine,][]{Izotov2021} with the LyC relative escape fraction ($f_{\rm esc,rel}$)\footnote{It is defined in a relative sense compared to the UV continuum (UVC) escape fraction (c.f., absolute escape fraction, $f_{\rm esc,abs}$), see the Equation \ref{eq:fescrel}}. being as high as $\sim73\%$ \citep{Izotov2016a,Izotov2016b,Izotov2018a,Izotov2018b}.
Recently, the Low-Redshift Lyman Continuum Survey \citep[LzLCS,][]{Flury2022a} used HST/COS to identify 35 $z=0.2$--0.4 LyC emitters among 66 SFGs, of which absolute escape fractions can be as high as $f_{\rm esc,abs} = 0.5$.
However, the escape fractions in the LzLCS sample span a wide range, and similarly, many local starburst galaxies have been found to exhibit relatively low $f_{\rm esc}$ values \citep[e.g.,][]{Leitherer1995,Grimes2009}.

Although the IGM attenuation increases with redshift, observations of individual LyC emitters remain feasible at high redshifts.
Those at one of the highest redshifts explored so far (i.e., $3.21<z<4.0$) include the SFGs, Ion 1 \citep{Vanzella2010b,Vanzella2012,Vanzella2015,Vanzella2020,Ji2020}, Ion 2 \citep{Vanzella2015,Vanzella2016,deBarros2016,Vanzella2020} and Ion 3 \citep{Vanzella2018}, all of which have high relative LyC escape fractions 32--64\%.
Objects with even higher (relative or absolute) escape fractions (75 to 100\%) include D3 \citep{Shapley2006}, MD5b \citep{Mostardi2015}, the Sunburst arc \citep[e.g.,][]{Rivera-Thorsen2017,Rivera-Thorsen2019}, and J1316-2614 \citep{Marques-Chaves2022} at $z=$ 2.37--3.61.
The relatively high escape fractions of high-$z$ galaxies, despite attenuation in the IGM, raise the question of selection biases in the sightline and/or galaxy sample.

The challenges in directly measuring high-$z$ LyC emission and estimating its escape fraction can be partly mitigated by using indirect probes such as Ly$\alpha$ spectral line profile shape, UV continuum slope, Mg II line ratios, C IV, and H$\beta$ EW \citep[e.g.,][]{Izotov2018b,Izotov2020,Chisholm2020,Flury2022b,Chang2024}.
However, these indirect indicators ultimately need to be validated against direct measurements, highlighting the importance of securing direct detections.
One effective approach is to obtain direct measurements from a large, uniformly selected sample of galaxies, stacking their fluxes.
Stacking objects not only enhances a detection's significance by improving its signal-to-noise (S/N) ratio, but it also provides a more reliable estimate of the average properties of a given population.
There are a few studies that encompass more than a hundred of galaxies at high redshifts.
These include the stacking of $\sim$600 star-forming dwarf galaxies at $z\sim1$ with $f_{\rm esc,abs}<2.1\%$ \citep{Rutkowski2016}, 
588 H$\alpha$- and 160 Ly$\alpha$-emitting galaxies (LAEs) at $z\sim2$ with $f_{\rm esc,abs}<2.8(6.4)\%$ \citep{Matthee2017}, 124 SFGs at $z\sim3$ with $f_{\rm esc,abs}=9\pm1\%$ \citep{Steidel2018},
148 SFGs at $z\sim3.5$ with $f_{\rm esc,abs}=7\pm2\%$ \citep{Begley2022},
102 Lyman break galaxies (LBGs) at $z=$3.4--4.5 with $f_{\rm esc,abs}<5$--20\% \citep{Vanzella2010b},
201 SFGs at $z=$3.5--4.3 with the LyC-to-UVC flux ratio ($F_{\rm LyC}/F_{\rm UVC}$) of $\sim$0.04 \citep{Marchi2018}.
Finally two integral field unit surveys have stacked LAEs: the Multi-Unit Spectroscopic Explorer (MUSE) instrument combined the spectra of 621 LAEs between $3.0 < z < 4.5$ and found an escape fraction of $f_{\rm esc,abs} \sim 12\%$ \citep{Kerutt2024}, and the Hobby Eberly Telescope Dark Energy Experiment (HETDEX) stacked 214 LAEs at $3.0 < z < 3.5$ and measured a LyC emission $\gtrsim 3\sigma$ confidence \citep{Davis2021}.

While the estimates on LyC escape fraction from these studies suggest a potential redshift evolution--for instance, higher escape fractions at higher redshifts-- definitive conclusions remain elusive due to differences in survey depth, sample selection, and dataset completeness.
The One-hundred-deg$^2$ DECam Imaging in Narrowband survey \citep[ODIN;][]{Lee2024} provides a unique opportunity to address these issues, as it offers large, uniformly selected samples of LAEs at three distinct redshifts, $z=2.4$, 3.1, and 4.5.
Leveraging this, our study aims to investigate LyC leakage of LAEs at $z=4.5$, where public imaging data covering the LyC range of galaxies at $z=4.5$ are already available.
This represents the highest redshift data for which such a systematic LyC analysis is feasible.
While previous works \citep[e.g.,][]{Vanzella2010b,Marchi2018,Kerutt2024} have reached similar redshifts, their sample includes only a handful of galaxies at $z\sim4.5$, whereas our sample is concentrated around this redshift.
The challenge imposed by the low IGM transmission of LyC photons from this redshift can be mitigated by targeting LAEs, as the escape of LyC and Ly$\alpha$ photons appear closely linked \citep[e.g.,][]{Micheva2017,Marchi2018,Steidel2018,Fletcher2019,Naidu2020,Naidu2022,Reddy2022,Flury2022a}.
As the survey is still ongoing, we utilized the early release of the LAE catalog in the COSMOS and XMM-LSS fields \citep{Firestone2024}, and will incorporate LAEs from the entire survey area in a subsequent work.

This paper is organized as follows. 
We describe our target LAEs in Section~\ref{sec:data}.
In Section~\ref{sec:method}, we outline the cleaning procedures to exclude possible contamination and detail the background modeling and associated sanity checks.
In Section~\ref{sec:results}, we present LyC flux estimates from individual objects and from stacks of the final sample.
We also provide a list of LyC leaker candidates with their photometric measurements.
An investigation with existing HST data is in Section~\ref{sec:hst} and the summary of this paper is in Section~\ref{sec:summary}.
Throughout this paper, we use AB magnitudes and physical distances, assuming a flat $\Lambda$CDM cosmology with $H_0=70~{\rm km\,s^{-1}\,Mpc^{-1}}$ and $\Omega_m=0.27$.

\begin{figure*}[ht!]
\includegraphics[width=\linewidth]{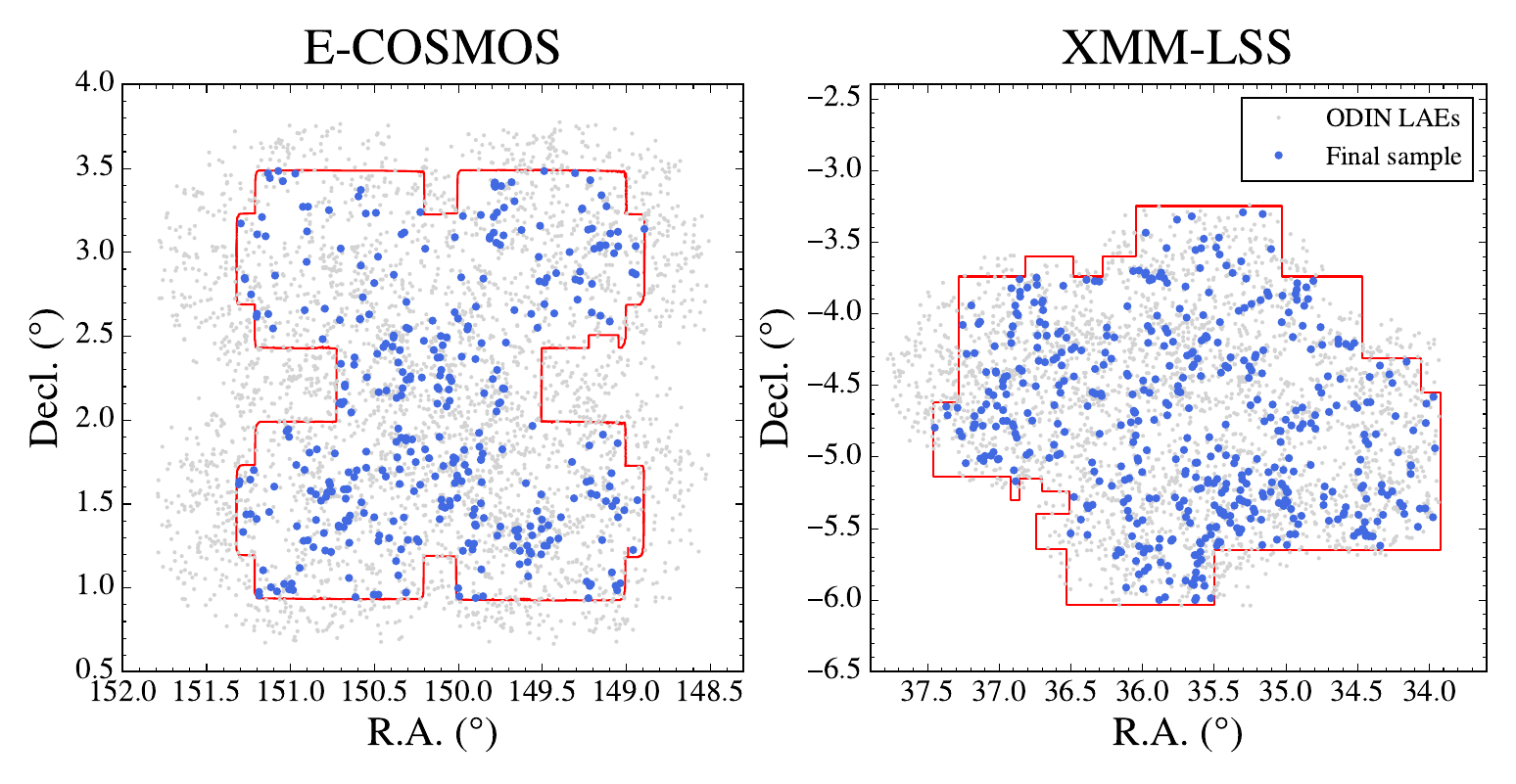}
\caption{The R.A. and Decl. distribution of the ODIN LAEs (gray dots) and of those in the final sample (blue circle, see Section \ref{sec:LyCflux}) in the E-COSMOS (left) and XMM-LSS (right) fields. Each field covers an area of $\sim$~9 deg$^2$. The CLAUDS footprint, where the $u/u^*$-band data is available, is denoted by red lines.
}
\label{fig:LAEskydistr}
\end{figure*}

\begin{figure*}[ht!]
\includegraphics[width=\linewidth]{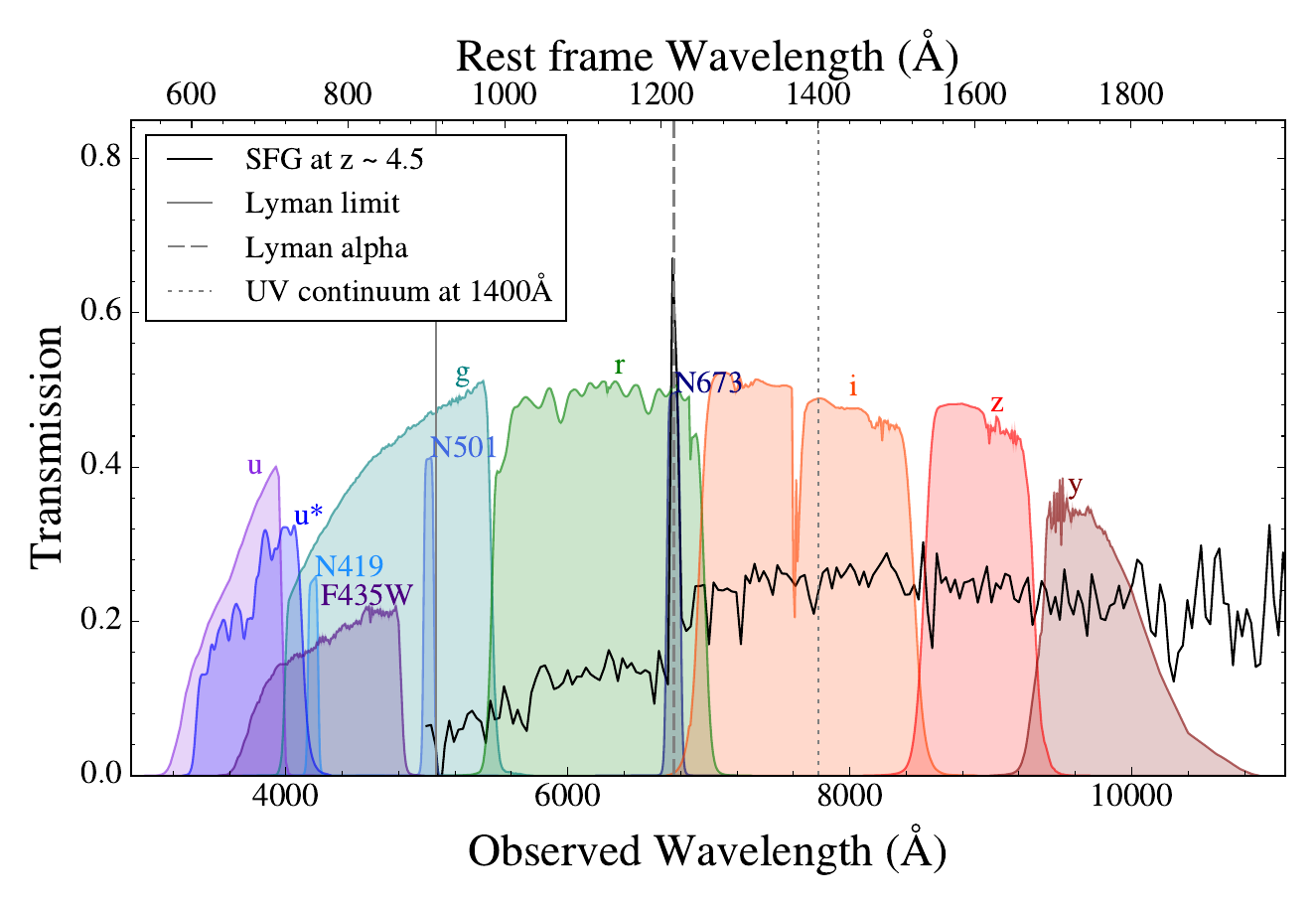}
\caption{Filter transmission for three narrowbands (N419, N501 and N673) and the eight broadbands ($u$, $u^*$ from CFHT/MegaCam, $g$, $r$, $i$, $z$ and $y$ from HSC, and F435W from HST) as a function of wavelength.
A composite spectrum of $z=3$ SFGs from \citet{Shapley2003}, shifted to $z=4.5$, is illustrated in black solid line.
The gray solid, dashed and dotted lines mark the rest-frame Lyman limit (912~\AA), Ly$\alpha$ (1216~\AA), and the UV continuum at 1400~\AA, respectively.}
\label{fig:filter-shapley}
\end{figure*}

\section{Data}\label{sec:data}
The ODIN survey has been conducted with the Dark Energy Camera on the Blanco 4m telescope at Cerro Tololo Inter-American Observatory (Prop. ID 2020B-0201; PI: K.-S. Lee).
The ODIN narrowband images, obtained with three custom-built narrowband filters (N419, N501, and N673) and combined with deeper broadband data from the Hyper Suprime-Cam Subaru Strategic Program \citep[HSC-SSP,][]{Aihara2019}, allow the selection of LAEs at $z=2.4$, 3.1, and 4.5.
With our current data, we can only measure rest-frame LyC using the the CFHT Large Area $u$-band Deep Survey \citep[CLAUDS,][]{Sawicki2019} $u/u^*$-band for LAEs at redshift $z=4.5$.
Specifically, the ODIN LAEs at $z=4.5$ are selected based on the flux density excess in N673 with respect to hybrid-weighted double-broadband continuum estimates using HSC $g$- and $i$-band images. 
Full details of the LAE selection method can be found in \citet{Firestone2024}.

A total of 4,101 and 3,397 $z=4.5$ LAEs are identified in the extended COSMOS (E-COSMOS) and XMM-LSS fields, respectively, each covering an area of $\sim$9 deg$^2$.
Figure~\ref{fig:LAEskydistr} shows the sky distribution of LAEs (gray dots).
The apparent gaps in the distribution correspond to regions masked due to saturated stars and CCD blooming (pixel oversaturation), which were removed during the starmasking step \citep{Firestone2024}. Starmasks were obtained from HSC-SSP \citep{Coupon2018}.
We detected the rest-frame LyC emission of ODIN LAEs using the CLAUDS $u$ (E-COSMOS) and $u^*$ (XMM-LSS) band data. 
The selection takes image quality into account: we removed cases where the depth is worse than the average or changes rapidly. 
Finally, the number of LAEs with good quality $u/u^*$-band data is  2,774 (E-COSMOS) and 2,989 (XMM-LSS).

Figure~\ref{fig:filter-shapley} shows the response curves of the filters which we use to construct the spectral energy distributions (SEDs) of LAEs: one from CLAUDS, five from HSC-SSP, three from ODIN.
The F435W filter of the Wide Field Camera 3 (WFC3) on the Hubble Space Telescope (HST) is included for reference, because the corresponding F435W images are used to measure the LyC emission of some LAEs (Section~\ref{sec:hst}).
The HST data used in this study are available at MAST (\dataset[10.17909/8s31-f778]{https://doi.org/10.17909/8s31-f778}, \dataset[10.17909/t9-pe8e-s980]{https://doi.org/10.17909/t9-pe8e-s980}).
The summary of the imaging data are presented in Table~\ref{tab:filter}.

The CFHT/$u$ and $u^{*}$ bands correspond to the rest-frame LyC emission at $\sim$~640 and 680 \r{A}, respectively, and, as illustrated in Figure~\ref{fig:filter-shapley}, their transmission cuts off well blueward of 912 \r{A} \citep[even considering the red leak around 5000 \r{A} in the $u^{*}$ filter; see Figure 2 in ][]{Iwata2022}.
Although the N419, N501, and F435W filters also sample LyC but closer to the Lyman limit than the $u/u^*$ band, the N419 and N501 imaging is shallower, and the F435W coverage is limited to subregions of the fields.
Therefore, the $u$ and $u^*$ filters are the only available option for probing LyC emission from $z=4.5$ galaxies without contamination from photons redward of the Lyman break.
That said, the mean IGM transmission at rest-frame wavelengths of 640--680~\AA\ from $z=4.5$ is very low ($\sim0.3\%$ based on \citealt{Inoue2014}), implying that the probability of detecting LyC emission in these bands is inherently low (see Section~\ref{sec:hst} and Figure \ref{fig:LyC-UVC}).
For the rest-frame UVC, we use the HSC/$i$-band data, enabling us to infer the systems' star-formation rates and therefore a mean LyC escape fraction.
Therefore, having access to both the $u/u^*$- and $i$-band photometric data is essential for our analyses.

Photometry was performed using a circular aperture of $1\arcsec$ radius.
Because a fixed aperture may underestimate the total flux, particularly for marginally extended sources, we apply an aperture correction derived from the growth curves of isolated point sources in each band \citep[see Figure~3 of][]{Firestone2024}.
The correction factors are approximately 0.79 and 0.86 in $u$ and $i$ bands, respectively.

\begin{deluxetable*}{cccccc}  
\tablecaption{Filter properties referred to in \citet{Lee2024,Sawicki2019,Aihara2019,Prichard2022,Wang2025}\label{tab:filter}}
\tablehead{
\colhead{Filter} & \colhead{$\lambda_{\rm c}$\tablenotemark{\footnotesize a}} & \colhead{$\lambda_{\rm c, rest}$\tablenotemark{\footnotesize b}} & \colhead{Filter width} & \colhead{Seeing\tablenotemark{\footnotesize c}} & \colhead{$5\sigma$ depth\tablenotemark{\footnotesize d}} 
\\
 & (\r{A}) & (\r{A}) & (\r{A}) & (arcsec) & (mag) \\
}
\startdata
N419&4193&762&75&1.1&25.5\\
N501&5014&912&76&0.9&25.7\\
N673&6750&1227&100&1.0&25.9\\
$u$&3538&643&868&0.92&27.1 (27.7)\\
$u^*$&3743&681&758&0.92&27.1 (27.7)\\
$g$&4811&875&1395&0.81&27.3\\
$r$&6223&1131&1503&0.74&26.9\\
$i$&7675&1395&1574&0.62&26.7\\
$z$&8908&1620&766&0.63&26.3\\
$y$&9785&1779&783&0.71&25.3\\
F435W&4297&781&900&0.1&28.0-29.8\\
\hline
\enddata
\tablenotetext{a}{The central wavelength of each filter}
\tablenotetext{b}{The rest-frame wavelength at $z=4.5$ that corresponds to the central wavelength of each filter in the observed frame}
\tablenotetext{c}{Averaged seeing (the full width half maximum of point spread function) over the entire survey area}
\tablenotetext{d}{Averaged depth over the entire survey area. For the $u$ and $u^*$, a small portion of the survey area has a deeper depth, whose value is provided in the parenthesis.}
\end{deluxetable*}

\section{Method}\label{sec:method}
\subsection{Sample cleaning} \label{sec:cleaning}
To ensure reliable $z=4.5$ LyC flux measurements, we first performed sample cleaning to mitigate contamination from foreground or neighbouring galaxies. 
This cleaning step is crucial, as the faint LyC signal at $z=4.5$ is highly susceptible to contamination from foreground and neighboring sources.
We applied a series of procedures to construct a clean sample.

\begin{figure*}
\includegraphics[width=\linewidth]{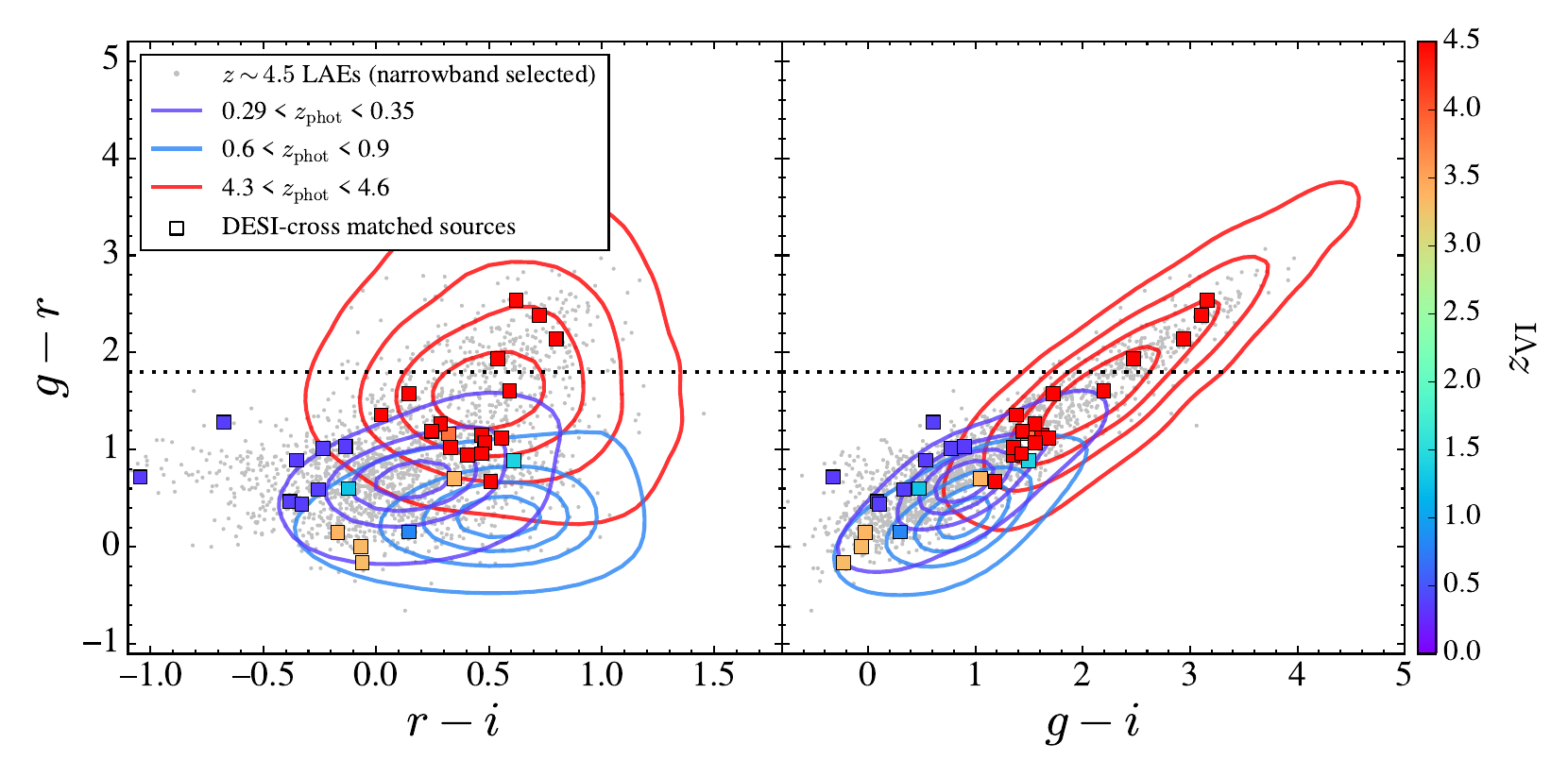}
\caption{Color-color diagrams of our $z\sim4.5$ LAE sample (gray dots) and sources from the COSMOS2025 catalog\citep{Shuntov2025} (colored contours). The colored squares correspond to LAEs with spectroscopic redshifts confirmed by DESI DR1, color-coded according to the scale on the right color bar. The horizontal dashed line at $g-r=1.8$ marks the color-cut criteria adopted to select a Clean sample of $z\sim4.5$ LAEs.}
\label{fig:colorcut}
\end{figure*}

We first rejected those that are likely foreground interlopers.
Although the LAE selection method employed by the ODIN survey shows high success rates in identifying galaxies at the targeted redshifts of the survey \textbf{(about 93 $\%$ at $z=2.4$, 96 $\%$ at $z=3.1$, and 92 $\%$ at $z=4.5$; \citealt{ODIN-DESI})}, even a small fraction of low-redshift contaminants could significantly affect our analysis.
We examined the distributions of $g-r$, $r-i$, and $g-i$ colors of galaxies in the COSMOS2025 catalog \citep{Shuntov2025}, grouped by their photometric redshifts to establish effective contaminant rejection criteria (Figure~\ref{fig:colorcut}).
We specifically focused on the low-$z$ groups known to contaminate $z=4.5$ Ly$\alpha$ searches, i.e., $z=0.81$ [OII] emitters and $z=0.35$ [OIII] emitters (see also \citealt{Firestone2024}).
Based on these distributions, we adopted a color cut of $g-r>1.8$ to efficiently exclude these low-$z$ interlopers.
We note that this criterion is applied only to those with $\geq 3\sigma$ detections in both $g$ and $r$ bands.
This conservative approach was chosen to prioritize contaminant rejection, even with the inherent risk of inadvertently excluding some bona fide LyC leakers.

To quantitatively assess the impact of this additional color selection, we cross-matched the LAE sample prior to applying the $g-r>1.8$ criterion with the publicly available DESI DR1 spectroscopic catalog (colored squares in Figure~\ref{fig:colorcut}).
Because DESI redshifts are automatically assigned by the pipeline and may occasionally be uncertain, we visually inspected all cross-matched spectra to confirm their reliability.
After this inspection, we find that no spectroscopically confirmed low-$z$ interlopers remain in the examined subsample after applying the $g-r>1.8$ cut.
Within this limited spectroscopic subsample, the cut removes 12 spectroscopically confirmed $z\sim4.5$ LAEs, leaving 4 objects and corresponding to a relative completeness of $\sim0.25$.
LyC measurements are highly sensitive to even rare foreground contaminants, which can introduce catastrophic bias in the inferred escape fractions.
In contrast, incompleteness primarily reduces statistical power.
We therefore deliberately adopt $g-r>1.8$ as a conservative criterion that prioritizes sample purity over completeness.
Although a slightly bluer threshold (e.g., $g-r>1.6$) may appear to separate $z\sim4.5$ LAEs from low-$z$ galaxies within the limited DESI-matched subsample, the broader COSMOS photometric-redshift distributions indicate that $g-r>1.8$ provides a cleaner separation with a larger safety margin against photometric scatter and photo-$z$ uncertainties.

We also argue that this color criterion is robust against LyC leakage bias.
At $z=4.5$, the $g$ band straddles the Lyman break and is more sensitive to Ly$\alpha$ forest absorption, which is the dominant factor causing the $g-r$ color to be extremely red for genuine high-$z$ LAEs.
While strong LyC leakage would blue the $g-r$ color by weakening the Lyman break, the dominant effect of Ly$\alpha$ forest absorption in the $g$ band ensures that most $z=4.5$ galaxies maintain a sufficiently red $g-r$ color to satisfy our $g-r>1.8$ cut\footnote{We acknowledge that a significant fraction (more than a half) of our LyC leaker candidates (see Section \ref{sec:LyCflux}) exhibit $g-r<1.8$, if we do not apply the condition that sources are detected in both the $g$ and $r$ bands with $\ge3\sigma$.
Crucially, we found that no clear correlation is observed between the LyC flux (or LyC-to-UVC flux ratio) and the $g-r$ color, which suggests that the bluer $g-r$ colors observed are not systematically driven by intrinsically higher LyC leakage, but rather by photometric scatter or noise.
Therefore, we conclude that the $g-r$ color selection did not introduce a significant bias against the strongest LyC leakers in our specific analysis.}.

Although photometry was performed using an aperture of 1\arcsec\ radius (2\arcsec\ diameter), which is approximately twice the seeing (FWHM of 1\arcsec), we conservatively excluded objects that have neighboring sources within six times the seeing in the N673, $i$-, $u/u^*$-, and $r$-band images.
The exclusion radius of six times the seeing is motivated by PSF considerations: for a point source, the majority of the flux is enclosed within approximately three times seeing, and requiring a centroid separation larger than twice this scale ensures negligible PSF overlap between the target and neighboring sources.
In addition, to account for the possibility of extended low-redshift contaminants, we inspected the photometric aperture region in the science images and excluded any object showing measurable flux overlap or extended emission encroaching into the aperture.
Objects affected by saturated stars had already been removed during the earlier cleaning stage.

Consequently, these cleaning procedures leave 1,819 (786 for E-COSMOS and 1,033 for XMM-LSS) LAEs.
We refer to this sample as the ``Clean'' sample.

\begin{figure*}
\includegraphics[width=\linewidth]{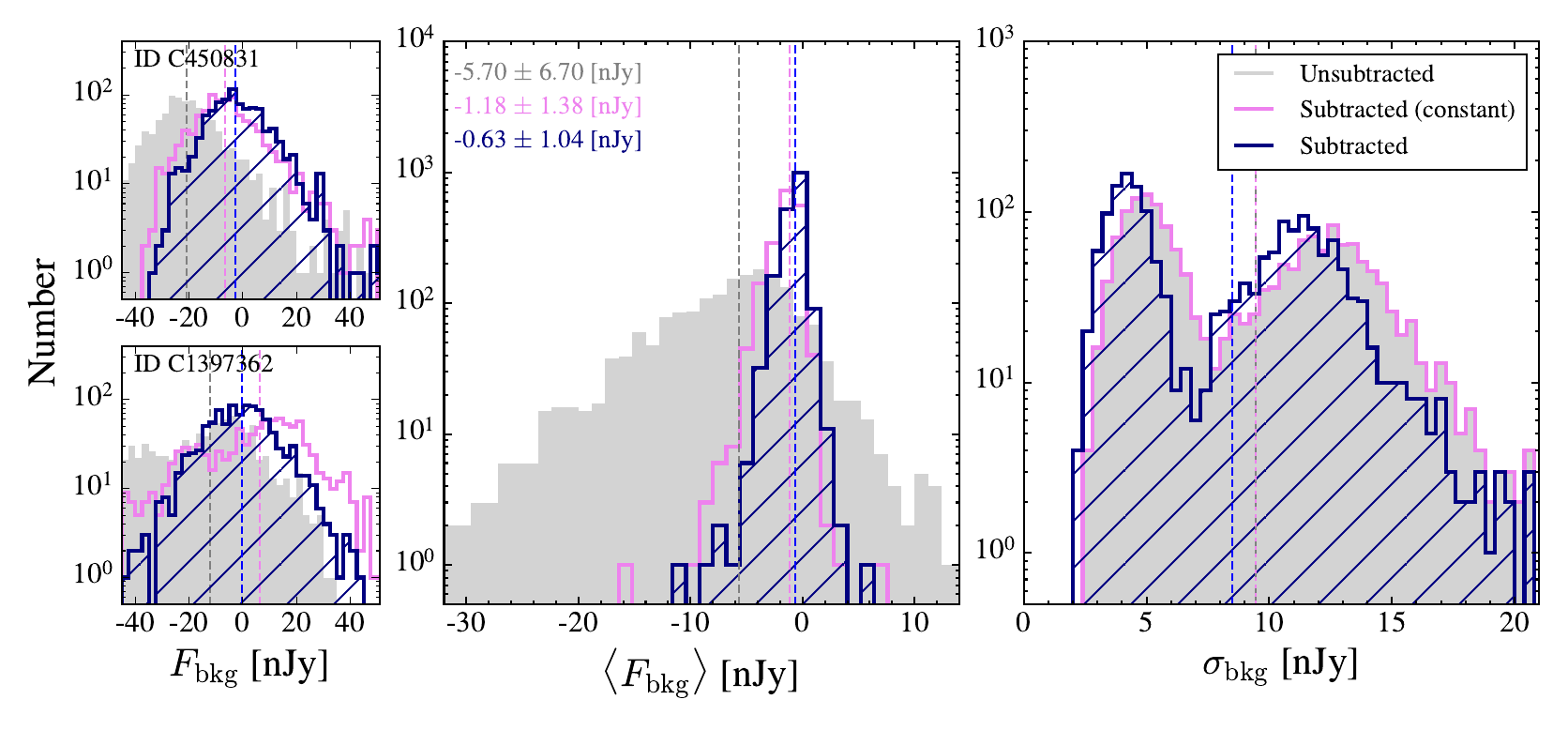}
\caption{The left panels show the flux distributions of background pixels free from sources ($F_{\rm bkg}$) in $u/u^*$-band cutouts (27$\arcsec$ on a side) around two example LAEs. 
The background fluxes are measured with a 1\arcsec-radius aperture at 1000 random locations within the cutout.
The distributions of background fluxes are shown before (gray filled histogram) and after background subtraction.
This subtraction models include a constant (pink open) and variable (blue hatched) background. 
The dashed vertical lines indicate the median of each distribution.
The blue hatched histogram appears narrower and more symmetric around zero.
The median ($\langle F_{\rm bkg} \rangle$) and standard deviation ($\sigma_{\rm bkg}$) of background fluxes are measured for each LAE in the Clean sample, and their distributions are shown in the middle and right panels, respectively. 
These distributions show that, before background subtraction, the background distribution is generally negatively skewed with large variation.
This issue is resolved by background subtraction, with a variable background model yielding the best result.}
\label{fig:bkgdist}
\end{figure*}

\subsection{Local background modeling} \label{sec:background}

Precise background modeling is critical for robust flux measurements, especially when detecting faint LyC signals.
The background subtraction applied before coadding images \citep[see][for more details]{Sawicki2019} mainly addresses global background trends and is often suboptimal for capturing small-scale, source-level variations.
Therefore, we performed additional background modeling tailored to our measurements.

We checked the background level in each $u/u^*$-band cutout (27\arcsec~= 100 pixels on a side) centered on an LAE candidate in the Clean sample by measuring background in $1\arcsec$-radius apertures at 1000 random locations within the cutout.
As examples, the background flux ($F_{\rm bkg}$) distributions for two LAEs are presented as gray histograms in the left panels in Figure~\ref{fig:bkgdist}.
The distributions are skewed toward negative values, indicating excessive background subtraction applied.
For each cutout, we measured the median and standard deviation of unmasked pixels, denoted $\langle F_{\rm bkg} \rangle$ and $\sigma_{\rm bkg}$, respectively. 
The distributions of these quantities from all LAEs are shown as the gray histograms in the middle and right panels.
The histogram of $\langle F_{\rm bkg} \rangle$ is skewed toward negative fluxes, indicating that the default background subtraction is likely to be overly aggressive in many cases.
The distribution of $\sigma_{\rm bkg}$ is double peaked, which arises from differences in the sky background variation between the two fields.

It is worth noting that the average $u/u^*$-band sky brightness is $\sim$9500 nJy in a 1\arcsec-radius aperture (22.70 ${\rm mag\,arcsec^{-2}}$)\footnote{https://www.cfht.hawaii.edu/Instruments/Imaging/MegaPrime/}.
The residual background noise remaining in the $u/u^*$-band data is on the order of several tens of nJy as seen in Figure \ref{fig:bkgdist}, which represents only a sub-percent level of the sky brightness.
While this level of residual noise is negligible for most science cases, it is critical for our analysis.
Even this small, sub-percent level of residual background noise can have a significant impact on the estimation of the LyC flux at high redshifts, given than the LyC flux we are expecting is on the order of $\sim0.1-10$ nJy \citep{Ji2020,Vanzella2018,Prichard2022}.

Our refined background modeling begins with aggressive masking to prevent contamination by any residual sources.
This requires adjusting the source detection parameters in \texttt{SExtractor} \citep{Bertin1996,Barbary2016} to generate an initial mask of detected sources, and subsequently manually masking additional source-contaminated pixels that were not included in the automated mask within each cutout.
We also masked the locations of LAEs (the central 1\arcsec-radius aperture of each cutout), even when there is no clear signal, to prevent any potential LyC flux from being included in the background modeling.

Various combinations of parameters are then explored within a Python routine \texttt{photutils.Background2D} to model the background.
The background level is determined in each box of a given size (\texttt{box\_size}) as the median of unmasked pixel intensities, resulting in a smoothed background image.
This image is once more median-filtered with a window of a given size (\texttt{filter\_size}). 
These two levels of median filtering serve to suppress local fluctuations of noise. 
The \texttt{box\_size} is set to 5, 10, 25, 50, and 100 pixels, and the resulting background images with corresponding resolutions of 20, 10, 4, 2, and 1 pixels, respectively.
The feasible values for \texttt{filter\_size} are odd integers between 1 and 19.
The background images are then resized to the original cutout size through interpolation.

In selecting the optimal parameter set of \texttt{box\_size} and \texttt{filter\_size}, our goal is to minimize excessive fine-tuning in background modeling while ensuring a smooth background through the subtraction process.
To mitigate fine-tuning, we evaluated a metric defined as the variation in the modeled background relative to that of the final background after subtraction; this metric tends to increase for smaller box and filter sizes.
We excluded parameter sets where the metric rapidly jumps to an unusually high value, as this sudden increase indicates an excessive and unstable level of fine-tuning compared to the smooth trend observed for other parameter sets.
Finally, we chose the parameter set that leads to the smoothest background after subtraction for each object individually.
Accordingly, optimal values vary across objects, but are typically 5 or 10 for \texttt{box\_size} and 3 or larger for \texttt{filter\_size} in most cases.

The blue histograms in the left panels of Figure~\ref{fig:bkgdist} show the resulting $F_{\rm bkg}$ for the two example cases.
The distributions are more symmetric around zero and narrower compared to those before the background level was readjusted  (i.e., the gray histograms).
The corresponding distributions of $\langle F_{\rm bkg} \rangle$ and $\sigma_{\rm bkg}$ values for all sources are denoted as blue histograms in the middle and right panels.
These demonstrate that the background modeling we additionally applied is generally successful.

As a limiting case, we consider a constant background model corresponding to \texttt{box\_size} of 100 and \texttt{filter\_size} of 1, which was initially our preferred model as it is simple and free from fine-tuning.
Subtracted background distributions are shown as pink histograms in Figure~\ref{fig:bkgdist}.
Although this model effectively resolves the over-subtraction issue, the improvement obtained by using a spatially-varying background model with the optimal \texttt{box\_size} and \texttt{filter\_size} values is non-negligible.
The middle panel in Figure~\ref{fig:bkgdist} illustrates that the median of the blue distribution is nearly half that of the pink distribution ($-0.63$ versus $-1.18$) and the blue distribution exhibits a marginally smaller dispersion (1.04 versus 1.38).
Furthermore, the background variation in each subtracted cutout appears smaller: the blue histogram is shifted toward left compared to the pink histogram in the right panel of Figure~\ref{fig:bkgdist}.

As a sanity check, we examined if there is any correlation between the LyC flux measured for an LAE and the model background at the location of the LyC measurement.
The presence of an anti-correlation indicates that a positive (negative) LyC flux is likely an artifact of background under-subtraction (over-subtraction).
However, no evidence of such an anti-correlation is found.
Therefore, we opt for the spatially varying background model over the constant one.

\begin{figure}
\includegraphics[width=\linewidth]{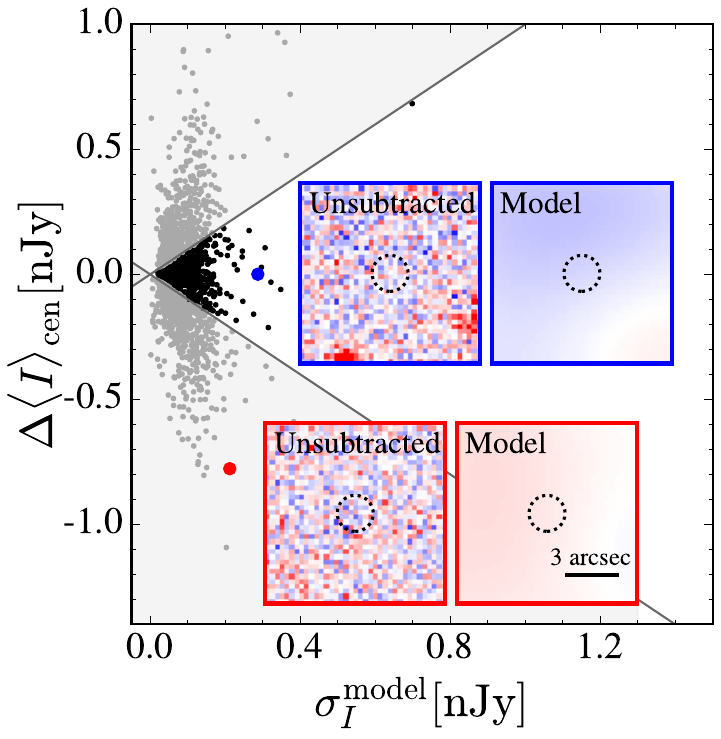}
\caption{The difference between the median intensities within a central 1$\arcsec$-radius aperture in the unsubtracted and model background cutouts ($\Delta \langle I\rangle_{\rm cen}$) is shown as a function of model variation ($\sigma_{I}^{\rm model}$). 
The gray solid lines denote the relations $\Delta \langle I\rangle_{\rm cen} = \pm\ \sigma_{I}^{\rm model}$.
The gray shaded region, where $|\Delta \langle I\rangle_{\rm cen}|>\sigma_{I}^{\rm model}$, indicates background modeling failure, given that the LyC signal we seek is expected to be rarely discernible from background noise, and those in that region are rejected.
Example unsubtracted and model cutouts that are kept (successful modeling) are showcased in the insets outlined in blue, while rejected examples (modeling failure) are outlined in red. 
The color scale of the background model images is four times narrower than that of the unsubtracted images.}
\label{fig:model1}
\end{figure}

\subsection{Background modeling quality control} \label{sec:validity}

Despite the careful background modeling described in Section~\ref{sec:background}, inaccuracies in the background estimation remain an issue, leading to, for example, negative LyC fluxes in some cutouts. 
Since the central region is masked during background modeling, such overestimation (underestimation) are likely caused by neighboring, clustered bright (faint) pixels.
As a result, the background-subtracted images exhibit artificially negative (positive) central fluxes, rather than representing the true signal.
Although such artificial signals are statistically insignificant in a single cutout, they could introduce biases when accumulated in the stack.

Since such background over- or under-estimation are effectively uncorrectable, our approach was to exclude these cases.
To identify them, we employed three different metrics, each developed to overcome limitations identified in the previous one.
While the first two metrics adopted a more direct and conservative approach to eliminate false signals, the last metric was designed to maximize the inclusion of potential LyC leakers at the expense of a higher risk of contamination.
As these three metrics are based on different purposes, they are complementary to one another and we therefore describe all three metrics here.

The first metric is the difference between the median intensity values of pixels within $1\arcsec$ from an LAE in the unsubtracted and model background images ($\Delta \langle I\rangle_{\rm cen}$).
Given that LyC emission from a $z=4.5$ galaxy is expected to be extremely faint \citep{Bassett2021}, its flux should not significantly deviate from background fluxes (i.e., $|\Delta \langle I\rangle_{\rm cen}|$ should not be large).
$\Delta \langle I\rangle_{\rm cen}$ is evaluated against the variation in the model background ($\sigma_{I}^{\rm model}$), which serves as a limit for acceptable differences.
Thus, we excluded cases of $|\Delta \langle I\rangle_{\rm cen}^{\rm unsub}|>\sigma_{I}^{\rm model}$, those in the upper and lower gray-shaded regions in Figure~\ref{fig:model1}.
For illustration, we present example cutouts of the unsubtracted and modeled background images for an excluded case (red dot and insets outlined in red) and a retained case (denoted in blue).
Based on this metric, we retain 689 LAEs and designate this group as the ``Direct'' sample.

\begin{figure}
\includegraphics[width=\linewidth]{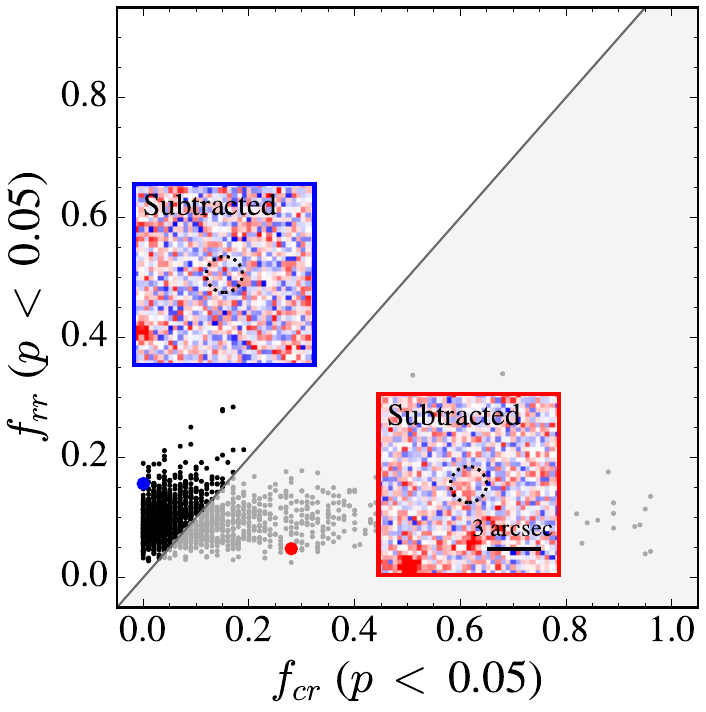}
\caption{A Kolmogorov-Smirnov test is performed to compare intensity distributions (within a 1\arcsec-radius aperture) in a subtracted cutout image.
We establish a central-random set by pairing the central distribution with 100 randomly-selected background distributions, and a random-random set consisting of 100 pairs of random background distributions.
The resulting fraction of central-random pairs with $p<0.05$ ($f_{cr}(p<0.05)$) is compared to the fraction for random-random pairs ($f_{rr}(p<0.05)$).
Cases where $f_{cr}(p<0.05)>f_{rr}(p<0.05)$ (gray shaded region) are rejected.
This rejection is based on the premise that the LyC signal we seek is expected to be rarely discernible from background noise; a significant deviation from background can imply an artificial signal from background modeling failure.
Example cases kept and rejected are shown in the insets outlined in blue and red, respectively.}
\label{fig:model2}
\end{figure}

\begin{figure}
\includegraphics[width=\linewidth]{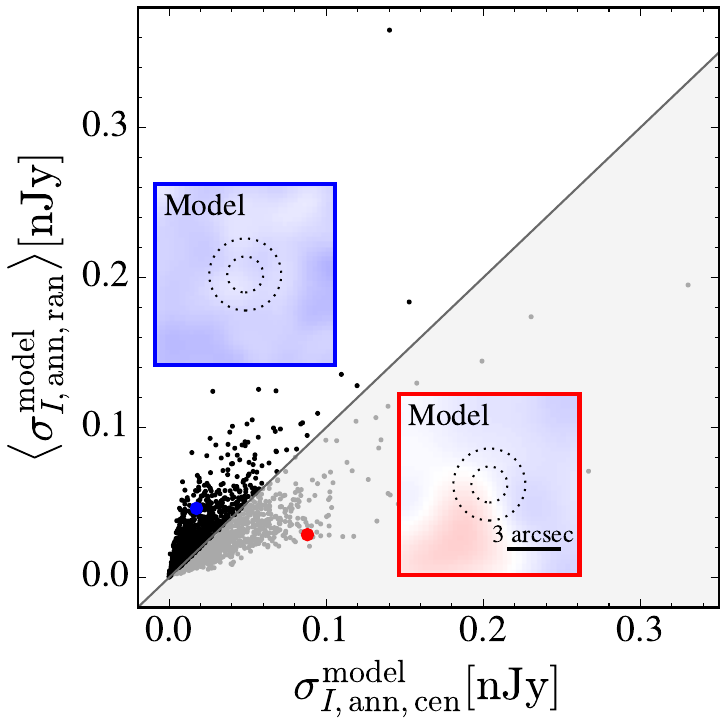}
\caption{The variation of the model background within a central 1--2\arcsec annulus ($\sigma_{I,{\rm ann,cen}}^{\rm model}$) is compared to the median of the variations at 100 random background locations ($\langle\sigma_{I,{\rm ann,ran}}^{\rm model}\rangle$).
Cases where $\sigma_{I,{\rm ann,cen}}^{\rm model}>\langle\sigma_{I,{\rm ann,ran}}^{\rm model}\rangle$ (gray shaded region) are rejected, as background modeling is likely to fail for the masked center because of large variation near the masked center.
This third metric is designed to minimize the risk of rejecting potential LyC leakers--a limitation of previous metrics (Figures~\ref{fig:model1} and \ref{fig:model2}) that compared intensities at the center.
By comparing modeled background around the center (in the annulus) to that at random locations, we assess the background model's reliability without interference from a potential central LyC signal.
The insets show example cases that are kept (outlined in blue) and rejected (red).}
\label{fig:model3}
\end{figure}

One limitation of the first metric is its tendency to overestimate the significance of $\Delta \langle I \rangle_{\rm cen}$.
Since $\sigma^{\rm model}_{I}$ is relatively small by construction--as it is derived from a smoothed image (see the insets of Figure~\ref{fig:model1})--the resulting normalized difference may appear more pronounced than it actually is.
To better evaluate the failure of background modeling at the center, we introduced a second metric that compares the central intensity distribution with those at randomly chosen source-free positions for each subtracted cutout.
Specifically, we performed the Kolmogorov-Smirnov (K-S) test on the pixel-wise intensity distributions within a 1\arcsec-radius aperture at the center and a randomly selected background location in a cutout (27\arcsec on a side), thereby quantifying the deviation between the two.
This test is repeated for 100 center-random pairs, and we calculated the fraction of pairs with a $p$-value below 0.05, denoted as $f_{cr}(p<0.05)$, which reflects the fraction of the center-random pairs exhibiting significantly different distributions.
For comparison, we computed $f_{rr}(p<0.05)$, the fraction of significantly different cases among 100 pairs of random-random locations.
If $f_{cr}(p<0.05)$ exceeds $f_{rr}(p<0.05)$, we interpreted this as an indication that the modeled central background is inconsistent with the general background, and we excluded such cases (see the lower-right region below the one-to-one line in Figure~\ref{fig:model2}).
As a result, 1,166 LAEs remain, yielding a larger sample than that obtained using the first metric.
This sample is referred to as the ``Direct-KS'' sample.

While these two metrics are effective in excluding cases with anomalous central fluxes based on the premise that the LyC signal we seek is expected to be rarely discernible from background noise, they may also reject potential LyC leakers.
Therefore, we introduce a third metric to distinguish between anomalous central fluxes caused by background modeling failures and those arising from genuine LyC emission, ensuring that only the former are excluded.
In this approach, we compared the distributions of modeled background intensities within annuli of 1--2\arcsec radii, sufficiently large compared to the PSF, around the center and at random locations.
If the intensity variation within the central annulus ($\sigma_{I,{\rm ann,cen}}^{\rm model}$) exceeds a typical value at random background locations ($\langle \sigma_{I,{\rm ann,ran}}^{\rm model}\rangle$), the background modeling at the center is likely unreliable. 
We therefore excluded such cases.
Here, $\langle\sigma_{I,{\rm ann,ran}}^{\rm model}\rangle$ is computed as the median of measurements at 100 different locations.
Figure~\ref{fig:model3} shows the distribution of $\langle\sigma_{I,{\rm ann,ran}}^{\rm model}\rangle$ versus $\sigma_{I,{\rm ann,cen}}^{\rm model}$ measured in individual cutouts.
LAEs with $\langle\sigma_{I,{\rm ann,ran}}^{\rm model}\rangle>\sigma_{I,{\rm ann,cen}}^{\rm model}$ are retained, resulting in a final count of 961 LAEs, which we designate as the ``Annulus'' sample.

\begin{figure}
\includegraphics[width=\linewidth]{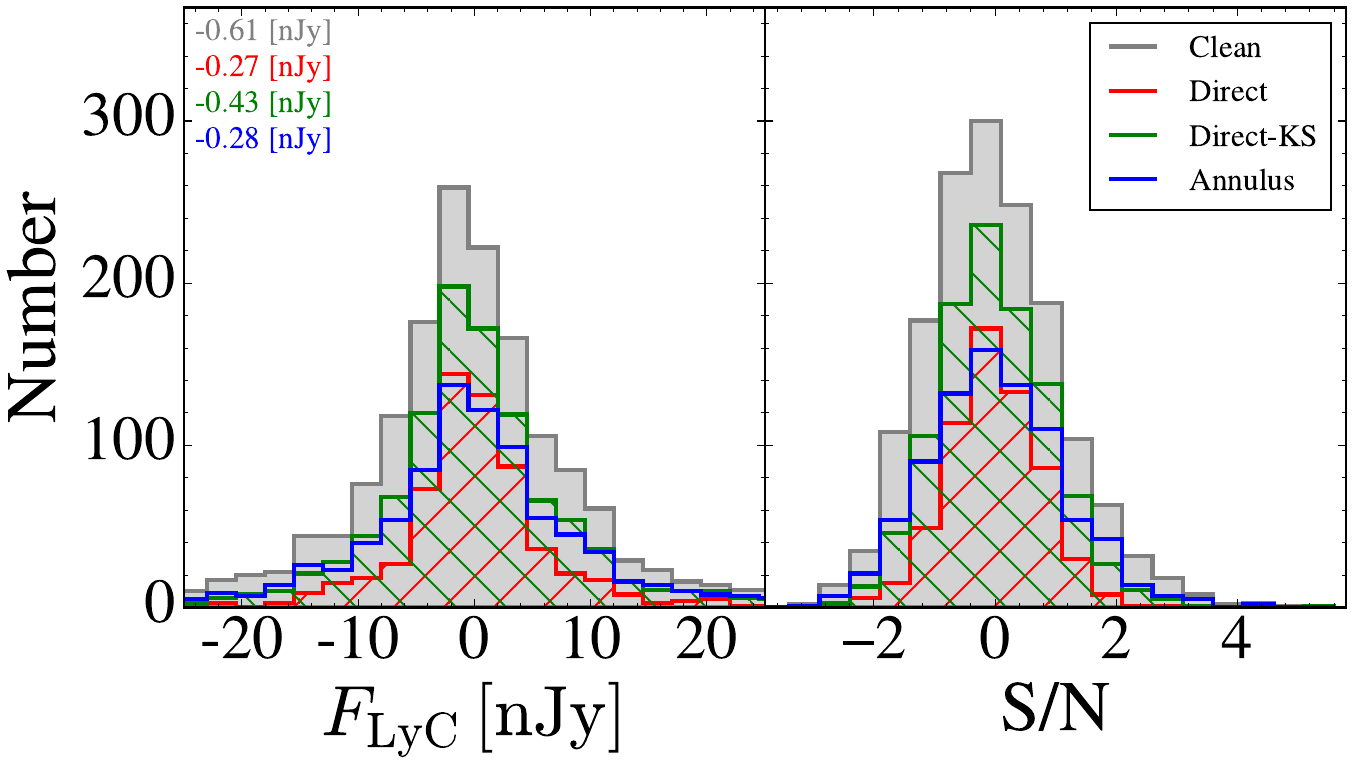}
\caption{The distributions of the LyC flux ($F_{\rm LyC}$, left panel) and signal-to-noise ratio (S/N, right panel) are shown for LAEs in the initial Clean sample (gray) and for those that survive three independent background modeling quality tests: the Direct test ($\Delta \langle I\rangle_{\rm cen}$ versus $\sigma_{I}^{\rm model}$), the Direct-KS test ($f_{cr}$ versus $f_{rr}$), and the Annulus test ($\sigma_{I,{\rm ann,cen}}^{\rm model}$ versus $\langle\sigma_{I,{\rm ann,ran}}^{\rm model}\rangle$) in red, green, and blue, respectively.
While the Clean sample exhibits a distribution skewed toward negative values, indicating that background oversubtraction is a general issue, this issue is substantially mitigated in the filtered samples.
The distributions of the Direct, Direct-KS, and Annulus samples are similar, though the Annulus sample contains the highest abundance of potential LyC leakers ($S/N>2$), as anticipated.}
\label{fig:modelhist}
\end{figure}

\section{Results and discussion} 
\label{sec:results}
\subsection{LyC signal of $z=4.5$ LAEs}\label{sec:LyCflux}

By applying three different metrics for background modeling quality control, we obtained three samples, consisting of 689, 1,166, and 961 LAEs, respectively.
As an additional sanity check, we visually inspected the galaxies' SED and their six broadband and one narrowband cutouts (as seen in Figures~\ref{fig:cutout-gold1} and \ref{fig:cutout-gold2}) selecting only those most likely to be genuine LAEs at $z=4.5$.

To ensure objectivity in this process, we adopted explicit quantitative criteria during the inspection.
First, we required clear Ly$\alpha$ emission in the photometric SED, defined as N673 flux exceeding the $1\sigma$ uncertainty of the adjacent UV continuum bands ($r$, $i$, $z$, and $y$).
Objects that do not exhibit statistically significant Ly$\alpha$ emission were excluded to maintain a pure LAE sample and to avoid mixing LAEs with continuum-selected LBGs.
Second, objects were excluded if no statistically significant LyC break was present, i.e., if the flux difference between wavelengths shortward and longward of the Lyman limit was consistent within the $1\sigma$ noise level.
Third, we excluded sources showing an unphysical continuum shape across the Lyman limit, specifically, those in which the $u/u^*$-band flux exceeds the $g$-band flux by more than $1\sigma$, as a strongly rising continuum toward shorter wavelengths is unlikely for genuine $z\sim4.5$ LyC emitters and may indicate foreground contamination or photometric artifacts.

Additionally, if an object is likely an LAE at $z=4.5$ but its $u/u^*$-band flux appears contaminated by neighboring sources (not excluded in the cleaning step), it is also excluded.
This visual inspection was conducted independently by six people,
resulting in samples of 615, 1,027, and 836 LAEs in the Direct, Direct-KS, Annulus samples, respectively.
The distributions of LyC flux ($F_{\rm LyC}$) and the signal-to-noise (S/N) of each sample are shown in Figure~\ref{fig:modelhist}.

As expected, only a few objects exhibit LyC emission detected at S/N$>2$ (4, 25, and 36 objects in each sample, respectively, with some overlap).
Although marginal, we constructed a LyC leaker candidate sample of LAEs with S/N $>$ 2 by combining those in the three samples.
We designated 12 LAEs with ${\rm S/N}>3$ as gold candidates and 39 LAEs with $2<{\rm S/N}<3$ as silver candidates (after removing overlaps).
Detailed information for the gold and silver candidates is provided in Appendix~\ref{sec:goldsilver} (Tables~\ref{tab:gold} and \ref{tab:silver}).
Figures~\ref{fig:cutout-gold1} and \ref{fig:cutout-gold2} show their multi-band cutouts and SEDs for the 12 gold candidates.

It is worth noting that the numbers of the gold and silver candidates exceed what is expected from a Gaussian distribution (1 and 18, respectively, for the final sample of 851 LAEs), suggesting that these detections are likely real rather than arising from random fluctuations.
We further examined the N501 band images to investigate LyC signals closer to the Lyman limit.
While N501 probes the rest-frame wavelengths where IGM transmission is statistically higher, it is significantly shallower (25.7 mag) than the $u/u^*$ (27.1 mag; see Table~\ref{tab:filter}).
Among the final sample, we identified 3 and 31 LAEs with signals at $>3\sigma$ and 2--3$\sigma$ in N501, respectively, and we find minimal overlap between these N501-detected sources and our $u/u^*$-selected LyC candidates.
This discrepancy is primarily driven by the depth difference; most of our LyC candidates detected in the deep $u/u^*$ images are fainter than the detection limit of N501.
Additionally, the highly stochastic and wavelength-dependent nature of IGM absorption at $z\sim4.5$ can cause variations in transmission between these bands.
Consequently, while N501 provides a useful auxiliary check for the brightest leakers near the Lyman limit, its limited sensitivity prevents its use as a primary tool for independent confirmation of our deeper LyC sample.
\textbf{Nevertheless, sources detected in N501 may also represent additional bright LyC candidate systems, as the N501 band probes wavelengths very close to the Lyman limit at $z\sim4.5$, provided that the detected flux indeed originates from LyC emission rather than contamination from non-ioninzing continuum.}

\begin{figure}
\includegraphics[width=\linewidth]{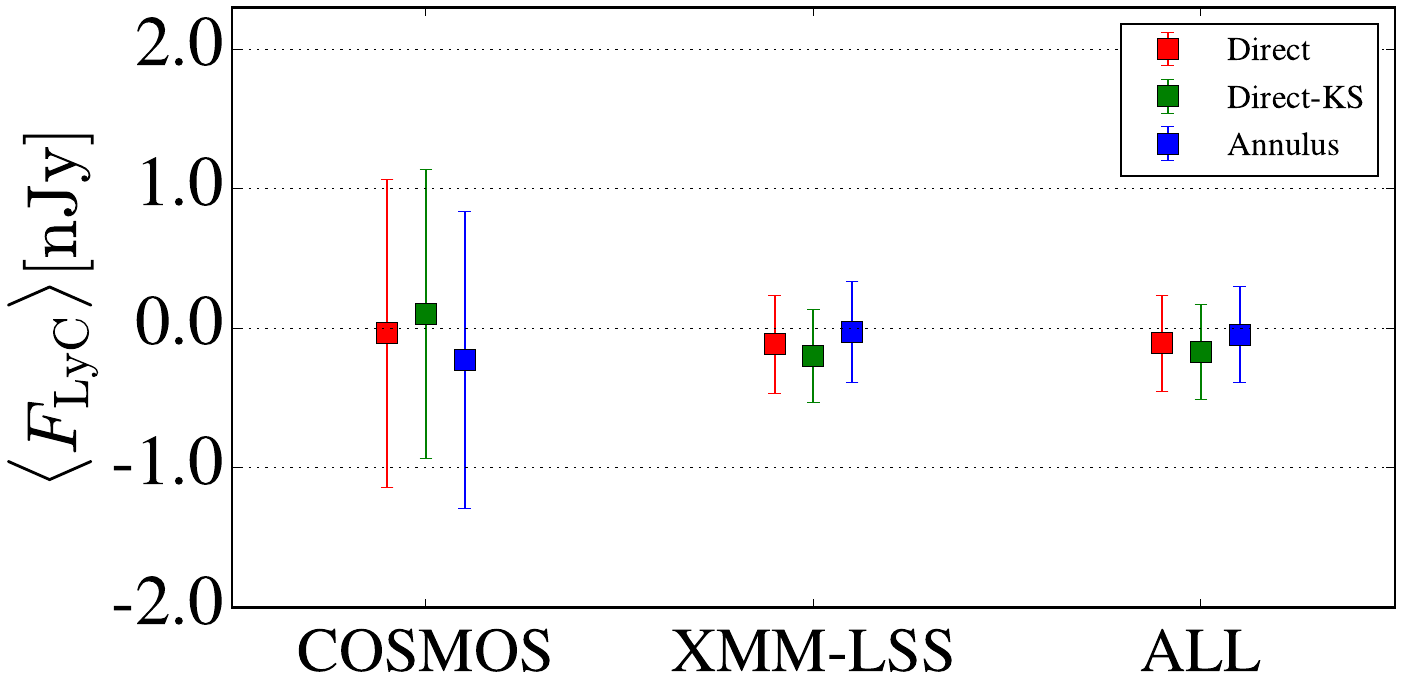}
\caption{LyC fluxes are measured from the weighted-mean stacks of LAEs in the COSMOS, XMM-LSS, and combined fields.
The flux errors are calculated as the quadratic sum of two components: the photometric error (estimated from the background variation in the stack) and the statistical error (estimated from bootstrap resampling used in constructing the stacks).
Stacks created for the Direct (red), Direct-KS (green), and Annulus (blue) samples yield consistently non-detection of LyC flux.}
\label{fig:stack_flux}
\end{figure}

Since most LAEs have low S/N, we measured the average LyC flux by stacking the $u/u^*$-band cutouts of individual LAEs, with weighted mean stack to account for differences in the two fields' noise level ($11.84 \pm 5.75$ and $4.47 \pm 2.25$ nJy for the mean noise level in the E-COSMOS and XMM-LSS fields, respectively).
The weights are given as the inverse variance measured within each cutout.
The measurements of the average LyC flux from all three samples consistently indicate non-detection (Figure~\ref{fig:stack_flux}).
Fluxes and flux errors are measured with an 1\arcsec-radius aperture. 
The measured fluxes were corrected using a factor of 0.79 to account for the $u$-band aperture loss.
Flux errors are calculated as the quadratic sum of the photometric uncertainty and the uncertainty due to sample variance. The photometric error is determined as the standard deviation (by applying a 3$\sigma$ clipping method) of fluxes measured (with a 1\arcsec-radius aperture) at 1000 locations free from any sources as mentioned earlier. The sample variance is estimated as the standard deviation of fluxes measured in 100 stacks constructed with bootstrap resampling.

We note that the absolute fluxes measured in the stacks reflect the average signal levels of the samples.
Because the sources span a wide range in UVC magnitudes and Ly$\alpha$ line fluxes, the stacked values cannot be straightforwardly translated into physical quantities such as escape fractions or covering fractions.
Instead, these measurements primarily serve to demonstrate the consistency of our background model quality check metrics, while also indicating that the LyC signal of $z=4.5$ LAEs lies below the detection limit of the CLAUDS $u/u^*$-band data.

\begin{figure*}[ht!]
\includegraphics[width=\linewidth]{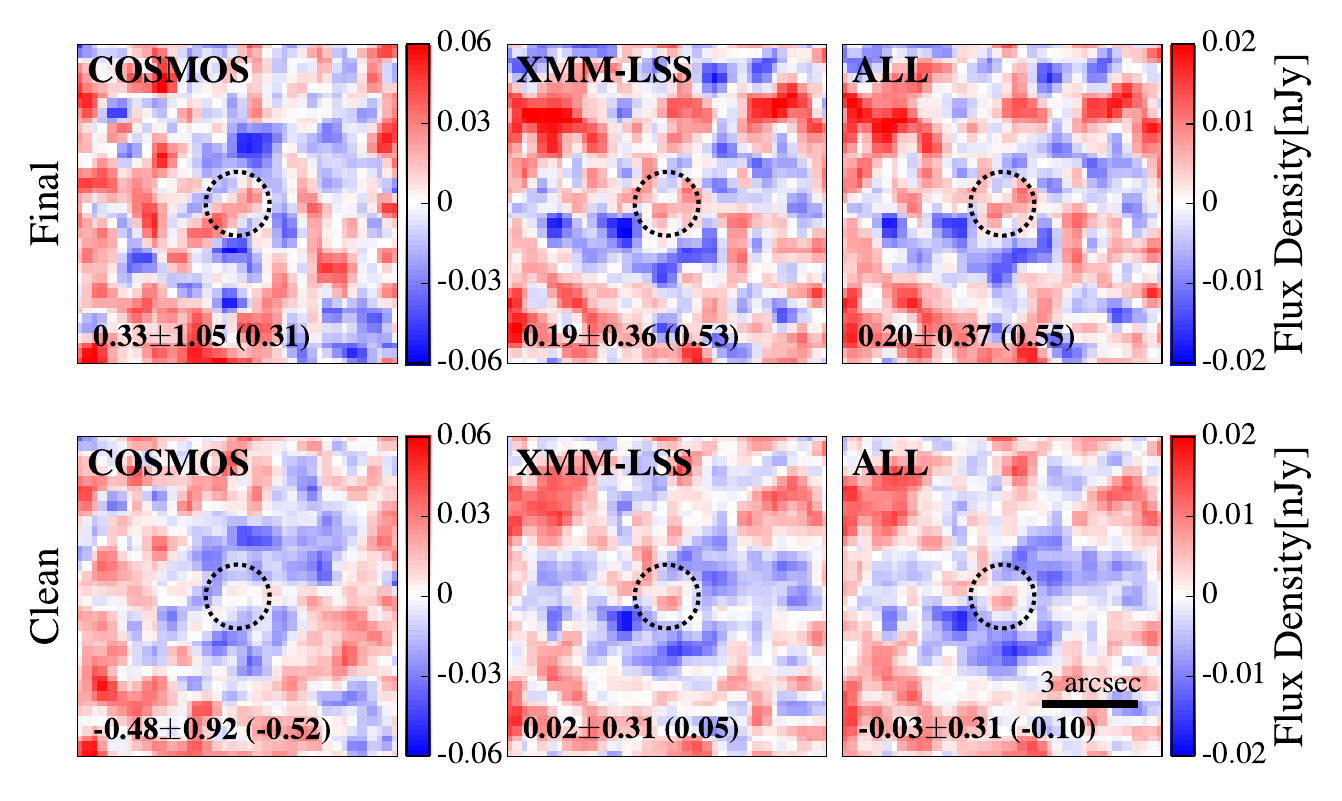}
\caption{Weighted-mean $u/u^*$-band stack images of the final LAE sample (top panels) and the ``Clean" sample (bottom) are displayed for the COSMOS (left panel), XMM-LSS (middle), and combined (right) fields.
The measured LyC flux and its error, given in nJy, are indicated at the bottom of each panel, with the corresponding signal-to-noise ratio shown in parenthesis. 
The dotted circle denotes the 1\arcsec-radius aperture used for measuring the flux and its error.
The bluish annular feature around the central aperture visible in the Clean sample is no longer present in the final sample, indicating that our background quality-control procedures effectively mitigate the oversubtraction issue.
}
\label{fig:stack}
\end{figure*}

The non-detection of LyC emission in the stacks may also be attributed to the rarity of strong LyC emitters in the overall sample.
Only a small fraction of galaxies may host strong LyC-emitting regions, leading to severe dilution of the average signal in the stack.
Although LyC emission is known to be intrinsically clumpy \citep{Micheva2017,Iwata2019,LeReste2025}, any intrinsic spatial offsets across galaxies would not significantly reduce the stacking efficiency given the seeing size of the $u/u^*$-band data.
Nevertheless, the overall effect of signal dilution due to the scarcity of bright sources underscores the importance of searching for individual, strong LyC emitters.

The fact that three samples of ``Direct", ``Direct-KS", and ``Annulus" yielded consistent measurements indicates that our results are robust against the choice of background modeling quality metric.
Among them, we ultimately adopted the Annulus sample, as its annular-based metric minimizes the risk of rejecting potential LyC leakers compared to the other metrics.
Our final sample of 851 LAEs is composed of the Annulus sample (836 LAEs) plus 15 additional LyC leaker candidates identified from the other two samples.
Figure~\ref{fig:stack} presents the weighted mean stacks of the final sample (top row).

To further validate the effectiveness of our background quality control, we also presented the weighted mean stacks of the initial ``Clean'' sample (the sample before any background quality control) in Figure~\ref{fig:stack} (bottom row).
While the stacks of the final and initial samples result in non-detections at the source position, the Clean sample exhibits a subtle but systematic negative flux bias in the local background.
In contrast, this bias is effectively mitigated in the final sample, where the local background level becomes consistent with zero.
This quantitative comparison demonstrates that our refinement procedure successfully removes systematic offsets, ensuring that our results and candidate selection are not compromised by background oversubtraction artifacts.

\begin{figure*}[ht!]
\includegraphics[width=0.9\linewidth]{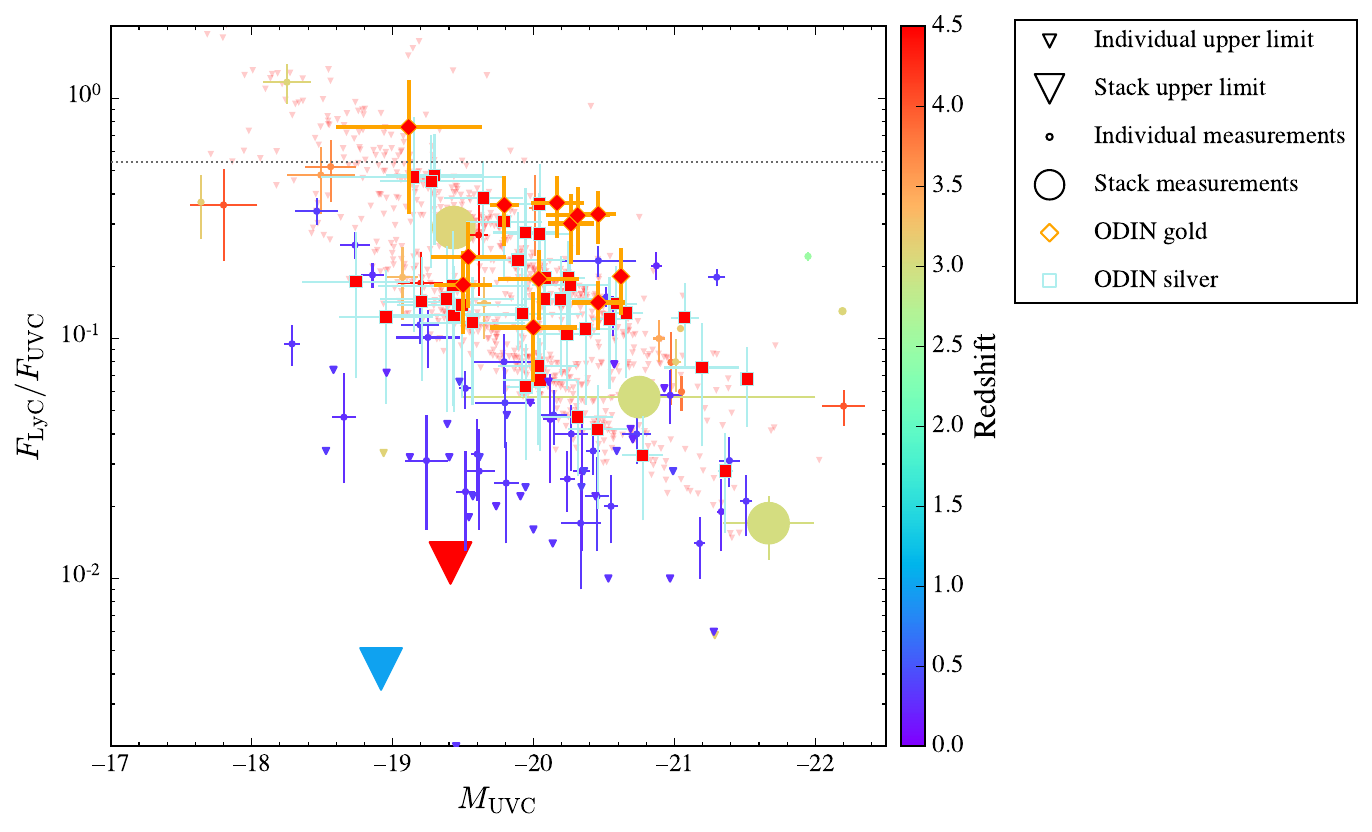}
\caption{The observed LyC-to-UVC flux ratio ($F_{\rm LyC}/F_{\rm UVC}$) as a function of UVC absolute magnitude ($M_{\rm UVC}$), including our measurements and data compiled from the literature.
Individual and stack measurements are denoted by small and large circles, respectively, with non-detections indicated as upside-down triangles.
Our gold and silver LyC leaker candidates are represented as diamonds and squares, respectively.
Symbols are color-coded by redshift as indicated in the colorbar.
Literature measurements include individual detections at $z=2.5$--4 \citep{Vanzella2010a,Vanzella2012,Vanzella2015,Siana2015,deBarros2016,Shapley2016,Vanzella2016,Bian2017,Vanzella2018,Ji2020,Marques-Chaves2021,Kerutt2024}, low-redshifts ($z=0.2$--0.4) galaxies from LzLCS \citep{Flury2022a}, and stack measurements at $z=3$ \citep{Shapley2006,Nestor2011,Steidel2018} and at $z=1$ \citep{Rutkowski2016}.
LyC and UVC fluxes are sampled at different rest-frame wavelengths across studies, but no wavelength homogenization has been applied; values are shown as reported in the original studies.
The dotted horizontal line represents the maximum predicted ratio ($\sim0.54$) for a simple stellar population with an age of 1 Myr, half-solar metallicity, and a Chabrier IMF, assuming 100\% LyC escape through the ISM and the IGM.
This value serves as an illustrative case based on extreme assumptions.
The observed anti-correlation between $F_{\rm LyC}/F_{\rm UVC}$ and $M_{\rm UVC}$ is likely an artifact of photometric errors that increase towards lower UVC luminosities (see text).}
\label{fig:key}
\end{figure*}

\subsection{LyC-to-UVC flux ratios}\label{sec:LyCfesc}

The escape fraction of LyC from a galaxy (through the ISM and CGM) is another important physical quantity related to LyC leakage, apart from the LyC flux itself.
It is often referred to the relative escape fraction, given as
\begin{equation}
\label{eq:fescrel}
    f_{\rm esc,rel}=\frac{F_{\rm LyC}/F_{\rm UVC}}{L_{\rm LyC}/L_{\rm UVC}}\frac{1}{\exp{(-\tau_{{\rm IGM},z})}},
\end{equation}
where $F$ and $L$ represent an observed flux and intrinsic luminosity at a given wavelength, and $\tau_{{\rm IGM},z}$ is the optical depth in the IGM for LyC radiation emitted at redshift $z$.
The corresponding IGM transmission can be written as $T_{\rm IGM}=\exp(-\tau_{{\rm IGM},z})$.
Determining the escape fraction, whether defined in an absolute or relative sense, requires not only stellar population synthesis modeling but also corrections for the IGM attenuation, both of which rely on underlying assumptions.
Combined with measurement uncertainties, these assumptions can lead to large discrepancies in the estimated LyC escape fractions for the same object \citep[e.g.,][]{Grimes2009,Leitet2011}.
In some cases, the estimated LyC escape fraction exceeds 100\%; this is partly due to uncertainties, particularly those related to the IGM transmission.
To minimize uncertainties arising from these assumptions, we focused on directly measured quantities rather than derived ones in this study.

Specifically, we examined the observed flux ratio $F_{\rm LyC}/F_{\rm UVC}$, as calculated using the fluxes measured in the $u/u^*$- and $i$-band cutouts within the same aperture.
The distribution of these measurements is shown in Figure~\ref{fig:key} as a function of rest-frame UVC absolute magnitude ($M_{\rm UVC}$).
Also plotted in Figure~\ref{fig:key} are measurements compiled from the literature across a wide redshift range.
For our sample, $M_{\rm UVC}$ is derived based on the $i$-band flux, assuming a redshift of $z=4.5$, including a standard $K$-correction of $2.5\log_{10} (1+z)$ under the assumption of a flat $f_\nu$ spectrum.
For the comparison sample in the literature, we adopt the published $M_{\rm UVC}$ values as reported in the respective studies.
We note that rest-frame UVC and LyC fluxes are measured at different wavelengths across different studies: 800--900~\r{AA} for LyC and 1100--1700~\r{AA} for UVC.
In our case, as summarized in Table \ref{tab:filter}, LyC is measured at 640--680~\r{AA} and UVC at 1400~\r{AA}.
The dotted horizontal line corresponds to a theoretical estimate of the maximum possible value, derived from a simple stellar population (SSP) model with an age of 1 Myr, half-solar metallicity, and the Chabrier initial mass function \citep[IMF,][]{Chabrier2003}, assuming 100\% of escape/transmission through the ISM, CGM, and IGM.
The SSP models used here are from the Flexible Stellar Populations Synthesis (FSPS) stellar populations code \citep{Conroy2009, Conroy2010}, utilizing the Padova evolution tracks \citep{Marigo2007,Marigo2008} and Basel 3.1 spectral library \citep{Lejeune1997,Lejeune1998,Westera2002}.
This line should be regarded not as a definitive upper limit, but rather as an illustrative reference under extreme assumptions.
A few objects lie above this line, but the excess is $\lesssim2\sigma$, making them marginal at best once photometric errors are considered.
These sources may still represent intriguing extreme-population candidates for follow-up spectroscopy.

Both the high- and low-redshift measurements show a decrease in $F_{\rm LyC}/F_{\rm UVC}$ with increasing UVC luminosity.
To test whether this trend could arise purely from observational effects, we generated mock data with a fixed intrinsic flux ratio and perturbed them with Gaussian noise consistent with the photometric uncertainties of the real data.
In this experiment, even a flux ratio fixed at zero reproduced a similar trend and spread, indicating that the apparent anti-correlation is largely driven by photometric uncertainties.

While the high-redshift sample tends to occupy slightly higher flux ratios at fixed $M_{\rm UVC}$ compared to the low-redshift measurements, this difference is difficult to interpret uniquely.
The two samples are selected using different criteria,\footnote{The low-redshift samples are selected based on high [O~{\sc iii}]/[O~{\sc ii}] ratios, blue UV slopes, and high star formation rate surface densities.
While all these factors are indicators of active star formation and a possibly optically-thin ISM, they do not necessarily point to the object being an LAE \citep{Steidel2018,Izotov2018a,Izotov2018b,Flury2022a}.
In contrast, the high-redshift samples primarily consist of LAEs, and such objects may have systematically higher LyC escape fractions.} and at $z\sim4.5$ the observed flux ratios are strongly modulated by stochastic IGM transmission and detection bias.

As demonstrated in our forward-modeling analysis (Appendix~\ref{sec:forwardmodeling}), the observed flux ratio depends on the combined effects of intrinsic LyC production efficiency, escape through the ISM/CGM, and highly stochastic line-of-sight IGM transmission.
Even before including observational noise and selection effects, the simulated flux ratio is most strongly correlated with $T_{\rm IGM}$, indicating that stochastic transmission already plays a dominant role in modulating the observed signal.
Once realistic noise and selection criterion are imposed, the detected subsample becomes further biased toward high-transmission sightlines, compounding this effect and complicating a direct physical interpretation.
Therefore, although the data may hint at differences in intrinsic LyC production or escape efficiency with redshift \citep{Kimm2014,Japelj2017,LeReste2025,Saldana-Lopez2025,Wang2025}, any such interpretation must be treated with caution and understood in a statistical sense.

\begin{figure*}[ht!]
\includegraphics[width=\linewidth]{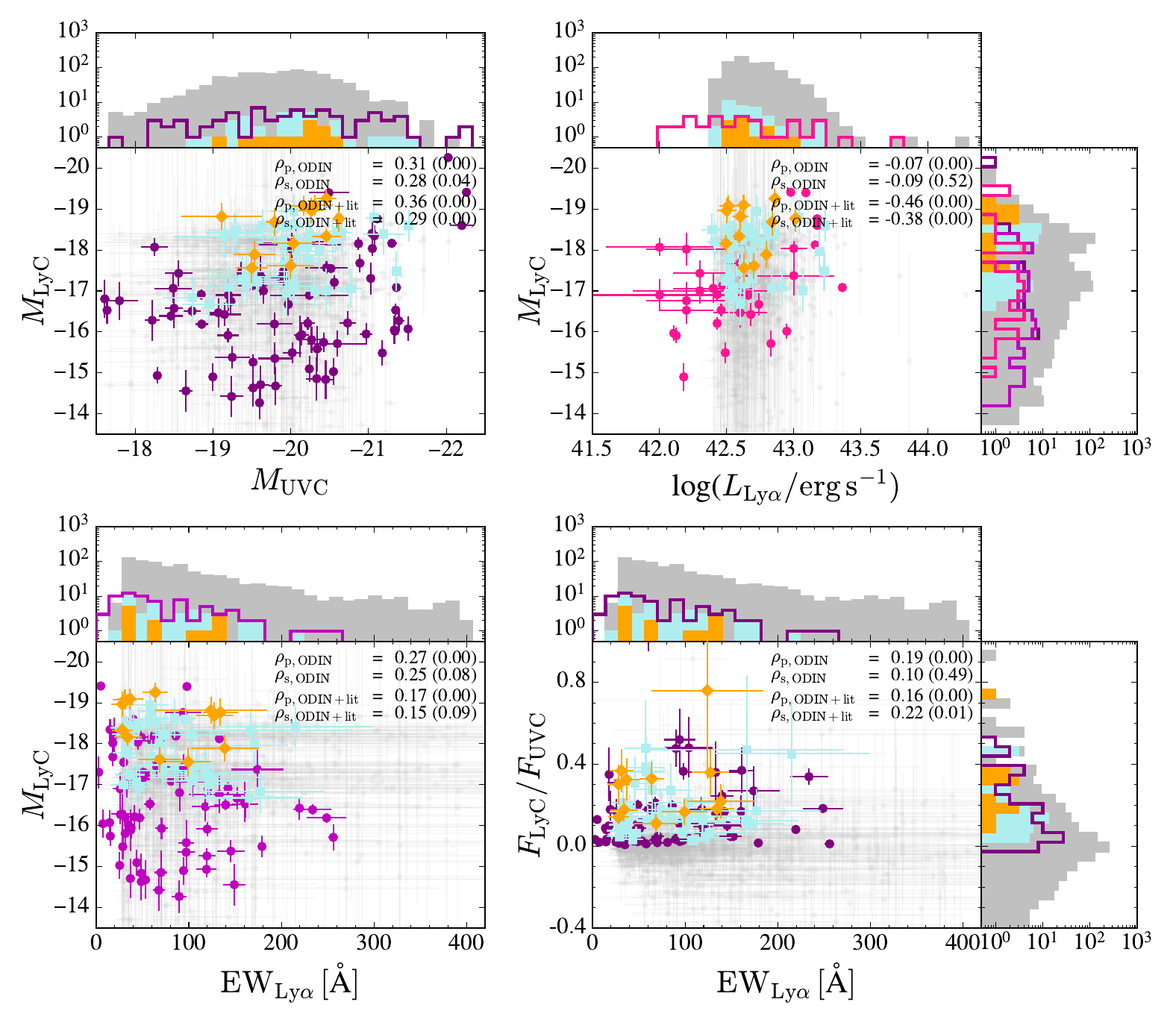}
\caption{The LyC absolute magnitude ($M_{\rm LyC}$) is plotted as a function of UVC absolute magnitude ($M_{\rm UVC}$) (top left), log($L_{\rm Ly\alpha}$) (top right), and Ly$\alpha$ equivalent width (EW) (bottom left) for 851 LAEs in the final sample.
The gold and silver LyC leaker candidates are represented by orange diamonds and cyan squares, respectively.
LyC detections reported in the literature are shown as dark purple, purple, and pink circles. Different colors are used to distinguish samples with different numbers in the 1D histograms.
The literature data includes samples such as \citep{Shapley2006,deBarros2016,Shapley2016,Bian2017,Micheva2017,Izotov2018a,Izotov2018b,Fletcher2019,Ji2020,Vanzella2020,Izotov2021,Marques-Chaves2021,Flury2022b,Kerutt2024,Liu2025}, covering a redshift range of $z=0.2-4.4$.
The LyC-to-UVC flux ratio ($F_{\rm LyC}/F_{\rm UVC}$) is shown as a function of Ly$\alpha$ EW (bottom right).
The correlation coefficients (the Pearson $\rho_{\rm p}$ and the Spearman's rank $\rho_{\rm s}$) and their corresponding $p$-values (in parentheses) are indicated at the top right of each panel for the ODIN gold and silver candidates (ODIN) and the combined samples of ODIN candidates and the literature data (ODIN+lit).}
\label{fig:property}
\end{figure*}

\subsection{Correlation between LyC and other observables}

One of our objectives is to investigate the emission and escape of LyC photons in relation to the physical properties of galaxies.
The relevant, measurable quantities include $M_{\rm UVC}$, $L_{\rm Ly\alpha}$ and $EW_{\rm Ly\alpha}$.
The correlations between these variables are quantified with the Pearson correlation coefficient $\rho_{\rm p}$ and Spearman's rank correlation coefficient $\rho_{\rm s}$.
Figure \ref{fig:property} presents the correlation analysis for the LyC-detected subsample (i.e., the gold and silver candidates), as meaningful trends can only be evaluated for galaxies with measurable LyC flux.
While no strong correlations are found, several marginal trends are apparent among these candidates.

The positive correlation between $M_{\rm LyC}$ and $M_{\rm UVC}$ (top left) is suggestive of a fundamental physical connection, as both flux components primarily originate from the massive young stars \citep{Guaita2016,Enders2023,Flury2025}.
Within our ODIN LyC-detected subsample, this relation exhibits the strongest signal among the explored correlations, consistent with its relatively direct physical origin.

The absence of a clear correlation between $M_{\rm LyC}$ and $L_{\rm Ly\alpha}$ (top right), however, implies a more complex, non-trivial interplay of physical processes \citep{Begley2022}.
This is likely due to the balance of two competing effects that determine the observed fluxes.
On one hand, the production of Ly$\alpha$ photons is a consequence of the absorption of LyC photons (via the ionization and recombination of hydrogen), which naturally induces a negative correlation between the escaping fluxes \citep[e.g.,][]{Fletcher2019,Puschnig2020,Kerutt2024}.
On the other hand, the synchronized escape of both photons through low-density channels created by stellar feedback promotes a positive correlation \citep[e.g.,][]{Reddy2016,Vanzella2020,Naidu2022,Farcy2025,Flury2025}.
However, further complication can arise due to resonant scattering of Ly$\alpha$ as it increases the effective path length of Ly$\alpha$ photons and subsequent absorption by dust \citep{Ji2020,Citro2024}.
This effect leads to differential attenuation of LyC and Ly$\alpha$ photons, possibly promoting a negative relationship \citep{Kornei2010,Mostardi2013,Hayes2014,Verhamme2015,Izotov2018b,Fletcher2019,Gazagnes2020}.

The correlation between $M_{\rm LyC}$ and $EW_{\rm Ly\alpha}$ (bottom left) can be naturally understood from the ratio definition of $EW_{\rm Ly\alpha}$.
Since $EW_{\rm Ly\alpha}$ scales as $L_{\rm Ly\alpha}/L_{\rm UVC}$, its correlation with $M_{\rm LyC}$ is governed by the relative strengths of the $M_{\rm LyC}$--$L_{\rm Ly\alpha}$ and $M_{\rm LyC}$--$L_{\rm UVC}$ relations.
Given that the former correlation is weak, the positive trend with $EW_{\rm Ly\alpha}$ is primarily driven by the stronger correlation between $M_{\rm LyC}$ and $M_{\rm UVC}$.

The marginal positive tendency between $F_{\rm LyC}/F_{\rm UVC}$ and $EW_{\rm Ly\alpha}$ (bottom right) may suggest the presence of the underlying synchronized escape mechanism.
However, the extremely weak nature of this correlation (i.e., $\rho_{\rm s}=0.1,\,p=0.49$) reflects again a near-perfect cancellation between this positive coupling and the negative coupling mentioned earlier, with the escape effect only trivially and insignificantly dominating the final outcome.

We emphasize that the correlations discussed above, when restricted to the ODIN gold and silver candidates, are derived from a relatively small sample of only a few tens of LyC leaker candidates.
Consequently, the statistical significance of these trends is limited, and the inferred relations should be regarded as tentative.
Future work incorporating additional LyC leaker candidates from the remaining ODIN survey data will be essential to improve the statistical robustness of these measurement.

Another possible factor contributing to the weak correlations in the ODIN-only sample is that we did not correct for the IGM attenuation, which varies significantly from sightline to sightline \citep{Vanzella2010a,Bassett2021}.
As a result, correlations based on the observed LyC flux may be substantially diluted compared to the intrinsic counterparts \citep{Saxena2022}.

Motivated by these limitations, we expanded the analysis by incorporating additional LyC candidates reported in the literature, thereby increasing the dynamic range and sample size.
In the combined dataset, most correlations become more pronounced compared to the ODIN-only results.
This strengthening likely reflects the increased broader luminosity baseline introduced by the literature compilation, which allows an underlying scaling relation to become more apparent.

We caution, however, that the literature compilation spans a range of redshifts and observational strategies, mixing sightlines with different IGM transmission statistics and heterogeneous selection functions.
The enhanced correlations in the combined dataset may therefore reflect not only intrinsic physical couplings but also systematic differences across surveys.
Accordingly, these results should be interpreted as exploratory empirical trends over an expended parameter space rather than definitive redshift-independent physical relations.
A more controlled comparison within homogeneous redshift bins, ideally with statistical IGM corrections applied, will be necessary to assess the intrinsic nature of these couplings.  

Nevertheless, the overall trends identified here provide useful guidance for future follow-up strategies aimed at maximizing LyC detectability--for instance, by prioritizing galaxies with high UVC luminosities rather than those selected solely on strong Ly$\alpha$ EW.

\begin{figure*}
\includegraphics[width=\linewidth]{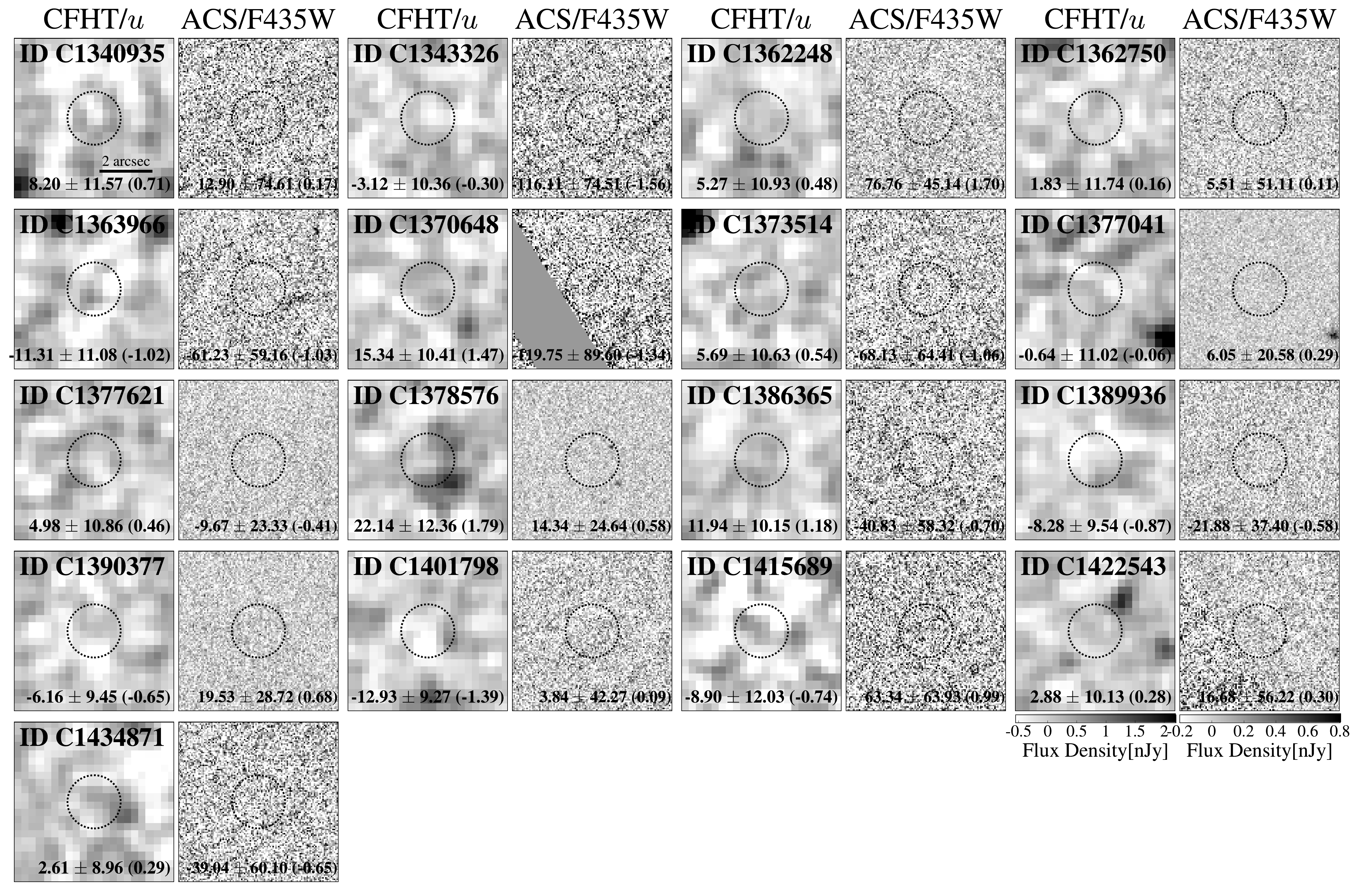}
\caption{Side-by-side cutouts (6$\arcsec$ on a side) are shown for the 17 LAEs from our final sample that are identified in the HST datasets.
Each pair displays the images in the CLAUDS $u$- (left) and UVCANDELS or LCGCOSMOS F435W (right) bands.
In each panel, the flux, its error, and the resulting signal-to-noise ratio (in parenthesis) are indicated.
The flux and error are measured with a 1\arcsec-radius aperture.
A colorbar for each band image is provided at the bottom right.
The flux values measured in both band images are consistent within their errors (non-detections).
This consistency validates our primary $u$-band result despite its relatively poor spatial resolution.
We note that the larger errors in the F435W-band are attributed to the smaller pixel scale ($0 \farcs 03$ versus $0 \farcs 27$ for the $u$-band image), which results in a larger number of pixels within the 1\arcsec-radius aperture and thus a large accumulation of noise, while the background noise in the F435W image is only about twice lower.}
\label{fig:cutout_HST}
\end{figure*}

\begin{figure*}[ht!]
\epsscale{1.1}
\plotone{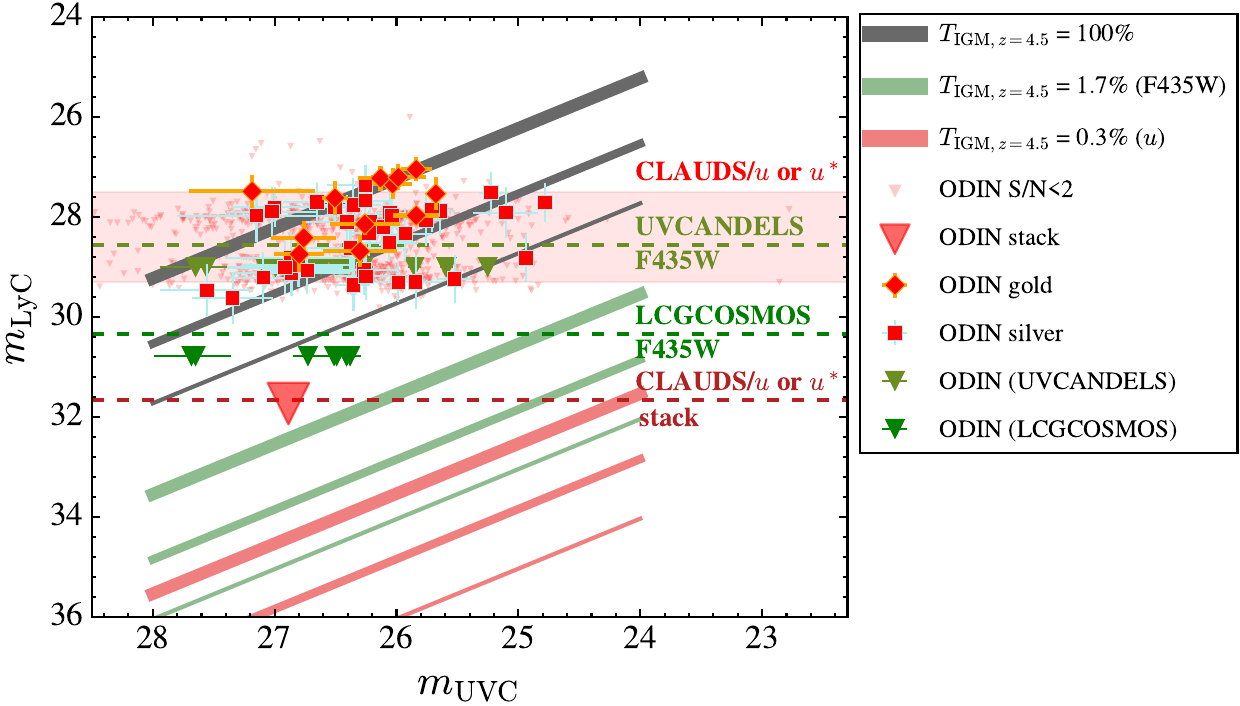}
\caption{
The observed LyC magnitude versus UVC magnitude for LAEs in our final sample is presented.
Observed values are denoted by colored symbols, with 2$\sigma$ upper limits represented by upside-down triangles.
Different colors correspond to different datasets: CLAUDS $u/u^*$ band (red), CLAUDS $u/u^*$-band stack (dark red), UVCANDELS F435W band (light green), and LCGCOSMOS F435W band (dark green).
The green symbols are those shown in Figure~\ref{fig:cutout_HST}.
LyC fluxes are measured using 1\arcsec-radius aperture.
The $3\sigma$ depth of each dataset is indicated by a horizontal line or shade in the corresponding color, with its data name written alongside.
Expected relations are presented using three different color lines.
These relations use the intrinsic UVC-to-LyC ratios of 3, and are calculated for an IGM transmission ($T_{\rm IGM}$) of 100\% (gray) and the typical values at $z=4.5$, which are 1.7\% for the F435W band (pale green) and 0.3\% for the $u$ band (pale red) calculated based on \citet{Inoue2014}.
Different line thicknesses represent three assumed escape fractions through the ISM of 100\%, 30\%, and 10\% (thicker for higher escape).
While most measurements are upper limits, which is reasonable given the image depths and the expected relation (i.e., the images are not deep enough to detect typical LyC fluxes, even in the stack), the gold and silver LyC leaker candidates lie significantly above the expected relations.
This suggests that the IGM transmission for these candidates is likely higher than the typical value, potentially near 100\%.}
\label{fig:LyC-UVC}
\end{figure*}

\subsection{Analyzing HST archival data}\label{sec:hst}

The COSMOS field has been extensively surveyed across various wavelengths, and offers a wealth of archival datasets including HST data.
We examined ACS/F435W data taken from UltraViolet Imaging of the Cosmic Assembly Near--infrared Deep Extragalactic Legacy Survey Fields \citep[UVCANDELS,][]{Wang2025} and Lyman Continuum Galaxy Candidates in COSMOS \citep[LCGCOSMOS,][]{Prichard2022}, which sample $\lambda_{\rm rest}=650$--880\AA\ for $z=4.5$ galaxies. The 5$\sigma$-limiting magnitudes of these surveys are 28--28.4 and 28.52--29.78, respectively.
We identified 14 and 7 objects from our sample within the coverage of UVCANDELS and LCGCOSMOS, respectively.
As four of them are found in both datasets, a total of 17 unique LAEs are available HST ACS/F435W data.
None of our gold and silver candidates lie in existing HST coverage.

The F435W cutouts of the 17 objects are presented in Figure~\ref{fig:cutout_HST}, along with their $u$-band counterpart cutouts (central wavelength 640~\AA\ at $z=4.5$).
The F435W cutouts, which have a higher spatial resolution (PSF $\sim 0\farcs 1$) and more uniform background, offer improved clarity to verify the authenticity of the $u$-band signals and to identify any potential contamination;
consistent with the $u$-band analysis, the deeper F435W images show no clear evidence of LyC leakage from these 17 LAEs.
The lack of signal in even the deeper HST observations robustly supports our conclusion that LyC emission is intrinsically faint or rare within our sample.

One LAE, ID C1378576, which exhibits a 1.79$\sigma$ signal in the $u$-band image, shows only a 0.58$\sigma$ signal in the F435W image.
However, although the overall signal within the central $1\arcsec$-radius aperture is weaker in F435W, it is notable that the signal seen in the $u$-band image appears to be due to two sub-arcsecond knots, located near the upper and lower right edges of the aperture relative to the object's $u$-band centroid\footnote{When measured with 0\farcs 2 apertures, the fluxes of these two knots are $10.68\pm2.33\,\textrm{nJy}$ (upper one, S/N $\sim5$) and $18.60\pm2.33\,\textrm{nJy}$ (lower one, S/N $\sim8$).}.
Given that the effective radius ($R_{\rm eff}$) of star-forming galaxies with stellar mass $5\times10^{10}\,M_\odot$ at $z\sim4.2$ is $\sim 2.6$ kpc \citep{Ward2024}, which corresponds to an angular size $0\arcsec.37$ at $z=4.5$, these knots are located away from the center at $\sim2.7R_{\rm eff}$, on the outskirt of a galaxy.
The possibility that the knots are associated with the $z=4.5$ LAE cannot be ruled out, as it has been shown that the spatial offset between ionizing and non-ionizing radiation can be as large as 10 kpc \citep[e.g.,][]{Micheva2017}.
This may provide a plausible clue to a scenario where ionizing photons escape not from the galactic center but from localized, off-nuclear star-forming clumps \citep{Steidel1996, Papovich2005, Conselice2005, Venemans2005, Bergvall2006, Pirzkal2007,Siana2015, Vanzella2016, Keenan2017, Rivera-Thorsen2017b, Ji2020}.
To confirm whether these knots are associated with the $z=4.5$ LAE rather than foreground sources, follow-up spectroscopy is required.
We refer the reader to Appendix~\ref{sec:goldsilver} for a detailed inspection of individual cutouts, where measurable spatial offsets between the LyC and non-ionizing emission are quantified and discussed for several gold candidates.

While the null detection in the $u$ band is confirmed by the F435W data, some signals seen in the $u$ band are absent in F435W (e.g., two knots on the right in the ID C1422543 cutout), highlighting the need to define a minimum flux in the $u$ band for reliable detection.
To this end, we identified visible emission spots (non-LAEs) in the $u$-band cutouts shown in Figure \ref{fig:cutout_HST}, and performed forced photometry at the same aperture positions in F435W.
From our comparison of $u$-band and F435W fluxes (figure not included), we find that $u$-band fluxes greater than $\sim0.03\mu$Jy are generally consistent with the corresponding F435W measurements within the photometric uncertainties.
In addition, when restricting the comparison to sources with approximately flat observed spectral slopes, the agreement between the two bands becomes noticeably tighter.
This indicates that the scatter in the full sample is primarily driven by intrinsic spectral variations rather than by systematic photometric offsets between the filters.
Therefore, $u$-band detections above this flux level are unlikely to arise from systematic calibration issues, although confirming their physical origin as genuine LyC emission ultimately requires higher-resolution follow-up observations.

We can assess the detectability of LyC emission in our sample by considering the expected LyC flux based on the observed UVC emission.
Figure~\ref{fig:LyC-UVC} shows the LyC versus UVC measurements for all LAEs in the final sample. 
Assuming an intrinsic UVC-to-LyC ratio of 3 as typical value for SFGs in the same model described in Section \ref{sec:LyCfesc}, the expected relationships between the same quantities with escape fraction values of 100, 30, and 10\% (through the ISM and CGM) are shown with solid lines of a given color but thickness increasing with escape fraction.
For each escape fraction, three levels of IGM transmission--100\% for the extreme case of no IGM absorption (gray), 1.7\% for F435W (pale green), and 0.3\% for the $u$ band (pale red)--are indicated by line color.
The values for F435W and $u$ band are calculated based on \citet{Inoue2014} for the average IGM transmission for $z=4.5$. 
Measured LyC magnitudes in a given filter  (symbols) should be compared to corresponding theoretical values (solid lines) of the same color.

Given the depth of LCGCOSMOS, only the brightest galaxies ($m_{\rm UVC}<\sim$24--25) would be detectable; all galaxies in our sample are fainter than this limit.
The gold and silver candidates, whose LyC fluxes are detected despite the shallower depths of CLAUDS, may indicate the presence of extreme stellar populations and a highly transparent pathway for LyC photons (especially through the IGM).

To quantitatively assess which factor primarily drives these detections, we utilize the forward-modeling framework described in Appendix~\ref{sec:forwardmodeling}.
This approach allows us to evaluate the posterior distributions of $L_{\rm LyC}/L_{\rm UVC}$, $f_{\rm esc}$, and $T_{\rm IGM}$ for the LyC-detected subsample.
We find that all three quantities are systematically biased toward higher values in the detected subsample relative to the parent population (see histograms in the top panels of Figure~\ref{fig:forwardmodeling}).
The mean $L_{\rm LyC}/L_{\rm UVC}$ increases from 0.33 in the parent distribution to 0.47 in the detected subsample, and the mean $f_{\rm esc}$ shifts from 0.20 to 0.35.
Most strikingly, the mean $T_{\rm IGM}$ increases from 0.22 in the parent distribution to 0.52 in the detected subsample.
This behavior is expected given the multiplicative dependence of the observed LyC flux on these three factors.
However, the relative shift is most pronounced for $T_{\rm IGM}$, whose intrinsically broader and highly skewed distribution at $z\sim4.5$ makes stochastic line-of-sight transmission particularly effective in enabling detections.
In other words, while LyC detections preferentially select galaxies with above-average $L_{\rm LyC}/L_{\rm UVC}$ and $f_{\rm esc}$, statistically reproducing the observed detection rate requires sightlines drawn from the high-transmission tail of the IGM distribution.

The stack of over 851 objects reaches a depth deeper than that of LCGCOSMOS ($\sim28+2.5\log\sqrt{851}\sim31.7$), yet this remains insufficient to detect the LyC flux given their average UVC brightness ($\sim$27~mag).
Even combining ODIN LAEs that are to be identified in the other fields may still be insufficient to reach the necessary depth for detection, given the low mean IGM transmission of $\sim 0.3\%$ at these wavelengths.
Instead, targeting individual LyC leaker candidates, like the gold and silver candidates identified in this work, appears to be a more effective strategy.
The large size of the full ODIN sample suggests a significant addition of LyC leaker candidates for future analyses.

\section{SUMMARY} \label{sec:summary}

We presented an investigation into the LyC leakage from LAEs at $z=4.5$ identified in the COSMOS and XMM-LSS fields as part of the ODIN survey.
Below is a summary of our key analyses and results.

\begin{itemize}
    \item To maximize the reliability and purity of our $z=4.5$ LyC flux measurements, we applied stringent cleaning procedures.
    We constructed a Clean sample of 1,819 LAEs by excluding sources potentially contaminated by low-redshift galaxies (using a $g-r$ color cut) or neighboring galaxies (with a large exclusion radius).
    
    \item Given the faintness of the LyC signal of high-$z$ galaxies, we performed additional background modeling to improve the accuracy of background subtraction and maximize the S/N of the faint LyC signal.
    This procedure involved aggressive masking to prevent contamination by any residual sources and the careful optimization of background model parameters for individual LAEs. 
    As a result, the background level was adjusted to be less biased and less variable (Figure \ref{fig:bkgdist}).
    
    \item The quality of the background modeling was assessed using three different metrics (so-called ``Direct", ``Direct-KS", and ``Annulus" metrics) to quantify the reliability of the background determination at the LAE locations.
    While all three metrics consistently resolve the existing background over- or under-subtraction issue (Figure \ref{fig:modelhist}), the Annulus metric was ultimately selected as it yields a sample least susceptible to the risk of rejecting potential LyC leakers.

    \item Based on the quality assessment of our background models and visual inspection of individual SEDs, we constructed our final sample of 851 LAEs.
    This sample comprises the Annulus sample (836 LAEs) supplemented by an additional LyC leaker candidates from the Direct and Direct-KS samples (15 LAEs).
    Within this final set, we designated 12 LAEs with $S/N>3$ as gold candidates and 39 LAEs with $2<S/N<3$ as silver candidates, while most LAEs exhibited non-detection of LyC (non-detection even in stacks, Figures \ref{fig:stack_flux} and \ref{fig:stack}).
    The overall low detection rate and non-detection even in stacked images indicate that the LyC signal of typical $z=4.5$ LAEs lies below the detection limit of the existing $u/u^*$-band data.

    \item The observed flux ratios, $F_{\rm LyC}/F_{\rm UVC}$, often used as a proxy for LyC escape, were compared with results from other studies across a range of redshifts as a function of $M_{\rm UVC}$ (Figure \ref{fig:key}).
    While no significant correlation is observed between $F_{\rm LyC}/F_{\rm UVC}$ and $M_{\rm UVC}$, the interpretation of the flux ratio at $z\sim4.5$ is complicated by stochastic IGM transmission and detection bias.
    Our forward-modeling analysis shows that the observed flux ratios are strongly influenced by rare high-transmission sightlines, and therefore must be interpreted statistically rather than as direct measurements of intrinsic escape fractions.
    
    \item Correlations of $M_{\rm LyC}$ with other properties were examined.
    While no clear correlation was found across the final 851 LAEs, marginal correlations were observed among the gold and silver candidates (Figure \ref{fig:property}).
    Specifically, $M_{\rm LyC}$ shows a positive correlation with $M_{\rm UVC}$ but a weak correlation with $EW_{\rm Ly\alpha}$.
    The positive $M_{\rm LyC}$--$M_{\rm UVC}$ correlation suggests a fundamental physical connection due to the co-production of LyC and UVC photons by massive young stars.
    Conversely, the weak $M_{\rm LyC}$--$EW_{\rm Ly\alpha}$ correlations, combined with the absence of strong correlations between $M_{\rm LyC}$ and $L_{\rm Ly\alpha}$ or $F_{\rm LyC}/F_{\rm UVC}$ and $EW_{\rm Ly\alpha}$, highlights a complex, non-trivial interplay of physical processes governing the production and escape of LyC and Ly$\alpha$ photons.
    By expanding the sample with additional literature data, the correlations between $M_{\rm LyC}$ and $L_{\rm Ly\alpha}$, as well as between $F_{\rm LyC}/F_{\rm UVC}$ and $EW_{\rm Ly\alpha}$, were strengthened in their expected directions.
    
    \item 17 unique LAEs from our final sample were re-analyzed using the archival HST/ACS F435W data from the UVCANDELS and LCGCOSMOS surveys (Figure \ref{fig:cutout_HST}).
    The HST data provides deeper limiting magnitudes and higher spatial resolution compared to the ground-based CLAUDS $u/u^*$-band data, allowing for a more stringent test of contamination.
    The null detection of LyC in the $u$ band was robustly confirmed by the F435W data, strengthening our conclusion that LyC emission is intrinsically faint or rare within our sample.
    
    \item Theoretical calculations of the expected LyC flux as a function of UVC flux reveal that the existing datasets are insufficiently deep for robust individual or stacked LyC detection  (Figure \ref{fig:LyC-UVC}).
    Our forward-modeling analysis further shows that LyC detections at $z\sim4.5$ are strongly biased toward rare high-transmission sightlines.
    Therefore, targeting individual LyC leaker candidates, like the gold and silver candidates identified in this work, represents the most effective strategy for future LyC detection survey.
    As a crucial next step, identifying additional LyC leaker candidates in the full ODIN sample and securing follow-up spectroscopic observations to confirm these candidates will be necessary to definitively estimate the LyC escape fraction.
\end{itemize}

\begin{acknowledgments}
We thank the anonymous referee for helpful comments that significantly improved the manuscript.
This work was supported by the National Research Foundation of Korea (NRF) grant funded by the Korea government (MSIT) (No. 2022R1A4A3031306) and Global - Learning \& Academic research institution for Master’s·PhD students, and Postdocs (G-LAMP) Program of the National Research Foundation of Korea (NRF) grant funded by the Ministry of Education (No. RS-2025-25442707).
The Institute for Gravitation and the Cosmos is supported by the Eberly College of Science and the Office of the Senior Vice President for Research at the Pennsylvania State University.
LG also gratefully acknowledges financial support from ANID - MILENIO - NCN2024\_112, ANID BASAL project FB210003, and FONDECYT regular project number 1230591.
EG acknowledges support from NSF grant AST-2206222.
HSH acknowledges the support of the National Research Foundation of Korea (NRF) grant funded by the Korea government (MSIT), NRF-2021R1A2C1094577, and Hyunsong Educational \& Cultural Foundation.
H.S. and J.L. acknowledges the support of the NRF of Korea grant funded by the Korea government (MSIT, 2022M3K3A1093827).
This study used data obtained with the Blanco 4m telescope's the Dark Energy Camera (DECam), which was constructed by the Dark Energy Survey (DES) collaboration.
On the basis of observations at Cerro Tololo Inter-American Observatory, NSF's NOIRLab (NOIRLab Prop. ID 2020B-0201; PI: K.-S. Lee), which is managed by the Association of Universities for Research in Astronomy (AURA) under a cooperative agreement with the National Science Foundation.
The $u/u^*$-band data were obtained and processed as part of the CFHT Large Area U-band Deep Survey (CLAUDS), which is a collaboration between astronomers from Canada, France, and China described in \citet{Sawicki2019}. CLAUDS data products can be accessed from https://www.clauds.net. CLAUDS is based on observations obtained with MegaPrime/ MegaCam, a joint project of CFHT and CEA/DAPNIA, at the CFHT which is operated by the National Research Council (NRC) of Canada, the Institut National des Science de l’Univers of the Centre National de la Recherche Scientifique (CNRS) of France, and the University of Hawaii. CLAUDS uses data obtained in part through the Telescope Access Program (TAP), which has been funded by the National Astronomical Observatories, Chinese Academy of Sciences, and the Special Fund for Astronomy from the Ministry of Finance of China. CLAUDS uses data products from TERAPIX and the Canadian Astronomy Data Centre (CADC) and was carried out using resources from Compute Canada and Canadian Advanced Network For Astrophysical Research (CANFAR).
\end{acknowledgments}

\bibliography{ref}

\begin{thebibliography}{}
\expandafter\ifx\csname natexlab\endcsname\relax\def\natexlab#1{#1}\fi
\providecommand{\url}[1]{\href{#1}{#1}}
\providecommand{\dodoi}[1]{doi:~\href{http://doi.org/#1}{\nolinkurl{#1}}}
\providecommand{\doeprint}[1]{\href{http://ascl.net/#1}{\nolinkurl{http://ascl.net/#1}}}
\providecommand{\doarXiv}[1]{\href{https://arxiv.org/abs/#1}{\nolinkurl{https://arxiv.org/abs/#1}}}

\bibitem[{{Aihara} {et~al.}(2019){Aihara}, {AlSayyad}, {Ando}, {Armstrong}, {Bosch}, {Egami}, {Furusawa}, {Furusawa}, {Goulding}, {Harikane}, {Hikage}, {Ho}, {Hsieh}, {Huang}, {Ikeda}, {Imanishi}, {Ito}, {Iwata}, {Jaelani}, {Kakuma}, {Kawana}, {Kikuta}, {Kobayashi}, {Koike}, {Komiyama}, {Li}, {Liang}, {Lin}, {Luo}, {Lupton}, {Lust}, {MacArthur}, {Matsuoka}, {Mineo}, {Miyatake}, {Miyazaki}, {More}, {Murata}, {Namiki}, {Nishizawa}, {Oguri}, {Okabe}, {Okamoto}, {Okura}, {Ono}, {Onodera}, {Onoue}, {Osato}, {Ouchi}, {Shibuya}, {Strauss}, {Sugiyama}, {Suto}, {Takada}, {Takagi}, {Takata}, {Takita}, {Tanaka}, {Terai}, {Toba}, {Uchiyama}, {Utsumi}, {Wang}, {Wang}, \& {Yamada}}]{Aihara2019}
{Aihara}, H., {AlSayyad}, Y., {Ando}, M., {et~al.} 2019, \pasj, 71, 114, \dodoi{10.1093/pasj/psz103}

\bibitem[{{Barbary}(2016)}]{Barbary2016}
{Barbary}, K. 2016, The Journal of Open Source Software, 1, 58, \dodoi{10.21105/joss.00058}

\bibitem[{{Bassett} {et~al.}(2021){Bassett}, {Ryan-Weber}, {Cooke}, {Me{\v{s}}tri{\'c}}, {Kakiichi}, {Prichard}, \& {Rafelski}}]{Bassett2021}
{Bassett}, R., {Ryan-Weber}, E.~V., {Cooke}, J., {et~al.} 2021, \mnras, 502, 108, \dodoi{10.1093/mnras/stab070}

\bibitem[{{Begley} {et~al.}(2022){Begley}, {Cullen}, {McLure}, {Dunlop}, {Hall}, {Carnall}, {Hamadouche}, {McLeod}, {Amor{\'\i}n}, {Calabr{\`o}}, {Fontana}, {Fynbo}, {Guaita}, {Hathi}, {Hibon}, {Ji}, {Llerena}, {Pentericci}, {Saldana-Lopez}, {Schaerer}, {Talia}, {Vanzella}, \& {Zamorani}}]{Begley2022}
{Begley}, R., {Cullen}, F., {McLure}, R.~J., {et~al.} 2022, \mnras, 513, 3510, \dodoi{10.1093/mnras/stac1067}

\bibitem[{{Bergvall} {et~al.}(2006){Bergvall}, {Zackrisson}, {Andersson}, {Arnberg}, {Masegosa}, \& {{\"O}stlin}}]{Bergvall2006}
{Bergvall}, N., {Zackrisson}, E., {Andersson}, B.~G., {et~al.} 2006, \aap, 448, 513, \dodoi{10.1051/0004-6361:20053788}

\bibitem[{{Bertin} \& {Arnouts}(1996)}]{Bertin1996}
{Bertin}, E., \& {Arnouts}, S. 1996, \aaps, 117, 393, \dodoi{10.1051/aas:1996164}

\bibitem[{{Bian} {et~al.}(2017){Bian}, {Fan}, {McGreer}, {Cai}, \& {Jiang}}]{Bian2017}
{Bian}, F., {Fan}, X., {McGreer}, I., {Cai}, Z., \& {Jiang}, L. 2017, \apjl, 837, L12, \dodoi{10.3847/2041-8213/aa5ff7}

\bibitem[{{Bouwens} {et~al.}(2007){Bouwens}, {Illingworth}, {Franx}, \& {Ford}}]{Bouwens2007}
{Bouwens}, R.~J., {Illingworth}, G.~D., {Franx}, M., \& {Ford}, H. 2007, \apj, 670, 928, \dodoi{10.1086/521811}

\bibitem[{{Cardamone} {et~al.}(2009){Cardamone}, {Schawinski}, {Sarzi}, {Bamford}, {Bennert}, {Urry}, {Lintott}, {Keel}, {Parejko}, {Nichol}, {Thomas}, {Andreescu}, {Murray}, {Raddick}, {Slosar}, {Szalay}, \& {Vandenberg}}]{Cardamone2009}
{Cardamone}, C., {Schawinski}, K., {Sarzi}, M., {et~al.} 2009, \mnras, 399, 1191, \dodoi{10.1111/j.1365-2966.2009.15383.x}

\bibitem[{{Chabrier}(2003)}]{Chabrier2003}
{Chabrier}, G. 2003, \pasp, 115, 763, \dodoi{10.1086/376392}

\bibitem[{{Chang} \& {Gronke}(2024)}]{Chang2024}
{Chang}, S.-J., \& {Gronke}, M. 2024, \mnras, 532, 3526, \dodoi{10.1093/mnras/stae1664}

\bibitem[{{Chisholm} {et~al.}(2020){Chisholm}, {Prochaska}, {Schaerer}, {Gazagnes}, \& {Henry}}]{Chisholm2020}
{Chisholm}, J., {Prochaska}, J.~X., {Schaerer}, D., {Gazagnes}, S., \& {Henry}, A. 2020, \mnras, 498, 2554, \dodoi{10.1093/mnras/staa2470}

\bibitem[{{Citro} {et~al.}(2024){Citro}, {Scarlata}, {Mantha}, {Williams}, {Rafelski}, {Revalski}, {Hayes}, {Henry}, {Rutkowski}, \& {Teplitz}}]{Citro2024}
{Citro}, A., {Scarlata}, C.~M., {Mantha}, K.~B., {et~al.} 2024, arXiv e-prints, arXiv:2406.07618, \dodoi{10.48550/arXiv.2406.07618}

\bibitem[{{Conroy} \& {Gunn}(2010)}]{Conroy2010}
{Conroy}, C., \& {Gunn}, J.~E. 2010, {FSPS: Flexible Stellar Population Synthesis}, Astrophysics Source Code Library, record ascl:1010.043

\bibitem[{{Conroy} {et~al.}(2009){Conroy}, {Gunn}, \& {White}}]{Conroy2009}
{Conroy}, C., {Gunn}, J.~E., \& {White}, M. 2009, \apj, 699, 486, \dodoi{10.1088/0004-637X/699/1/486}

\bibitem[{{Conselice} {et~al.}(2005){Conselice}, {Blackburne}, \& {Papovich}}]{Conselice2005}
{Conselice}, C.~J., {Blackburne}, J.~A., \& {Papovich}, C. 2005, \apj, 620, 564, \dodoi{10.1086/426102}

\bibitem[{{Coupon} {et~al.}(2018){Coupon}, {Czakon}, {Bosch}, {Komiyama}, {Medezinski}, {Miyazaki}, \& {Oguri}}]{Coupon2018}
{Coupon}, J., {Czakon}, N., {Bosch}, J., {et~al.} 2018, \pasj, 70, S7, \dodoi{10.1093/pasj/psx047}

\bibitem[{{Davis} {et~al.}(2021){Davis}, {Gebhardt}, {Mentuch Cooper}, {Chisholm}, {Ciardullo}, {Farrow}, {Finkelstein}, {Gronwall}, {Gawiser}, {Hill}, {Hopp}, {Jeong}, {Landriau}, {Liu}, {Lujan Niemeyer}, {Schneider}, {Snigula}, \& {Tuttle}}]{Davis2021}
{Davis}, D., {Gebhardt}, K., {Mentuch Cooper}, E., {et~al.} 2021, \apj, 920, 122, \dodoi{10.3847/1538-4357/ac1598}

\bibitem[{{de Barros} {et~al.}(2016){de Barros}, {Vanzella}, {Amor{\'\i}n}, {Castellano}, {Siana}, {Grazian}, {Suh}, {Balestra}, {Vignali}, {Verhamme}, {Zamorani}, {Mignoli}, {Hasinger}, {Comastri}, {Pentericci}, {P{\'e}rez-Montero}, {Fontana}, {Giavalisco}, \& {Gilli}}]{deBarros2016}
{de Barros}, S., {Vanzella}, E., {Amor{\'\i}n}, R., {et~al.} 2016, \aap, 585, A51, \dodoi{10.1051/0004-6361/201527046}

\bibitem[{{Enders} {et~al.}(2023){Enders}, {Bomans}, \& {Wittje}}]{Enders2023}
{Enders}, A.~U., {Bomans}, D.~J., \& {Wittje}, A. 2023, \aap, 672, A11, \dodoi{10.1051/0004-6361/202245167}

\bibitem[{{Fan} {et~al.}(2006){Fan}, {Carilli}, \& {Keating}}]{Fan2006}
{Fan}, X., {Carilli}, C.~L., \& {Keating}, B. 2006, \araa, 44, 415, \dodoi{10.1146/annurev.astro.44.051905.092514}

\bibitem[{{Farcy} {et~al.}(2025){Farcy}, {Rosdahl}, {Dubois}, {Blaizot}, {Martin-Alvarez}, {Haehnelt}, {Kimm}, \& {Teyssier}}]{Farcy2025}
{Farcy}, M., {Rosdahl}, J., {Dubois}, Y., {et~al.} 2025, \aap, 698, A89, \dodoi{10.1051/0004-6361/202553924}

\bibitem[{{Feng} {et~al.}(2016){Feng}, {Di-Matteo}, {Croft}, {Bird}, {Battaglia}, \& {Wilkins}}]{Feng2016}
{Feng}, Y., {Di-Matteo}, T., {Croft}, R.~A., {et~al.} 2016, \mnras, 455, 2778, \dodoi{10.1093/mnras/stv2484}

\bibitem[{{Ferrara} {et~al.}(2025){Ferrara}, {Giavalisco}, {Pentericci}, {Vanzella}, {Calabr{\`o}}, \& {Llerena}}]{Ferrara2025}
{Ferrara}, A., {Giavalisco}, M., {Pentericci}, L., {et~al.} 2025, The Open Journal of Astrophysics, 8, 125, \dodoi{10.33232/001c.143600}

\bibitem[{{Firestone} {et~al.}(2024){Firestone}, {Gawiser}, {Ramakrishnan}, {Lee}, {Valdes}, {Park}, {Yang}, {Ciardullo}, {Artale}, {Benda}, {Broussard}, {Eid}, {Farooq}, {Gronwall}, {Guaita}, {Gwyn}, {Hwang}, {Im}, {Jeong}, {Karthikeyan}, {Lang}, {Moon}, {Padilla}, {Sawicki}, {Seo}, {Singh}, {Song}, \& {Troncoso Iribarren}}]{Firestone2024}
{Firestone}, N.~M., {Gawiser}, E., {Ramakrishnan}, V., {et~al.} 2024, \apj, 974, 217, \dodoi{10.3847/1538-4357/ad71c9}

\bibitem[{{Fletcher} {et~al.}(2019){Fletcher}, {Tang}, {Robertson}, {Nakajima}, {Ellis}, {Stark}, \& {Inoue}}]{Fletcher2019}
{Fletcher}, T.~J., {Tang}, M., {Robertson}, B.~E., {et~al.} 2019, \apj, 878, 87, \dodoi{10.3847/1538-4357/ab2045}

\bibitem[{{Flury} {et~al.}(2022{\natexlab{a}}){Flury}, {Jaskot}, {Ferguson}, {Worseck}, {Makan}, {Chisholm}, {Saldana-Lopez}, {Schaerer}, {McCandliss}, {Wang}, {Ford}, {Heckman}, {Ji}, {Giavalisco}, {Amorin}, {Atek}, {Blaizot}, {Borthakur}, {Carr}, {Castellano}, {Cristiani}, {De Barros}, {Dickinson}, {Finkelstein}, {Fleming}, {Fontanot}, {Garel}, {Grazian}, {Hayes}, {Henry}, {Mauerhofer}, {Micheva}, {Oey}, {Ostlin}, {Papovich}, {Pentericci}, {Ravindranath}, {Rosdahl}, {Rutkowski}, {Santini}, {Scarlata}, {Teplitz}, {Thuan}, {Trebitsch}, {Vanzella}, {Verhamme}, \& {Xu}}]{Flury2022a}
{Flury}, S.~R., {Jaskot}, A.~E., {Ferguson}, H.~C., {et~al.} 2022{\natexlab{a}}, \apjs, 260, 1, \dodoi{10.3847/1538-4365/ac5331}

\bibitem[{{Flury} {et~al.}(2022{\natexlab{b}}){Flury}, {Jaskot}, {Ferguson}, {Worseck}, {Makan}, {Chisholm}, {Saldana-Lopez}, {Schaerer}, {McCandliss}, {Xu}, {Wang}, {Oey}, {Ford}, {Heckman}, {Ji}, {Giavalisco}, {Amor{\'\i}n}, {Atek}, {Blaizot}, {Borthakur}, {Carr}, {Castellano}, {De Barros}, {Dickinson}, {Finkelstein}, {Fleming}, {Fontanot}, {Garel}, {Grazian}, {Hayes}, {Henry}, {Mauerhofer}, {Micheva}, {Ostlin}, {Papovich}, {Pentericci}, {Ravindranath}, {Rosdahl}, {Rutkowski}, {Santini}, {Scarlata}, {Teplitz}, {Thuan}, {Trebitsch}, {Vanzella}, \& {Verhamme}}]{Flury2022b}
---. 2022{\natexlab{b}}, \apj, 930, 126, \dodoi{10.3847/1538-4357/ac61e4}

\bibitem[{{Flury} {et~al.}(2025){Flury}, {Jaskot}, {Saldana-Lopez}, {Oey}, {Chisholm}, {Amor{\'\i}n}, {Bait}, {Borthakur}, {Carr}, {Ferguson}, {Giavalisco}, {Hayes}, {Heckman}, {Henry}, {Ji}, {Komarova}, {Leclercq}, {Le Reste}, {McCandliss}, {Marques-Chaves}, {{\"O}stlin}, {Pentericci}, {Ravindranath}, {Rutkowski}, {Scarlata}, {Schaerer}, {Thuan}, {Trebitsch}, {Vanzella}, {Verhamme}, {Wang}, {Worseck}, \& {Xu}}]{Flury2025}
{Flury}, S.~R., {Jaskot}, A.~E., {Saldana-Lopez}, A., {et~al.} 2025, \apj, 985, 128, \dodoi{10.3847/1538-4357/adc305}

\bibitem[{{Gazagnes} {et~al.}(2020){Gazagnes}, {Chisholm}, {Schaerer}, {Verhamme}, \& {Izotov}}]{Gazagnes2020}
{Gazagnes}, S., {Chisholm}, J., {Schaerer}, D., {Verhamme}, A., \& {Izotov}, Y. 2020, \aap, 639, A85, \dodoi{10.1051/0004-6361/202038096}

\bibitem[{{Georgakakis} {et~al.}(2015){Georgakakis}, {Aird}, {Buchner}, {Salvato}, {Menzel}, {Brandt}, {McGreer}, {Dwelly}, {Mountrichas}, {Koki}, {Georgantopoulos}, {Hsu}, {Merloni}, {Liu}, {Nandra}, \& {Ross}}]{Georgakakis2015}
{Georgakakis}, A., {Aird}, J., {Buchner}, J., {et~al.} 2015, \mnras, 453, 1946, \dodoi{10.1093/mnras/stv1703}

\bibitem[{{Giallongo} {et~al.}(2015){Giallongo}, {Grazian}, {Fiore}, {Fontana}, {Pentericci}, {Vanzella}, {Dickinson}, {Kocevski}, {Castellano}, {Cristiani}, {Ferguson}, {Finkelstein}, {Grogin}, {Hathi}, {Koekemoer}, {Newman}, \& {Salvato}}]{Giallongo2015}
{Giallongo}, E., {Grazian}, A., {Fiore}, F., {et~al.} 2015, \aap, 578, A83, \dodoi{10.1051/0004-6361/201425334}

\bibitem[{{Grazian} {et~al.}(2016){Grazian}, {Giallongo}, {Gerbasi}, {Fiore}, {Fontana}, {Le F{\`e}vre}, {Pentericci}, {Vanzella}, {Zamorani}, {Cassata}, {Garilli}, {Le Brun}, {Maccagni}, {Tasca}, {Thomas}, {Zucca}, {Amor{\'\i}n}, {Bardelli}, {Cassar{\`a}}, {Castellano}, {Cimatti}, {Cucciati}, {Durkalec}, {Giavalisco}, {Hathi}, {Ilbert}, {Lemaux}, {Paltani}, {Ribeiro}, {Schaerer}, {Scodeggio}, {Sommariva}, {Talia}, {Tresse}, {Vergani}, {Bonchi}, {Boutsia}, {Capak}, {Charlot}, {Contini}, {de la Torre}, {Dunlop}, {Fotopoulou}, {Guaita}, {Koekemoer}, {L{\'o}pez-Sanjuan}, {Mellier}, {Merlin}, {Paris}, {Pforr}, {Pilo}, {Santini}, {Scoville}, {Taniguchi}, \& {Wang}}]{Grazian2016}
{Grazian}, A., {Giallongo}, E., {Gerbasi}, R., {et~al.} 2016, \aap, 585, A48, \dodoi{10.1051/0004-6361/201526396}

\bibitem[{{Grazian} {et~al.}(2017){Grazian}, {Giallongo}, {Paris}, {Boutsia}, {Dickinson}, {Santini}, {Windhorst}, {Jansen}, {Cohen}, {Ashcraft}, {Scarlata}, {Rutkowski}, {Vanzella}, {Cusano}, {Cristiani}, {Giavalisco}, {Ferguson}, {Koekemoer}, {Grogin}, {Castellano}, {Fiore}, {Fontana}, {Marchi}, {Pedichini}, {Pentericci}, {Amor{\'\i}n}, {Barro}, {Bonchi}, {Bongiorno}, {Faber}, {Fumana}, {Galametz}, {Guaita}, {Kocevski}, {Merlin}, {Nonino}, {O'Connell}, {Pilo}, {Ryan}, {Sani}, {Speziali}, {Testa}, {Weiner}, \& {Yan}}]{Grazian2017}
{Grazian}, A., {Giallongo}, E., {Paris}, D., {et~al.} 2017, \aap, 602, A18, \dodoi{10.1051/0004-6361/201730447}

\bibitem[{{Grimes} {et~al.}(2009){Grimes}, {Heckman}, {Aloisi}, {Calzetti}, {Leitherer}, {Martin}, {Meurer}, {Sembach}, \& {Strickland}}]{Grimes2009}
{Grimes}, J.~P., {Heckman}, T., {Aloisi}, A., {et~al.} 2009, \apjs, 181, 272, \dodoi{10.1088/0067-0049/181/1/272}

\bibitem[{{Guaita} {et~al.}(2016){Guaita}, {Pentericci}, {Grazian}, {Vanzella}, {Nonino}, {Giavalisco}, {Zamorani}, {Bongiorno}, {Cassata}, {Castellano}, {Garilli}, {Gawiser}, {Le Brun}, {Le F{\`e}vre}, {Lemaux}, {Maccagni}, {Merlin}, {Santini}, {Tasca}, {Thomas}, {Zucca}, {De Barros}, {Hathi}, {Amorin}, {Bardelli}, \& {Fontana}}]{Guaita2016}
{Guaita}, L., {Pentericci}, L., {Grazian}, A., {et~al.} 2016, \aap, 587, A133, \dodoi{10.1051/0004-6361/201527597}

\bibitem[{{Hayes} {et~al.}(2014){Hayes}, {{\"O}stlin}, {Duval}, {Sandberg}, {Guaita}, {Melinder}, {Adamo}, {Schaerer}, {Verhamme}, {Orlitov{\'a}}, {Mas-Hesse}, {Cannon}, {Atek}, {Kunth}, {Laursen}, {Ot{\'\i}-Floranes}, {Pardy}, {Rivera-Thorsen}, \& {Herenz}}]{Hayes2014}
{Hayes}, M., {{\"O}stlin}, G., {Duval}, F., {et~al.} 2014, \apj, 782, 6, \dodoi{10.1088/0004-637X/782/1/6}

\bibitem[{{Inoue} {et~al.}(2005){Inoue}, {Iwata}, {Deharveng}, {Buat}, \& {Burgarella}}]{Inoue2005}
{Inoue}, A.~K., {Iwata}, I., {Deharveng}, J.-M., {Buat}, V., \& {Burgarella}, D. 2005, \aap, 435, 471, \dodoi{10.1051/0004-6361:20041769}

\bibitem[{{Inoue} {et~al.}(2014){Inoue}, {Shimizu}, {Iwata}, \& {Tanaka}}]{Inoue2014}
{Inoue}, A.~K., {Shimizu}, I., {Iwata}, I., \& {Tanaka}, M. 2014, \mnras, 442, 1805, \dodoi{10.1093/mnras/stu936}

\bibitem[{{Iwata} {et~al.}(2019){Iwata}, {Inoue}, {Micheva}, {Matsuda}, \& {Yamada}}]{Iwata2019}
{Iwata}, I., {Inoue}, A.~K., {Micheva}, G., {Matsuda}, Y., \& {Yamada}, T. 2019, \mnras, 488, 5671, \dodoi{10.1093/mnras/stz2081}

\bibitem[{{Iwata} {et~al.}(2022){Iwata}, {Sawicki}, {Inoue}, {Akiyama}, {Micheva}, {Kawaguchi}, {Kashikawa}, {Gwyn}, {Arnouts}, {Coupon}, \& {Desprez}}]{Iwata2022}
{Iwata}, I., {Sawicki}, M., {Inoue}, A.~K., {et~al.} 2022, \mnras, 509, 1820, \dodoi{10.1093/mnras/stab2742}

\bibitem[{{Izotov} {et~al.}(2016{\natexlab{a}}){Izotov}, {Orlitov{\'a}}, {Schaerer}, {Thuan}, {Verhamme}, {Guseva}, \& {Worseck}}]{Izotov2016a}
{Izotov}, Y.~I., {Orlitov{\'a}}, I., {Schaerer}, D., {et~al.} 2016{\natexlab{a}}, \nat, 529, 178, \dodoi{10.1038/nature16456}

\bibitem[{{Izotov} {et~al.}(2016{\natexlab{b}}){Izotov}, {Schaerer}, {Thuan}, {Worseck}, {Guseva}, {Orlitov{\'a}}, \& {Verhamme}}]{Izotov2016b}
{Izotov}, Y.~I., {Schaerer}, D., {Thuan}, T.~X., {et~al.} 2016{\natexlab{b}}, \mnras, 461, 3683, \dodoi{10.1093/mnras/stw1205}

\bibitem[{{Izotov} {et~al.}(2018{\natexlab{a}}){Izotov}, {Schaerer}, {Worseck}, {Guseva}, {Thuan}, {Verhamme}, {Orlitov{\'a}}, \& {Fricke}}]{Izotov2018a}
{Izotov}, Y.~I., {Schaerer}, D., {Worseck}, G., {et~al.} 2018{\natexlab{a}}, \mnras, 474, 4514, \dodoi{10.1093/mnras/stx3115}

\bibitem[{{Izotov} {et~al.}(2020){Izotov}, {Schaerer}, {Worseck}, {Verhamme}, {Guseva}, {Thuan}, {Orlitov{\'a}}, \& {Fricke}}]{Izotov2020}
---. 2020, \mnras, 491, 468, \dodoi{10.1093/mnras/stz3041}

\bibitem[{{Izotov} {et~al.}(2021){Izotov}, {Worseck}, {Schaerer}, {Guseva}, {Chisholm}, {Thuan}, {Fricke}, \& {Verhamme}}]{Izotov2021}
{Izotov}, Y.~I., {Worseck}, G., {Schaerer}, D., {et~al.} 2021, \mnras, 503, 1734, \dodoi{10.1093/mnras/stab612}

\bibitem[{{Izotov} {et~al.}(2018{\natexlab{b}}){Izotov}, {Worseck}, {Schaerer}, {Guseva}, {Thuan}, {Fricke}, \& {Orlitov{\'a}}}]{Izotov2018b}
---. 2018{\natexlab{b}}, \mnras, 478, 4851, \dodoi{10.1093/mnras/sty1378}

\bibitem[{{Japelj} {et~al.}(2017){Japelj}, {Vanzella}, {Fontanot}, {Cristiani}, {Caminha}, {Tozzi}, {Balestra}, {Rosati}, \& {Meneghetti}}]{Japelj2017}
{Japelj}, J., {Vanzella}, E., {Fontanot}, F., {et~al.} 2017, \mnras, 468, 389, \dodoi{10.1093/mnras/stx477}

\bibitem[{{Ji} {et~al.}(2020){Ji}, {Giavalisco}, {Vanzella}, {Siana}, {Pentericci}, {Jaskot}, {Liu}, {Nonino}, {Ferguson}, {Castellano}, {Mannucci}, {Schaerer}, {Fynbo}, {Papovich}, {Carnall}, {Amorin}, {Simons}, {Hathi}, {Cullen}, \& {McLeod}}]{Ji2020}
{Ji}, Z., {Giavalisco}, M., {Vanzella}, E., {et~al.} 2020, \apj, 888, 109, \dodoi{10.3847/1538-4357/ab5fdc}

\bibitem[{{Jia} {et~al.}(2011){Jia}, {Ptak}, {Heckman}, {Overzier}, {Hornschemeier}, \& {LaMassa}}]{Jia2011}
{Jia}, J., {Ptak}, A., {Heckman}, T.~M., {et~al.} 2011, \apj, 731, 55, \dodoi{10.1088/0004-637X/731/1/55}

\bibitem[{{Keenan} {et~al.}(2017){Keenan}, {Oey}, {Jaskot}, \& {James}}]{Keenan2017}
{Keenan}, R.~P., {Oey}, M.~S., {Jaskot}, A.~E., \& {James}, B.~L. 2017, \apj, 848, 12, \dodoi{10.3847/1538-4357/aa8b77}

\bibitem[{{Kerutt} {et~al.}(2024){Kerutt}, {Oesch}, {Wisotzki}, {Verhamme}, {Atek}, {Herenz}, {Illingworth}, {Kusakabe}, {Matthee}, {Mauerhofer}, {Montes}, {Naidu}, {Nelson}, {Reddy}, {Schaye}, {Simmonds}, {Urrutia}, \& {Vitte}}]{Kerutt2024}
{Kerutt}, J., {Oesch}, P.~A., {Wisotzki}, L., {et~al.} 2024, \aap, 684, A42, \dodoi{10.1051/0004-6361/202346656}

\bibitem[{{Kimm} \& {Cen}(2014)}]{Kimm2014}
{Kimm}, T., \& {Cen}, R. 2014, \apj, 788, 121, \dodoi{10.1088/0004-637X/788/2/121}

\bibitem[{{Kornei} {et~al.}(2010){Kornei}, {Shapley}, {Erb}, {Steidel}, {Reddy}, {Pettini}, \& {Bogosavljevi{\'c}}}]{Kornei2010}
{Kornei}, K.~A., {Shapley}, A.~E., {Erb}, D.~K., {et~al.} 2010, \apj, 711, 693, \dodoi{10.1088/0004-637X/711/2/693}

\bibitem[{{Le Reste} {et~al.}(2025){Le Reste}, {Scarlata}, {Hayes}, {Melinder}, {Saldana-Lopez}, {Runnholm}, {Lin}, {Amor{\'\i}n}, {Atek}, {Borthakur}, {Carr}, {Flury}, {Giavalisco}, {Henry}, {Jaskot}, {Ji}, {Jung}, {Leclercq}, {Marques-Chaves}, {McCandliss}, {Oey}, {{\"O}stlin}, {Ravindranath}, {Schaerer}, {Thuan}, \& {Xu}}]{LeReste2025}
{Le Reste}, A., {Scarlata}, C., {Hayes}, M., {et~al.} 2025, arXiv e-prints, arXiv:2504.07056, \dodoi{10.48550/arXiv.2504.07056}

\bibitem[{{Lee} {et~al.}(2024){Lee}, {Gawiser}, {Park}, {Yang}, {Valdes}, {Lang}, {Ramakrishnan}, {Moon}, {Firestone}, {Appleby}, {Artale}, {Andrews}, {Bauer}, {Benda}, {Broussard}, {Chiang}, {Ciardullo}, {Dey}, {Farooq}, {Gronwall}, {Guaita}, {Huang}, {Hwang}, {Im}, {Jeong}, {Karthikeyan}, {Kim}, {Kim}, {Kumar}, {Nagaraj}, {Nantais}, {Padilla}, {Park}, {Pope}, {Popescu}, {Schlegel}, {Seo}, {Singh}, {Song}, {Troncoso}, {Vivas}, {Zabludoff}, \& {Zenteno}}]{Lee2024}
{Lee}, K.-S., {Gawiser}, E., {Park}, C., {et~al.} 2024, \apj, 962, 36, \dodoi{10.3847/1538-4357/ad165e}

\bibitem[{{Leitet} {et~al.}(2011){Leitet}, {Bergvall}, {Piskunov}, \& {Andersson}}]{Leitet2011}
{Leitet}, E., {Bergvall}, N., {Piskunov}, N., \& {Andersson}, B.~G. 2011, \aap, 532, A107, \dodoi{10.1051/0004-6361/201015654}

\bibitem[{{Leitherer} {et~al.}(1995){Leitherer}, {Ferguson}, {Heckman}, \& {Lowenthal}}]{Leitherer1995}
{Leitherer}, C., {Ferguson}, H.~C., {Heckman}, T.~M., \& {Lowenthal}, J.~D. 1995, \apjl, 454, L19, \dodoi{10.1086/309760}

\bibitem[{{Lejeune} {et~al.}(1997){Lejeune}, {Cuisinier}, \& {Buser}}]{Lejeune1997}
{Lejeune}, T., {Cuisinier}, F., \& {Buser}, R. 1997, \aaps, 125, 229, \dodoi{10.1051/aas:1997373}

\bibitem[{{Lejeune} {et~al.}(1998){Lejeune}, {Cuisinier}, \& {Buser}}]{Lejeune1998}
---. 1998, \aaps, 130, 65, \dodoi{10.1051/aas:1998405}

\bibitem[{{Liu} {et~al.}(2025){Liu}, {Mascia}, {Pentericci}, {Watson}, {Alavi}, {Bergamini}, {Brada{\v{c}}}, {Calabr{\`o}}, {Glazebrook}, {Henry}, {Llerena}, {Merlin}, {Metha}, {Nanayakkara}, {Napolitano}, {Roy}, {Siana}, {Vanzella}, {Vulcani}, \& {Wang}}]{Liu2025}
{Liu}, Y., {Mascia}, S., {Pentericci}, L., {et~al.} 2025, \aap, 704, A328, \dodoi{10.1051/0004-6361/202556410}

\bibitem[{{Madau} \& {Haardt}(2015)}]{Madau2015}
{Madau}, P., \& {Haardt}, F. 2015, \apjl, 813, L8, \dodoi{10.1088/2041-8205/813/1/L8}

\bibitem[{{Marchi} {et~al.}(2017){Marchi}, {Pentericci}, {Guaita}, {Ribeiro}, {Castellano}, {Schaerer}, {Hathi}, {Lemaux}, {Grazian}, {Le F{\`e}vre}, {Garilli}, {Maccagni}, {Amorin}, {Bardelli}, {Cassata}, {Fontana}, {Koekemoer}, {Le Brun}, {Tasca}, {Thomas}, {Vanzella}, {Zamorani}, \& {Zucca}}]{Marchi2017}
{Marchi}, F., {Pentericci}, L., {Guaita}, L., {et~al.} 2017, \aap, 601, A73, \dodoi{10.1051/0004-6361/201630054}

\bibitem[{{Marchi} {et~al.}(2018){Marchi}, {Pentericci}, {Guaita}, {Schaerer}, {Verhamme}, {Castellano}, {Ribeiro}, {Garilli}, {Le F{\`e}vre}, {Amorin}, {Bardelli}, {Cassata}, {Durkalec}, {Grazian}, {Hathi}, {Lemaux}, {Maccagni}, {Vanzella}, \& {Zucca}}]{Marchi2018}
---. 2018, \aap, 614, A11, \dodoi{10.1051/0004-6361/201732133}

\bibitem[{{Marigo} \& {Girardi}(2007)}]{Marigo2007}
{Marigo}, P., \& {Girardi}, L. 2007, \aap, 469, 239, \dodoi{10.1051/0004-6361:20066772}

\bibitem[{{Marigo} {et~al.}(2008){Marigo}, {Girardi}, {Bressan}, {Groenewegen}, {Silva}, \& {Granato}}]{Marigo2008}
{Marigo}, P., {Girardi}, L., {Bressan}, A., {et~al.} 2008, \aap, 482, 883, \dodoi{10.1051/0004-6361:20078467}

\bibitem[{{Marques-Chaves} {et~al.}(2021){Marques-Chaves}, {Schaerer}, {{\'A}lvarez-M{\'a}rquez}, {Colina}, {Dessauges-Zavadsky}, {P{\'e}rez-Fournon}, {Saldana-Lopez}, \& {Verhamme}}]{Marques-Chaves2021}
{Marques-Chaves}, R., {Schaerer}, D., {{\'A}lvarez-M{\'a}rquez}, J., {et~al.} 2021, \mnras, 507, 524, \dodoi{10.1093/mnras/stab2187}

\bibitem[{{Marques-Chaves} {et~al.}(2022){Marques-Chaves}, {Schaerer}, {Amor{\'\i}n}, {Atek}, {Borthakur}, {Chisholm}, {Fern{\'a}ndez}, {Flury}, {Giavalisco}, {Grazian}, {Hayes}, {Heckman}, {Henry}, {Izotov}, {Jaskot}, {Ji}, {McCandliss}, {Oey}, {{\"O}stlin}, {Ravindranath}, {Rutkowski}, {Saldana-Lopez}, {Teplitz}, {Thuan}, {Verhamme}, {Wang}, {Worseck}, \& {Xu}}]{Marques-Chaves2022}
{Marques-Chaves}, R., {Schaerer}, D., {Amor{\'\i}n}, R.~O., {et~al.} 2022, \aap, 663, L1, \dodoi{10.1051/0004-6361/202243598}

\bibitem[{{Matthee} {et~al.}(2017){Matthee}, {Sobral}, {Best}, {Khostovan}, {Oteo}, {Bouwens}, \& {R{\"o}ttgering}}]{Matthee2017}
{Matthee}, J., {Sobral}, D., {Best}, P., {et~al.} 2017, \mnras, 465, 3637, \dodoi{10.1093/mnras/stw2973}

\bibitem[{{Micheva} {et~al.}(2017){Micheva}, {Iwata}, {Inoue}, {Matsuda}, {Yamada}, \& {Hayashino}}]{Micheva2017}
{Micheva}, G., {Iwata}, I., {Inoue}, A.~K., {et~al.} 2017, \mnras, 465, 316, \dodoi{10.1093/mnras/stw2700}

\bibitem[{{Mitra} {et~al.}(2015){Mitra}, {Choudhury}, \& {Ferrara}}]{Mitra2015}
{Mitra}, S., {Choudhury}, T.~R., \& {Ferrara}, A. 2015, \mnras, 454, L76, \dodoi{10.1093/mnrasl/slv134}

\bibitem[{{Mostardi} {et~al.}(2013){Mostardi}, {Shapley}, {Nestor}, {Steidel}, {Reddy}, \& {Trainor}}]{Mostardi2013}
{Mostardi}, R.~E., {Shapley}, A.~E., {Nestor}, D.~B., {et~al.} 2013, \apj, 779, 65, \dodoi{10.1088/0004-637X/779/1/65}

\bibitem[{{Mostardi} {et~al.}(2015){Mostardi}, {Shapley}, {Steidel}, {Trainor}, {Reddy}, \& {Siana}}]{Mostardi2015}
{Mostardi}, R.~E., {Shapley}, A.~E., {Steidel}, C.~C., {et~al.} 2015, \apj, 810, 107, \dodoi{10.1088/0004-637X/810/2/107}

\bibitem[{{Naidu} {et~al.}(2020){Naidu}, {Tacchella}, {Mason}, {Bose}, {Oesch}, \& {Conroy}}]{Naidu2020}
{Naidu}, R.~P., {Tacchella}, S., {Mason}, C.~A., {et~al.} 2020, \apj, 892, 109, \dodoi{10.3847/1538-4357/ab7cc9}

\bibitem[{{Naidu} {et~al.}(2022){Naidu}, {Oesch}, {van Dokkum}, {Nelson}, {Suess}, {Brammer}, {Whitaker}, {Illingworth}, {Bouwens}, {Tacchella}, {Matthee}, {Allen}, {Bezanson}, {Conroy}, {Labbe}, {Leja}, {Leonova}, {Magee}, {Price}, {Setton}, {Strait}, {Stefanon}, {Toft}, {Weaver}, \& {Weibel}}]{Naidu2022}
{Naidu}, R.~P., {Oesch}, P.~A., {van Dokkum}, P., {et~al.} 2022, \apjl, 940, L14, \dodoi{10.3847/2041-8213/ac9b22}

\bibitem[{{Nestor} {et~al.}(2011){Nestor}, {Shapley}, {Steidel}, \& {Siana}}]{Nestor2011}
{Nestor}, D.~B., {Shapley}, A.~E., {Steidel}, C.~C., \& {Siana}, B. 2011, \apj, 736, 18, \dodoi{10.1088/0004-637X/736/1/18}

\bibitem[{{Ouchi} {et~al.}(2009{\natexlab{a}}){Ouchi}, {Mobasher}, {Shimasaku}, {Ferguson}, {Fall}, {Ono}, {Kashikawa}, {Morokuma}, {Nakajima}, {Okamura}, {Dickinson}, {Giavalisco}, \& {Ohta}}]{Ouchi2009a}
{Ouchi}, M., {Mobasher}, B., {Shimasaku}, K., {et~al.} 2009{\natexlab{a}}, \apj, 706, 1136, \dodoi{10.1088/0004-637X/706/2/1136}

\bibitem[{{Ouchi} {et~al.}(2009{\natexlab{b}}){Ouchi}, {Ono}, {Egami}, {Saito}, {Oguri}, {McCarthy}, {Farrah}, {Kashikawa}, {Momcheva}, {Shimasaku}, {Nakanishi}, {Furusawa}, {Akiyama}, {Dunlop}, {Mortier}, {Okamura}, {Hayashi}, {Cirasuolo}, {Dressler}, {Iye}, {Jarvis}, {Kodama}, {Martin}, {McLure}, {Ohta}, {Yamada}, \& {Yoshida}}]{Ouchi2009b}
{Ouchi}, M., {Ono}, Y., {Egami}, E., {et~al.} 2009{\natexlab{b}}, \apj, 696, 1164, \dodoi{10.1088/0004-637X/696/2/1164}

\bibitem[{{Papovich} {et~al.}(2005){Papovich}, {Dickinson}, {Giavalisco}, {Conselice}, \& {Ferguson}}]{Papovich2005}
{Papovich}, C., {Dickinson}, M., {Giavalisco}, M., {Conselice}, C.~J., \& {Ferguson}, H.~C. 2005, \apj, 631, 101, \dodoi{10.1086/429120}

\bibitem[{{Pawlik} {et~al.}(2009){Pawlik}, {Schaye}, \& {van Scherpenzeel}}]{Pawlik2009}
{Pawlik}, A.~H., {Schaye}, J., \& {van Scherpenzeel}, E. 2009, \mnras, 394, 1812, \dodoi{10.1111/j.1365-2966.2009.14486.x}

\bibitem[{{Pinarski} {et~al.}(2026){Pinarski}, {Ramgopal}, {Firestone}, {Lee}, {Gawiser}, {Dey}, {Raichoor}, {Valdes}, {Ciardullo}, {Aguilar}, {Ahlen}, {Bianchi}, {Brooks}, {Castander}, {Candela Cerdosino}, {Claybaugh}, {Cuceu}, {Dawson}, {de la Macorra}, {Doel}, {Ferraro}, {Font-Ribera}, {Forero-Romero}, {Gazta{\~n}aga}, {Gontcho}, {Guaita}, {Gutierrez}, {Gwyn}, {Herrera-Alcantar}, {Hwang}, {Joyce}, {Juneau}, {Kehoe}, {Kirkby}, {Kisner}, {Kremin}, {Kumar}, {Lamman}, {Landriau}, {Le Guillou}, {Levi}, {Luo}, {Manera}, {Martini}, {Meisner}, {Miquel}, {Moustakas}, {Myers}, {Nadathur}, {Nagaraj}, {Palanque-Delabrouille}, {Park}, {Percival}, {P{\'e}rez-R{\`a}fols}, {Prada}, {Rossi}, {Sanchez}, {Sawicki}, {Schlegel}, {Schubnell}, {Silber}, {Song}, {Sprayberry}, {Tarl{\'e}}, {Troncoso Iribarren}, {Weaver}, {Yang}, \& {Zabludoff}}]{ODIN-DESI}
{Pinarski}, E., {Ramgopal}, G., {Firestone}, N., {et~al.} 2026, arXiv e-prints, arXiv:2603.09905.
\newblock \doarXiv{2603.09905}

\bibitem[{{Pirzkal} {et~al.}(2007){Pirzkal}, {Malhotra}, {Rhoads}, \& {Xu}}]{Pirzkal2007}
{Pirzkal}, N., {Malhotra}, S., {Rhoads}, J.~E., \& {Xu}, C. 2007, \apj, 667, 49, \dodoi{10.1086/519485}

\bibitem[{{Prichard} {et~al.}(2022){Prichard}, {Rafelski}, {Cooke}, {Me{\v{s}}tri{\'c}}, {Bassett}, {Ryan-Weber}, {Sunnquist}, {Alavi}, {Hathi}, {Wang}, {Revalski}, {Bajaj}, {O'Meara}, \& {Spitler}}]{Prichard2022}
{Prichard}, L.~J., {Rafelski}, M., {Cooke}, J., {et~al.} 2022, \apj, 924, 14, \dodoi{10.3847/1538-4357/ac3004}

\bibitem[{{Puschnig} {et~al.}(2020){Puschnig}, {Hayes}, {{\"O}stlin}, {Cannon}, {Smirnova-Pinchukova}, {Husemann}, {Kunth}, {Bridge}, {Herenz}, {Messa}, \& {Oteo}}]{Puschnig2020}
{Puschnig}, J., {Hayes}, M., {{\"O}stlin}, G., {et~al.} 2020, \aap, 644, A10, \dodoi{10.1051/0004-6361/201936768}

\bibitem[{{Reddy} {et~al.}(2016){Reddy}, {Steidel}, {Pettini}, {Bogosavljevi{\'c}}, \& {Shapley}}]{Reddy2016}
{Reddy}, N.~A., {Steidel}, C.~C., {Pettini}, M., {Bogosavljevi{\'c}}, M., \& {Shapley}, A.~E. 2016, \apj, 828, 108, \dodoi{10.3847/0004-637X/828/2/108}

\bibitem[{{Reddy} {et~al.}(2022){Reddy}, {Topping}, {Shapley}, {Steidel}, {Sanders}, {Du}, {Coil}, {Mobasher}, {Price}, \& {Shivaei}}]{Reddy2022}
{Reddy}, N.~A., {Topping}, M.~W., {Shapley}, A.~E., {et~al.} 2022, \apj, 926, 31, \dodoi{10.3847/1538-4357/ac3b4c}

\bibitem[{{Rivera-Thorsen} {et~al.}(2017{\natexlab{a}}){Rivera-Thorsen}, {{\"O}stlin}, {Hayes}, \& {Puschnig}}]{Rivera-Thorsen2017}
{Rivera-Thorsen}, T.~E., {{\"O}stlin}, G., {Hayes}, M., \& {Puschnig}, J. 2017{\natexlab{a}}, \apj, 837, 29, \dodoi{10.3847/1538-4357/aa5d0a}

\bibitem[{{Rivera-Thorsen} {et~al.}(2017{\natexlab{b}}){Rivera-Thorsen}, {{\"O}stlin}, {Hayes}, \& {Puschnig}}]{Rivera-Thorsen2017b}
---. 2017{\natexlab{b}}, \apj, 837, 29, \dodoi{10.3847/1538-4357/aa5d0a}

\bibitem[{{Rivera-Thorsen} {et~al.}(2019){Rivera-Thorsen}, {Dahle}, {Chisholm}, {Florian}, {Gronke}, {Rigby}, {Gladders}, {Mahler}, {Sharon}, \& {Bayliss}}]{Rivera-Thorsen2019}
{Rivera-Thorsen}, T.~E., {Dahle}, H., {Chisholm}, J., {et~al.} 2019, Science, 366, 738, \dodoi{10.1126/science.aaw0978}

\bibitem[{{Robertson} {et~al.}(2015){Robertson}, {Ellis}, {Furlanetto}, \& {Dunlop}}]{Robertson2015}
{Robertson}, B.~E., {Ellis}, R.~S., {Furlanetto}, S.~R., \& {Dunlop}, J.~S. 2015, \apjl, 802, L19, \dodoi{10.1088/2041-8205/802/2/L19}

\bibitem[{{Robertson} {et~al.}(2013){Robertson}, {Furlanetto}, {Schneider}, {Charlot}, {Ellis}, {Stark}, {McLure}, {Dunlop}, {Koekemoer}, {Schenker}, {Ouchi}, {Ono}, {Curtis-Lake}, {Rogers}, {Bowler}, \& {Cirasuolo}}]{Robertson2013}
{Robertson}, B.~E., {Furlanetto}, S.~R., {Schneider}, E., {et~al.} 2013, \apj, 768, 71, \dodoi{10.1088/0004-637X/768/1/71}

\bibitem[{{Rutkowski} {et~al.}(2016){Rutkowski}, {Scarlata}, {Haardt}, {Siana}, {Henry}, {Rafelski}, {Hayes}, {Salvato}, {Pahl}, {Mehta}, {Beck}, {Malkan}, \& {Teplitz}}]{Rutkowski2016}
{Rutkowski}, M.~J., {Scarlata}, C., {Haardt}, F., {et~al.} 2016, \apj, 819, 81, \dodoi{10.3847/0004-637X/819/1/81}

\bibitem[{{Saldana-Lopez} {et~al.}(2025){Saldana-Lopez}, {Hayes}, {Le Reste}, {Scarlata}, {Melinder}, {Henry}, {Amorin}, {Atek}, {Bait}, {Chisholm}, {Jaskot}, {Jung}, {Ji}, {Komarova}, {Leclercq}, {Ostlin}, {Runnholm}, {Thuan}, \& {Xu}}]{Saldana-Lopez2025}
{Saldana-Lopez}, A., {Hayes}, M.~J., {Le Reste}, A., {et~al.} 2025, arXiv e-prints, arXiv:2504.07074, \dodoi{10.48550/arXiv.2504.07074}

\bibitem[{{Sawicki} {et~al.}(2019){Sawicki}, {Arnouts}, {Huang}, {Coupon}, {Golob}, {Gwyn}, {Foucaud}, {Moutard}, {Iwata}, {Liu}, {Chen}, {Desprez}, {Harikane}, {Ono}, {Strauss}, {Tanaka}, {Thibert}, {Balogh}, {Bundy}, {Chapman}, {Gunn}, {Hsieh}, {Ilbert}, {Jing}, {LeF{\`e}vre}, {Li}, {Matsuda}, {Miyazaki}, {Nagao}, {Nishizawa}, {Ouchi}, {Shimasaku}, {Silverman}, {de la Torre}, {Tresse}, {Wang}, {Willott}, {Yamada}, {Yang}, \& {Yee}}]{Sawicki2019}
{Sawicki}, M., {Arnouts}, S., {Huang}, J., {et~al.} 2019, \mnras, 489, 5202, \dodoi{10.1093/mnras/stz2522}

\bibitem[{{Saxena} {et~al.}(2022){Saxena}, {Pentericci}, {Ellis}, {Guaita}, {Calabr{\`o}}, {Schaerer}, {Vanzella}, {Amor{\'\i}n}, {Bolzonella}, {Castellano}, {Fontanot}, {Hathi}, {Hibon}, {Llerena}, {Mannucci}, {Saldana-Lopez}, {Talia}, \& {Zamorani}}]{Saxena2022}
{Saxena}, A., {Pentericci}, L., {Ellis}, R.~S., {et~al.} 2022, \mnras, 511, 120, \dodoi{10.1093/mnras/stab3728}

\bibitem[{{Shapley} {et~al.}(2003){Shapley}, {Steidel}, {Pettini}, \& {Adelberger}}]{Shapley2003}
{Shapley}, A.~E., {Steidel}, C.~C., {Pettini}, M., \& {Adelberger}, K.~L. 2003, \apj, 588, 65, \dodoi{10.1086/373922}

\bibitem[{{Shapley} {et~al.}(2006){Shapley}, {Steidel}, {Pettini}, {Adelberger}, \& {Erb}}]{Shapley2006}
{Shapley}, A.~E., {Steidel}, C.~C., {Pettini}, M., {Adelberger}, K.~L., \& {Erb}, D.~K. 2006, \apj, 651, 688, \dodoi{10.1086/507511}

\bibitem[{{Shapley} {et~al.}(2016){Shapley}, {Steidel}, {Strom}, {Bogosavljevi{\'c}}, {Reddy}, {Siana}, {Mostardi}, \& {Rudie}}]{Shapley2016}
{Shapley}, A.~E., {Steidel}, C.~C., {Strom}, A.~L., {et~al.} 2016, \apjl, 826, L24, \dodoi{10.3847/2041-8205/826/2/L24}

\bibitem[{{Shuntov} {et~al.}(2025){Shuntov}, {Akins}, {Paquereau}, {Casey}, {Ilbert}, {Arango-Toro}, {McCracken}, {Franco}, {Harish}, {Kartaltepe}, {Koekemoer}, {Yang}, {Huertas-Company}, {Berman}, {McCleary}, {Toft}, {Gavazzi}, {Achenbach}, {Bertin}, {Brinch}, {Champagne}, {Chartab}, {Drakos}, {Egami}, {Endsley}, {Faisst}, {Fan}, {Flayhart}, {Hartley}, {Hatamnia}, {Gozaliasl}, {Gentile}, {Jermann}, {Jin}, {Kakiichi}, {Khostovan}, {K{\"u}mmel}, {Laigle}, {Laishram}, {Lambrides}, {Liu}, {Lyu}, {Magdis}, {Mobasher}, {Moutard}, {Renzini}, {Robertson}, {Schefer}, {Scognamiglio}, {Scoville}, {Sattari}, {Sanders}, {Taamoli}, {Trakhtenbrot}, {Valentino}, {Wang}, {Weaver}, \& {Yang}}]{Shuntov2025}
{Shuntov}, M., {Akins}, H.~B., {Paquereau}, L., {et~al.} 2025, arXiv e-prints, arXiv:2506.03243, \dodoi{10.48550/arXiv.2506.03243}

\bibitem[{{Siana} {et~al.}(2015){Siana}, {Shapley}, {Kulas}, {Nestor}, {Steidel}, {Teplitz}, {Alavi}, {Brown}, {Conselice}, {Ferguson}, {Dickinson}, {Giavalisco}, {Colbert}, {Bridge}, {Gardner}, \& {de Mello}}]{Siana2015}
{Siana}, B., {Shapley}, A.~E., {Kulas}, K.~R., {et~al.} 2015, \apj, 804, 17, \dodoi{10.1088/0004-637X/804/1/17}

\bibitem[{{Steidel} {et~al.}(2018){Steidel}, {Bogosavljevi{\'c}}, {Shapley}, {Reddy}, {Rudie}, {Pettini}, {Trainor}, \& {Strom}}]{Steidel2018}
{Steidel}, C.~C., {Bogosavljevi{\'c}}, M., {Shapley}, A.~E., {et~al.} 2018, \apj, 869, 123, \dodoi{10.3847/1538-4357/aaed28}

\bibitem[{{Steidel} {et~al.}(1996){Steidel}, {Giavalisco}, {Dickinson}, \& {Adelberger}}]{Steidel1996}
{Steidel}, C.~C., {Giavalisco}, M., {Dickinson}, M., \& {Adelberger}, K.~L. 1996, \aj, 112, 352, \dodoi{10.1086/118019}

\bibitem[{{Steidel} {et~al.}(2001){Steidel}, {Pettini}, \& {Adelberger}}]{Steidel2001}
{Steidel}, C.~C., {Pettini}, M., \& {Adelberger}, K.~L. 2001, \apj, 546, 665, \dodoi{10.1086/318323}

\bibitem[{{Vanzella} {et~al.}(2010{\natexlab{a}}){Vanzella}, {Siana}, {Cristiani}, \& {Nonino}}]{Vanzella2010b}
{Vanzella}, E., {Siana}, B., {Cristiani}, S., \& {Nonino}, M. 2010{\natexlab{a}}, \mnras, 404, 1672, \dodoi{10.1111/j.1365-2966.2010.16408.x}

\bibitem[{{Vanzella} {et~al.}(2010{\natexlab{b}}){Vanzella}, {Giavalisco}, {Inoue}, {Nonino}, {Fontanot}, {Cristiani}, {Grazian}, {Dickinson}, {Stern}, {Tozzi}, {Giallongo}, {Ferguson}, {Spinrad}, {Boutsia}, {Fontana}, {Rosati}, \& {Pentericci}}]{Vanzella2010a}
{Vanzella}, E., {Giavalisco}, M., {Inoue}, A.~K., {et~al.} 2010{\natexlab{b}}, \apj, 725, 1011, \dodoi{10.1088/0004-637X/725/1/1011}

\bibitem[{{Vanzella} {et~al.}(2012){Vanzella}, {Guo}, {Giavalisco}, {Grazian}, {Castellano}, {Cristiani}, {Dickinson}, {Fontana}, {Nonino}, {Giallongo}, {Pentericci}, {Galametz}, {Faber}, {Ferguson}, {Grogin}, {Koekemoer}, {Newman}, \& {Siana}}]{Vanzella2012}
{Vanzella}, E., {Guo}, Y., {Giavalisco}, M., {et~al.} 2012, \apj, 751, 70, \dodoi{10.1088/0004-637X/751/1/70}

\bibitem[{{Vanzella} {et~al.}(2015){Vanzella}, {de Barros}, {Castellano}, {Grazian}, {Inoue}, {Schaerer}, {Guaita}, {Zamorani}, {Giavalisco}, {Siana}, {Pentericci}, {Giallongo}, {Fontana}, \& {Vignali}}]{Vanzella2015}
{Vanzella}, E., {de Barros}, S., {Castellano}, M., {et~al.} 2015, \aap, 576, A116, \dodoi{10.1051/0004-6361/201525651}

\bibitem[{{Vanzella} {et~al.}(2016){Vanzella}, {de Barros}, {Vasei}, {Alavi}, {Giavalisco}, {Siana}, {Grazian}, {Hasinger}, {Suh}, {Cappelluti}, {Vito}, {Amorin}, {Balestra}, {Brusa}, {Calura}, {Castellano}, {Comastri}, {Fontana}, {Gilli}, {Mignoli}, {Pentericci}, {Vignali}, \& {Zamorani}}]{Vanzella2016}
{Vanzella}, E., {de Barros}, S., {Vasei}, K., {et~al.} 2016, \apj, 825, 41, \dodoi{10.3847/0004-637X/825/1/41}

\bibitem[{{Vanzella} {et~al.}(2018){Vanzella}, {Nonino}, {Cupani}, {Castellano}, {Sani}, {Mignoli}, {Calura}, {Meneghetti}, {Gilli}, {Comastri}, {Mercurio}, {Caminha}, {Caputi}, {Rosati}, {Grillo}, {Cristiani}, {Balestra}, {Fontana}, \& {Giavalisco}}]{Vanzella2018}
{Vanzella}, E., {Nonino}, M., {Cupani}, G., {et~al.} 2018, \mnras, 476, L15, \dodoi{10.1093/mnrasl/sly023}

\bibitem[{{Vanzella} {et~al.}(2020){Vanzella}, {Caminha}, {Calura}, {Cupani}, {Meneghetti}, {Castellano}, {Rosati}, {Mercurio}, {Sani}, {Grillo}, {Gilli}, {Mignoli}, {Comastri}, {Nonino}, {Cristiani}, {Giavalisco}, \& {Caputi}}]{Vanzella2020}
{Vanzella}, E., {Caminha}, G.~B., {Calura}, F., {et~al.} 2020, \mnras, 491, 1093, \dodoi{10.1093/mnras/stz2286}

\bibitem[{{Venemans} {et~al.}(2005){Venemans}, {R{\"o}ttgering}, {Miley}, {Kurk}, {De Breuck}, {Overzier}, {van Breugel}, {Carilli}, {Ford}, {Heckman}, {Pentericci}, \& {McCarthy}}]{Venemans2005}
{Venemans}, B.~P., {R{\"o}ttgering}, H.~J.~A., {Miley}, G.~K., {et~al.} 2005, \aap, 431, 793, \dodoi{10.1051/0004-6361:20042038}

\bibitem[{{Verhamme} {et~al.}(2015){Verhamme}, {Orlitov{\'a}}, {Schaerer}, \& {Hayes}}]{Verhamme2015}
{Verhamme}, A., {Orlitov{\'a}}, I., {Schaerer}, D., \& {Hayes}, M. 2015, \aap, 578, A7, \dodoi{10.1051/0004-6361/201423978}

\bibitem[{{Wang} {et~al.}(2025){Wang}, {Teplitz}, {Smith}, {Windhorst}, {Rafelski}, {Mehta}, {Alavi}, {Ji}, {Brammer}, {Colbert}, {Grogin}, {Hathi}, {Koekemoer}, {Prichard}, {Scarlata}, {Sunnquist}, {Arrabal Haro}, {Conselice}, {Gawiser}, {Guo}, {Hayes}, {Jansen}, {Lucas}, {O'Connell}, {Robertson}, {Rutkowski}, {Siana}, {Vanzella}, {Ashcraft}, {Bagley}, {Baronchelli}, {Barro}, {Blanche}, {Broussard}, {Carleton}, {Chartab}, {Cheng}, {Codoreanu}, {Cohen}, {Dai}, {Darvish}, {Dav{\'e}}, {Degroot}, {de Mello}, {Dickinson}, {Emami}, {Ferguson}, {Ferreira}, {Finkelstein}, {Finkelstein}, {Gardner}, {Gburek}, {Giavalisco}, {Grazian}, {Gronwall}, {Hemmati}, {Howell}, {Iyer}, {Kaviraj}, {Kurczynski}, {Lazar}, {MacKenty}, {Mantha}, {Martin}, {Martin}, {McCabe}, {Mobasher}, {Nedkova}, {Olsen}, {Otteson}, {Ravindranath}, {Redshaw}, {Sattari}, {Soto}, {Yung}, {Zabelle}, \& {UVCANDELS Team}}]{Wang2025}
{Wang}, X., {Teplitz}, H.~I., {Smith}, B.~M., {et~al.} 2025, \apj, 980, 74, \dodoi{10.3847/1538-4357/ada4ab}

\bibitem[{{Ward} {et~al.}(2024){Ward}, {de la Vega}, {Mobasher}, {McGrath}, {Iyer}, {Calabr{\`o}}, {Costantin}, {Dickinson}, {Holwerda}, {Huertas-Company}, {Hirschmann}, {Lucas}, {Pandya}, {Wilkins}, {Yung}, {Arrabal Haro}, {Bagley}, {Finkelstein}, {Kartaltepe}, {Koekemoer}, {Papovich}, \& {Pirzkal}}]{Ward2024}
{Ward}, E., {de la Vega}, A., {Mobasher}, B., {et~al.} 2024, \apj, 962, 176, \dodoi{10.3847/1538-4357/ad20ed}

\bibitem[{{Westera} {et~al.}(2002){Westera}, {Lejeune}, {Buser}, {Cuisinier}, \& {Bruzual}}]{Westera2002}
{Westera}, P., {Lejeune}, T., {Buser}, R., {Cuisinier}, F., \& {Bruzual}, G. 2002, \aap, 381, 524, \dodoi{10.1051/0004-6361:20011493}

\bibitem[{{Wise}(2019)}]{Wise2019}
{Wise}, J.~H. 2019, Contemporary Physics, 60, 145, \dodoi{10.1080/00107514.2019.1631548}

\end{thebibliography}
\bibliographystyle{aasjournal}

\clearpage
\appendix
\counterwithin{figure}{section}
\counterwithin{table}{section}

\section{The gold and silver candidates}\label{sec:goldsilver}
This appendix presents detailed information on the gold and silver candidates identified in our study.
Tables~\ref{tab:gold} and \ref{tab:silver} list RA, Dec, rest frame Ly$\alpha$ EW, LyC flux measured in the $u/u^*$ band, UVC flux from the $i$ band, and their flux ratios for the gold and silver candidates.

\begin{figure*}
\includegraphics[width=\linewidth]{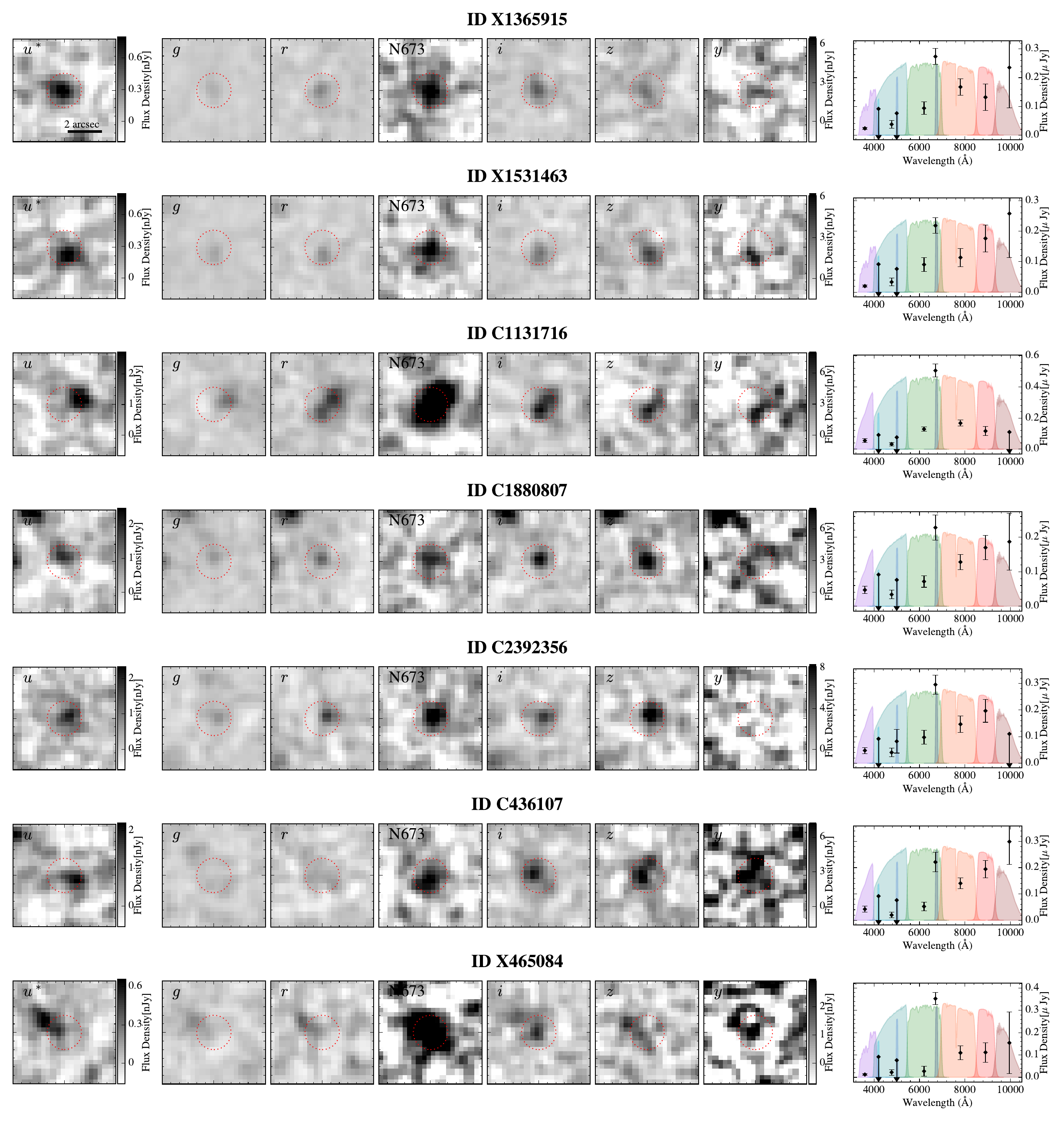}
\caption{Postage stamps (6$\arcsec$ on a side) of six broadbands and one narrowband for seven of the 12 gold candidates and their SEDs. 
The red dotted circle in each panel represents 1$\arcsec$-radius aperture. 
In the rightmost column for SEDs, the shaded curves denote the filter transmission for three narrowbands of N419, N501, and N673 and six broadbands of $u$ (or $u^{*}$), $g$, $r$, $i$, $z$, and $y$ filters; these are the same as Figure~\ref{fig:filter-shapley}. 
When the flux in a band is undetected, it is marked as a $2\sigma$ upper limit. 
For the $u/u^*$ band, this limit was determined by our additional precise background modeling and noise measurement.
Conversely, for the other bands, the limiting magnitude reported by the original survey was used as a reference without any background processing.}
\label{fig:cutout-gold1}
\end{figure*}

\begin{figure*}
\includegraphics[width=\linewidth]{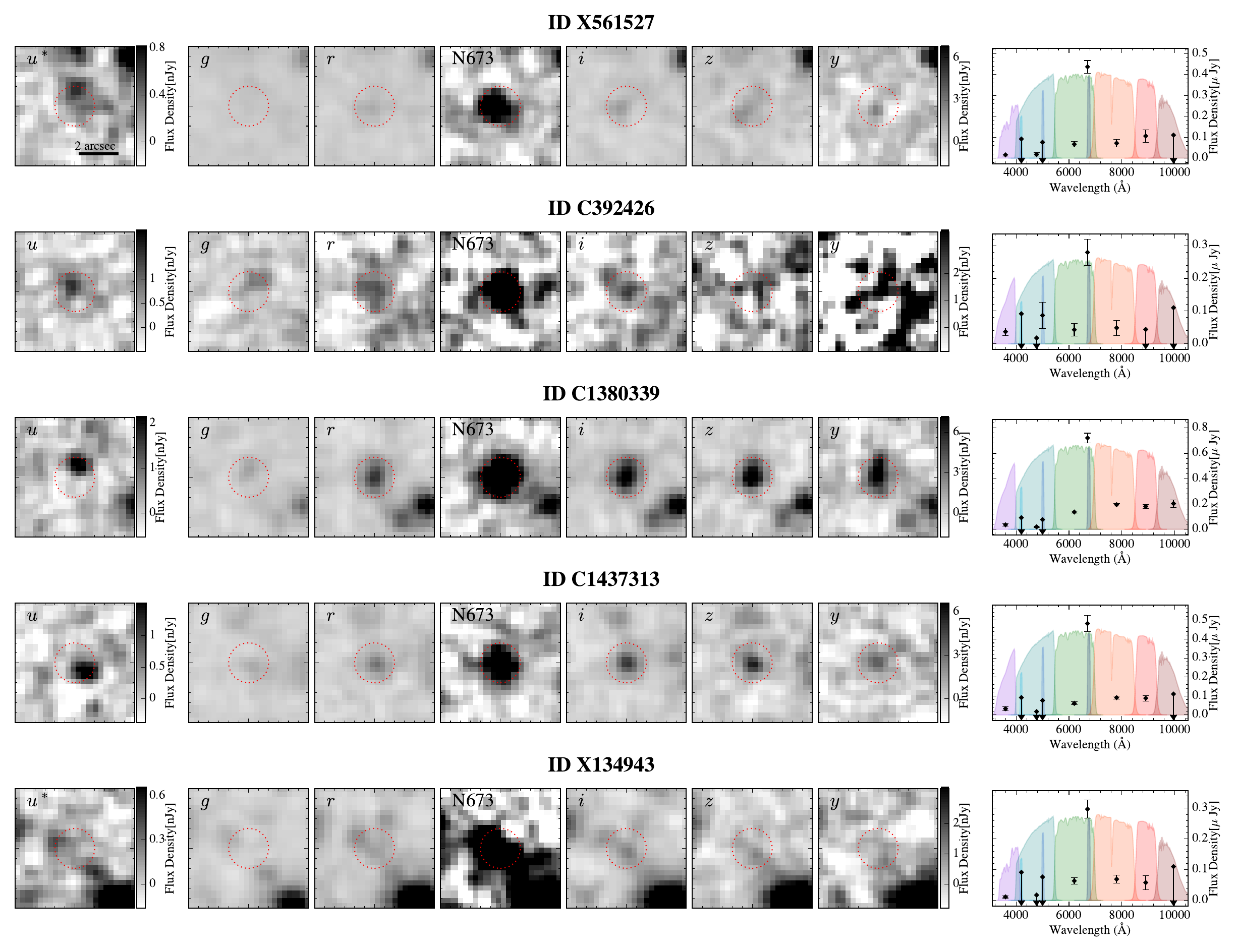}
\caption{Same as \ref{fig:cutout-gold1}, but for the remaining five gold candidates.}
\label{fig:cutout-gold2}
\end{figure*}

For the gold candidates, postage stamps and SEDs are presented in Figures \ref{fig:cutout-gold1} and \ref{fig:cutout-gold2}.
The strong emission in N673 highlights their prominent Ly$\alpha$ emission features, while the $u/u^*$-band emission remains relatively faint, as expected.
The color scale is kept identical across all bands except for the $u/u^*$ band, where it is adjusted to enhance faint signals.
We omit the N419 and N501 cutouts since their shallower depths result in excessive noise; the deeper $u/u^*$-band data provide a more reliable representation of the LyC emission.
The postage stamps show the $u/u^*$-band emission is not always centered on the Ly$\alpha$ or UVC emission.

In particular, contamination from neighbouring sources is likely in the cases of IDs C1131716, X465084, and X134943.
Clear spatial offsets between the $u/u^*$ band and other bands are also seen in IDs C436107, X561527, C392426, C1380339, and C147313.
We note that no obvious sources are present at the offset positions in the other bands, suggesting that the $u/u^*$-band detections are less likely to be due to contamination from neighbouring objects.
The measured spatial offsets range from approximately 0\farcs3 to 1\farcs0 ($\sim$2.0--6.8 kpc at $z\sim4.5$).
These values are comparable to previously reported offsets between ionizing and non-ionizing emission in high-redshift galaxies \citep[e.g.,][]{Micheva2017}, supporting a scenario in which LyC photons originate from spatially localized substructures rather than the main galaxy centroid, as discussed in Section \ref{sec:hst}.
Ultimately, our ground-based imaging is seeing-limited, which prevents us from unambiguously distinguishing between clumpy star-forming regions, minor mergers, or projection effects.
Therefore, while the observed offsets are suggestive of spatially localized LyC escape, higher-resolution imaging and spectroscopy are required to determine the physical origin of the LyC emission.

\begin{deluxetable*}{lcccccc}
\tablecaption{Overview of LyC leaker gold candidates \label{tab:gold}} 
\tablehead{
\colhead{ID}& \colhead{RA [\textdegree ]} & \colhead{Dec [\textdegree ]} & \colhead{EW$_{\rm Ly\alpha}$ [\r{A}]} & \colhead{$F_{\rm LyC}$ [nJy] (S/N)} & \colhead{$F_{\rm UVC}$ [nJy]} & \colhead{$F_{\rm LyC}/F_{\rm UVC}$}
}
\startdata
\textbf{Gold candidates}\\
X1365915& 35.422408& -4.363631& 28.14 $\pm$ 11.28& 23.69 $\pm$ 3.76 (6.31)& 167.67 $\pm$ 28.84& 0.14 $\pm$ 0.03\\
X1531463& 36.024662& -3.698088& 34.01 $\pm$ 14.83& 20.13 $\pm$ 3.84 (5.24)& 113.66 $\pm$ 30.00& 0.18 $\pm$ 0.06\\
C1131716& 149.472678& 2.847194& 63.86 $\pm$ 13.44& 55.29 $\pm$ 12.21 (4.53)& 167.74 $\pm$ 19.71& 0.33 $\pm$ 0.08\\
C1880807& 149.513469& 3.158484& 31.09 $\pm$ 12.90& 47.14 $\pm$ 10.87 (4.34)& 128.12 $\pm$ 22.15& 0.37 $\pm$ 0.11\\
C2392356& 151.199032& 3.107247& 36.68 $\pm$ 13.83& 47.78 $\pm$ 11.31 (4.23)& 146.53 $\pm$ 30.56& 0.33 $\pm$ 0.10\\
C436107& 149.471997& 1.251824& 28.23 $\pm$ 12.20& 42.32 $\pm$ 12.05 (3.51)& 140.39 $\pm$ 20.82& 0.30 $\pm$ 0.10\\
X465084& 35.600483& -5.596006& 68.68 $\pm$ 23.69& 12.21 $\pm$ 3.61 (3.38)& 109.76 $\pm$ 30.97& 0.11 $\pm$ 0.05\\
X561527& 35.304643& -5.157709& 138.85 $\pm$ 36.70& 15.71 $\pm$ 4.75 (3.31)& 71.71 $\pm$ 17.70& 0.22 $\pm$ 0.09\\
C392426& 150.357824& 1.074718& 123.92 $\pm$ 60.56& 36.89 $\pm$ 11.23 (3.28)& 48.54 $\pm$ 23.14& 0.76 $\pm$ 0.43\\
C1380339& 150.699746& 2.335836& 134.14 $\pm$ 18.93& 35.32 $\pm$ 10.87 (3.25)& 194.61 $\pm$ 8.96& 0.18 $\pm$ 0.06\\
C1437313& 149.941651& 2.559852& 127.26 $\pm$ 20.10& 32.69 $\pm$ 10.06 (3.25)& 90.78 $\pm$ 8.67& 0.36 $\pm$ 0.12\\
X134943& 34.593306& -5.389103& 99.38 $\pm$ 24.75& 11.60 $\pm$ 3.80 (3.05)& 69.32 $\pm$ 13.10& 0.17 $\pm$ 0.06\\
\hline
\enddata
\end{deluxetable*}

\begin{deluxetable*}{lcccccc}
\tablecaption{Overview of LyC leaker silver candidates \label{tab:silver}} 
\tablehead{
\colhead{ID}& \colhead{RA [\textdegree ]} & \colhead{Dec [\textdegree ]} & \colhead{EW$_{\rm Ly\alpha}$ [\r{A}]} & \colhead{$F_{\rm LyC}$ [nJy] (S/N)} & \colhead{$F_{\rm UVC}$ [nJy]} & \colhead{$F_{\rm LyC}/F_{\rm UVC}$}
}
\startdata
\textbf{Silver candidates}\\
X1831991& 36.553345& -3.804408& 31.20 $\pm$ 12.96& 23.36 $\pm$ 7.83 (2.98)& 139.83 $\pm$ 34.75& 0.17 $\pm$ 0.07\\
C401192& 149.863386& 1.110779& 45.69 $\pm$ 18.02& 27.80 $\pm$ 9.63 (2.89)& 90.65 $\pm$ 22.51& 0.31 $\pm$ 0.13\\
C436212& 149.488299& 1.251774& 57.32 $\pm$ 21.05& 30.43 $\pm$ 10.61 (2.87)& 79.13 $\pm$ 20.27& 0.38 $\pm$ 0.17\\
X1795029& 36.880217& -3.995438& 31.71 $\pm$ 13.68& 14.31 $\pm$ 4.99 (2.87)& 136.88 $\pm$ 29.71& 0.10 $\pm$ 0.04\\
X113687& 34.078602& -5.487443& 29.54 $\pm$ 13.97& 6.90 $\pm$ 2.52 (2.73)& 146.58 $\pm$ 33.87& 0.05 $\pm$ 0.02\\
C1503106& 150.501421& 2.817368& 96.54 $\pm$ 10.85& 30.08 $\pm$ 11.01 (2.73)& 444.12 $\pm$ 10.41& 0.07 $\pm$ 0.02\\
C781096& 150.879894& 1.578834& 57.48 $\pm$ 22.80& 27.48 $\pm$ 10.13 (2.71)& 57.51 $\pm$ 18.36& 0.48 $\pm$ 0.23\\
C2447714& 151.134583& 3.470776& 57.90 $\pm$ 20.39& 41.54 $\pm$ 15.39 (2.70)& 114.30 $\pm$ 33.09& 0.36 $\pm$ 0.17\\
C687584& 151.080965& 0.979590& 58.84 $\pm$ 11.19& 36.08 $\pm$ 13.56 (2.66)& 295.13 $\pm$ 26.66& 0.12 $\pm$ 0.05\\
X478949& 36.501913& -5.533621& 59.88 $\pm$ 28.20& 12.94 $\pm$ 5.01 (2.58)& 102.06 $\pm$ 40.02& 0.13 $\pm$ 0.07\\
C1203319& 150.770351& 1.631930& 40.50 $\pm$ 11.69& 24.70 $\pm$ 9.95 (2.48)& 137.83 $\pm$ 14.71& 0.18 $\pm$ 0.07\\
X729045& 36.289724& -4.556644& 71.67 $\pm$ 34.51& 9.49 $\pm$ 3.89 (2.44)& 68.90 $\pm$ 29.15& 0.14 $\pm$ 0.08\\
X636938& 36.543445& -4.824855& 119.05 $\pm$ 62.73& 8.15 $\pm$ 3.36 (2.43)& 65.35 $\pm$ 28.85& 0.12 $\pm$ 0.08\\
C696734& 151.038069& 1.024010& 37.55 $\pm$ 12.69& 26.37 $\pm$ 11.00 (2.40)& 188.84 $\pm$ 25.88& 0.14 $\pm$ 0.06\\
X1388995& 36.260620& -4.271983& 101.57 $\pm$ 42.36& 8.58 $\pm$ 3.64 (2.36)& 73.71 $\pm$ 25.41& 0.12 $\pm$ 0.06\\
X1104887& 34.824627& -4.247819& 33.18 $\pm$ 12.19& 21.04 $\pm$ 9.00 (2.34)& 99.38 $\pm$ 13.69& 0.21 $\pm$ 0.10\\
X1123857& 34.951700& -4.162150& 67.88 $\pm$ 16.03& 17.20 $\pm$ 7.36 (2.34)& 117.53 $\pm$ 17.38& 0.15 $\pm$ 0.07\\
X533046& 35.904482& -5.288338& 126.11 $\pm$ 35.82& 7.71 $\pm$ 3.31 (2.33)& 114.63 $\pm$ 27.53& 0.07 $\pm$ 0.03\\
C2167994& 149.780780& 3.390500& 84.62 $\pm$ 22.52& 31.16 $\pm$ 13.76 (2.26)& 114.22 $\pm$ 23.09& 0.27 $\pm$ 0.13\\
X588383& 35.618375& -5.039728& 43.58 $\pm$ 11.46& 7.31 $\pm$ 3.23 (2.26)& 224.16 $\pm$ 30.02& 0.03 $\pm$ 0.02\\
X141294& 34.065190& -5.360431& 108.47 $\pm$ 14.10& 10.79 $\pm$ 4.79 (2.25)& 384.91 $\pm$ 19.48& 0.03 $\pm$ 0.01\\
C515863& 149.083290& 1.487517& 36.73 $\pm$ 11.63& 25.72 $\pm$ 11.56 (2.23)& 201.45 $\pm$ 22.57& 0.13 $\pm$ 0.06\\
C1524586& 149.751189& 3.044464& 215.03 $\pm$ 84.14& 25.44 $\pm$ 11.53 (2.21)& 56.36 $\pm$ 19.05& 0.45 $\pm$ 0.26\\
X1146961& 35.029179& -4.059045& 36.16 $\pm$ 14.25& 19.12 $\pm$ 8.93 (2.14)& 131.19 $\pm$ 29.07& 0.15 $\pm$ 0.08\\
X501442& 35.244831& -5.430170& 36.56 $\pm$ 15.57& 8.68 $\pm$ 4.05 (2.14)& 112.99 $\pm$ 29.06& 0.08 $\pm$ 0.04\\
X151037& 34.988808& -5.315525& 103.35 $\pm$ 30.67& 7.53 $\pm$ 3.52 (2.14)& 52.85 $\pm$ 13.39& 0.14 $\pm$ 0.08\\
C130193& 148.933319& 1.522984& 29.14 $\pm$ 13.54& 24.98 $\pm$ 11.75 (2.13)& 329.95 $\pm$ 81.19& 0.08 $\pm$ 0.04\\
X293391& 34.929435& -4.675471& 165.43 $\pm$ 56.82& 5.16 $\pm$ 2.44 (2.11)& 41.87 $\pm$ 12.90& 0.12 $\pm$ 0.07\\
C459991& 149.643400& 1.347856& 27.92 $\pm$ 10.57& 21.79 $\pm$ 10.33 (2.11)& 180.48 $\pm$ 20.62& 0.12 $\pm$ 0.06\\
X608011& 35.711274& -4.950988& 84.84 $\pm$ 41.49& 10.79 $\pm$ 5.15 (2.10)& 64.91 $\pm$ 32.04& 0.17 $\pm$ 0.11\\
X465839& 35.469374& -5.592572& 170.74 $\pm$ 36.42& 16.91 $\pm$ 8.11 (2.09)& 154.43 $\pm$ 31.03& 0.11 $\pm$ 0.06\\
X974625& 37.260131& -4.856079& 166.66 $\pm$ 105.10& 23.58 $\pm$ 11.41 (2.07)& 50.16 $\pm$ 30.23& 0.47 $\pm$ 0.36\\
X227038& 34.216876& -4.967846& 47.05 $\pm$ 13.43& 6.57 $\pm$ 3.20 (2.05)& 104.41 $\pm$ 13.06& 0.06 $\pm$ 0.03\\
C2128677& 149.866044& 3.222852& 64.50 $\pm$ 20.57& 28.82 $\pm$ 14.09 (2.05)& 104.25 $\pm$ 22.05& 0.28 $\pm$ 0.15\\
X941952& 37.003528& -5.012601& 103.53 $\pm$ 37.33& 21.13 $\pm$ 10.38 (2.04)& 99.12 $\pm$ 30.34& 0.21 $\pm$ 0.12\\
X215798& 34.479484& -5.018344& 177.37 $\pm$ 68.89& 5.95 $\pm$ 2.93 (2.03)& 34.45 $\pm$ 12.16& 0.17 $\pm$ 0.10\\
X409872& 35.583885& -5.846610& 121.66 $\pm$ 28.06& 6.98 $\pm$ 3.46 (2.02)& 167.06 $\pm$ 32.56& 0.04 $\pm$ 0.02\\
C1231364& 150.031148& 1.746568& 73.96 $\pm$ 13.24& 21.24 $\pm$ 10.54 (2.02)& 118.65 $\pm$ 8.42& 0.18 $\pm$ 0.09\\
X997680& 36.946482& -4.747710& 118.09 $\pm$ 50.88& 9.08 $\pm$ 4.51 (2.01)& 62.27 $\pm$ 27.31& 0.15 $\pm$ 0.10\\
\hline
\enddata
\end{deluxetable*}

\clearpage
\section{Forward modeling of LyC detection bias}\label{sec:forwardmodeling}
To quantify the impact of stochastic IGM transmission, intrinsic parameter distributions, and observational selection effects on the observed LyC-to-UVC flux ratios, we performed simple forward-modeling simulations.

For each mock galaxy, we generated intrinsic LyC-to-UVC luminosity ratios, escape fraction, and IGM transmission values from simple parametric distribution intended to capture their broad expected behavior.

The intrinsic luminosity ratio $L_{\rm LyC}/L_{\rm UVC}$ was drawn from a Gaussian distribution with mean 0.3 and standard deviation 0.2, truncated to the physical range $[0,\infty)$, representative of young, moderately metal-poor stellar populations \citep{Steidel2001,Inoue2005,Shapley2006,Grazian2016,Grazian2017,Marchi2017}.
The intrinsic escape fraction $f_{\rm esc}$ was drawn from a Gaussian distribution with mean 0.1 and standard deviation 0.2, truncated to $[0,1]$, consistent with typical values inferred for star-forming galaxies \citep{Robertson2013,Begley2022,Ferrara2025}.

The line-of-sight IGM transmission $T_{\rm IGM}$ was drawn from an exponential-like functional form with an extended high-transmission tail.
When we fix this distribution to the \citet{Inoue2014}-motivated baseline (with a very low mean transmission at 640--680\AA), our simple forward model underpredicts the observed LyC detection fraction.
Therefore, we explored a phenomenological variant in which the scale (mean and width) of the parent transmission distribution is allowed to vary to reproduce the observed detection fraction.

For each realization, the observed LyC flux was computed as
\begin{equation}
    F_{\rm LyC}=F_{\rm UVC}\times (L_{\rm LyC}/L_{\rm UVC}) \times f_{\rm esc} \times T_{\rm IGM}.    
\end{equation}
We adopted the observed UVC flux distribution of our sample and generated mock LyC fluxes accordingly.
Gaussian noise consistent with the measured photometric uncertainties in the CLAUDS $u/u^*$ data was added, and the same LyC detection threshold (e.g., $>3\sigma$) as in the real data was applied.

In this exercise, the best-fitting effective parent mean transmission is 0.22.
We emphasize this value should not be interpreted as the cosmic-mean IGM transmission; rather, it is an empirical parameter that likely absorbs residual uncertainties in the selection function and in the joint (possibly correlated and/or heavy-tailed) distribution of intrinsic LyC production/escape and line-of-sight transmission.
We note that the modest field-to-field variations in the ionizing background and absorber incidence may introduce additional scatter around the cosmic mean within our survey footprint; however, we do not interpret the effective mean value inferred above as a direct measurement of such field-level deviations.

\begin{figure*}
\includegraphics[width=\linewidth]{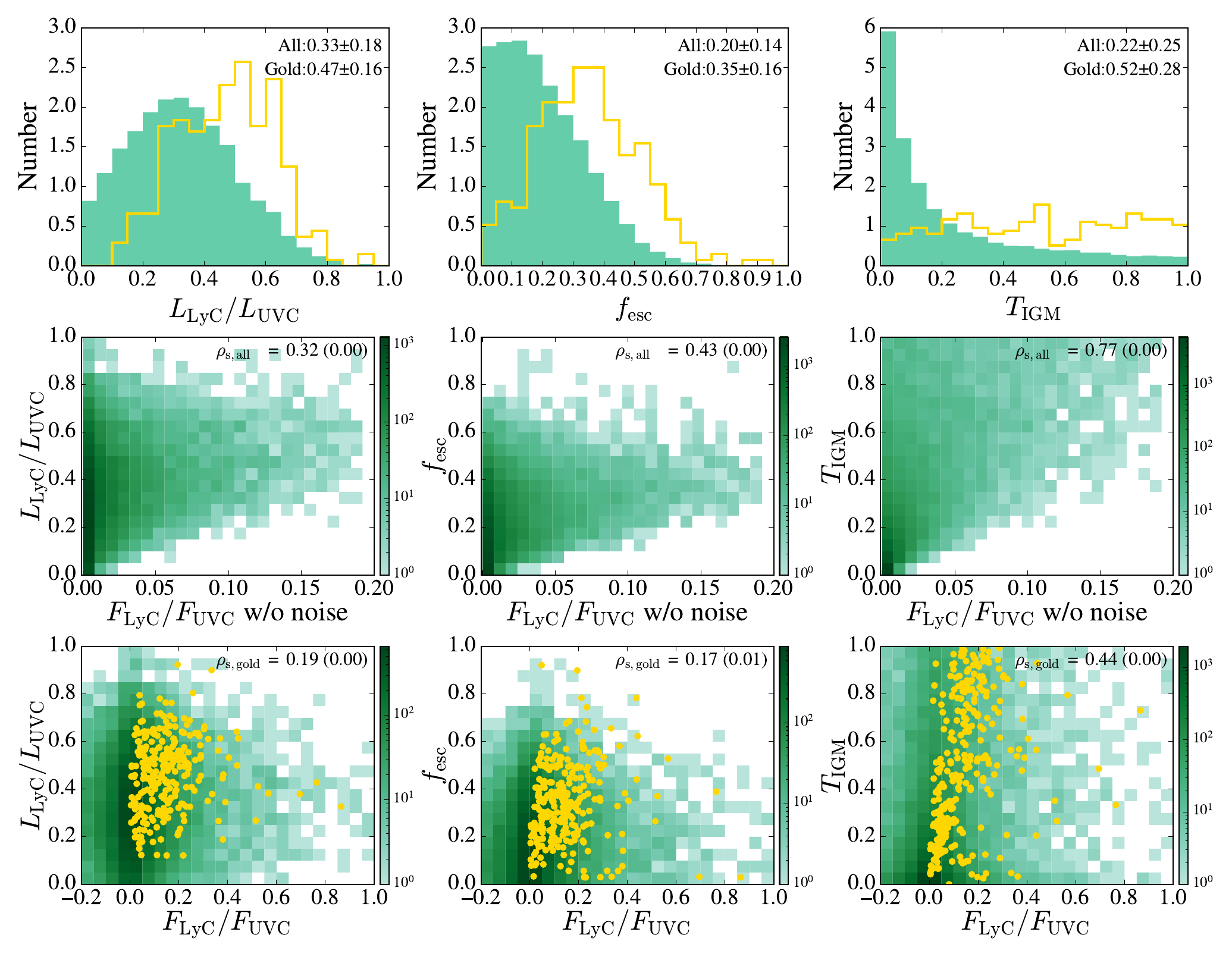}
\caption{Distributions $L_{\rm LyC}/L_{\rm UVC}$, $f_{\rm esc}$, and $T_{\rm IGM}$ for the full mock sample (green filled histograms) and the detected (gold) subsample (empty yellow histograms) are shown in the top panels. The mean and standard deviation are indicated in each panel.
The middle panels present the correlations between the noise-free flux ratio $F_{\rm LyC}/F_{\rm UVC}$ and $L_{\rm LyC}/L_{\rm UVC}$, $f_{\rm esc}$, and $T_{\rm IGM}$ for the full sample. The Spearman rank correlation coefficient ($\rho_{\rm s}$) is shown in each panel, with the corresponding $p$-value provided in parentheses.
The bottom panels show the same correlations, also for the gold subsample, using the $F_{\rm LyC}/F_{\rm UVC}$ after applying observational effects.}
\label{fig:forwardmodeling}
\end{figure*}

Each adopted input distribution is shown as a filled green histogram in the first row of Figure~\ref{fig:forwardmodeling}.
We emphasize that our goal is not to infer the precise IGM transmission distribution, but to illustrate how stochastic transmission and selection bias affect the interpretation of the observed flux ratios.

Before including observational noise and selection effects, the simulated flux ratio $F_{\rm LyC}/F_{\rm UVC}$ shows a very strong correlation with $T_{\rm IGM}$ (correlation coefficient $\rho_{\rm s}\approx0.77$), and only a moderate correlation with $f_{\rm esc}$ ($\rho_{\rm s}\approx0.43$), as shown in the middle row in Figure~\ref{fig:forwardmodeling}.
This indicates that, even in the idealized noiseless case, the flux ratio is primarily modulated by stochastic IGM transmission at $z\sim4.5$.

Once realistic photometric noise and the LyC detection criterion are imposed, the detected subsample becomes systematically biased toward higher values of all three governing parameters (empty yellow histogram).
Photometric LyC surveys have demonstrated that LyC-detected samples are preferentially drawn from unusually transparent sightlines \citep[e.g.,][]{Bassett2021}.
Similarly, the effective $T_{\rm IGM}$ distribution of detected sources in our simulations is strongly skewed toward the high-transmission tail, with detections requiring sampling approximately the top $\sim14\%$ of the parent IGM transmission distribution.

$f_{\rm esc}$ and $L_{\rm LyC}/L_{\rm UVC}$ are also shifted toward higher values in the detected subsample, but their relative shifts are less extreme compared to that of $T_{\rm IGM}$.
This reflects the multiplicative dependence of the observed LyC flux on these quantities, combined with the intrinsically broader and more highly skewed distribution of IGM transmission at $z\sim4.5$.

These simulations demonstrate that, at $z\sim4.5$ and rest-frame 640--680\AA, the observed LyC-to-UVC flux ratio is primarily regulated by stochastic IGM transmission and detection bias.
While the flux ratio retains statistical meaning at the population level, it cannot be interpreted as a direct tracer of the intrinsic escape fraction for individual galaxies.

\end{document}